\documentclass[aip,amsmath,amssymb,reprint]{revtex4-1}

\usepackage{bm}
\usepackage{braket}
\usepackage{graphicx}
\usepackage{hyperref}
\usepackage{mathptmx}
\usepackage{xcolor}

\usepackage[utf8]{inputenc}
\usepackage[T1]{fontenc}

\makeatletter
\def\@email#1#2{
\endgroup
\patchcmd{\titleblock@produce}
{\frontmatter@RRAPformat}
{\frontmatter@RRAPformat{\produce@RRAP{*#1\href{mailto:#2}{#2}}}\frontmatter@RRAPformat}
{}{}
}
\makeatother

\begin{document}

\title[MBPT with hybrid DFT accelerated by ACE]{Many-body perturbation theory with hybrid density functional theory starting points accelerated by adaptively compressed exchange}

\author{Victor~Wen-zhe~Yu}
\affiliation{Materials Science Division, Argonne National Laboratory, Lemont, Illinois 60439, USA}

\author{Marco Govoni}
\email[Corresponding author e-mail:~]{mgovoni@unimore.it}
\affiliation{Department of Physics, Computer Science, and Mathematics, University of Modena and Reggio Emilia, Modena, 41125, Italy}
\affiliation{Materials Science Division, Argonne National Laboratory, Lemont, Illinois 60439, USA}
\affiliation{Pritzker School of Molecular Engineering, The University of Chicago, Chicago, Illinois 60637, USA}

\date{\today}

\begin{abstract}
We report on the use of the adaptively compressed exchange (ACE) operator to accelerate many-body perturbation theory (MBPT) calculations, including G$_0$W$_0$ and the Bethe Salpeter equation (BSE), for hybrid density functional theory starting points. We show that by approximating the exact exchange operator with the low-rank ACE operator, substantial computational savings can be achieved with systematically controllable errors in the quasiparticle energies computed with full-frequency G$_0$W$_0$ and the optical absorption spectra and vertical excitation energies computed by solving the BSE within density matrix perturbation theory. Our implementation makes use of the ACE-accelerated electronic Hamiltonian to carry out both G$_0$W$_0$ and BSE without explicitly computing empty states. We show the robustness of the approach and present the computational gains obtained on both the central processing unit and graphics processing unit nodes. Our work will facilitate the exploration and evaluation of fine-tuned hybrid starting points aimed at enhancing the accuracy of MBPT calculations without involving computationally demanding self-consistency in Hedin's equations.

\end{abstract}

\maketitle

\section{Introduction}

First-principles simulations based on Kohn-Sham density functional theory (KS-DFT)~\cite{dft_hohenberg_1964,dft_kohn_1965} serve as a computational workhorse for predicting the structures and properties of molecules and materials in their ground state. However, KS-DFT fails to quantitatively describe electrons in excited states. The framework of many-body perturbation theory (MBPT) has emerged as a powerful method for characterizing electronic excitations starting from the output of KS-DFT. Within MBPT, the GW method~\cite{gw_hybertsen_1985,gw_hybertsen_1986,gw_strinati_1988,gw_aryasetiawan_1998,gw_golze_2019}, where G is the single-particle Green's function and W is the screened Coulomb interaction, describes charged excitations as measured in direct or inverse photoemission spectroscopy. The Bethe Salpeter equation (BSE)~\cite{bse_salpeter_1951,bse_albrecht_1998,bse_onida_2002} describes neutral excitations inclusive of excitonic effects that can be measured in ultraviolet-visible (UV/VIS) spectroscopy. The application of conventional implementations of MBPT to complex heterogenous systems is hindered by the scaling with respect to system size $N$: $\mathcal{O}(N^4)$ and $\mathcal{O}(N^6)$ for GW and BSE, respectively.

Among efforts to push GW~\cite{gw_bruneval_2008,gw_umari_2009,gw_umari_2010,gw_giustino_2010,gw_berger_2010,gw_kang_2010,gw_foerster_2011,gw_berger_2012,pdep_nguyen_2012,pdep_pham_2013,gw_lambert_2013,stochastic_neuhauser_2014,gw_janssen_2015,gw_bruneval_2016,gw_liu_2016,stochastic_vlcek_2017,stochastic_vlcek_2018,gw_wilhelm_2018,berkelygw_delben_2020,gw_forster_2020,gw_forster_2021,gw_duchemin_2021,gw_wilhelm_2021,gw_ma_2021,gw_leon_2021,gw_leon_2023} and BSE~\cite{bse_rocca_2010,bse_rocca_2012,bse_ping_2012,bse_marsili_2017,bse_krause_2017,bse_bradbury_2022,bse_franzke_2022,bse_merkel_2024} to large systems, we have developed the WEST (Without Empty STates) software~\cite{west_website} to solve MBPT within a plane wave pseudopotential implementation of DFT. The code is capable of computing electron self-energies, e.g., within the full-frequency G$_0$W$_0$ method~\cite{west_govoni_2015,soc_scherpelz_2016,gw100_govoni_2018,gw_ma_2019,gw_yang_2019,west_yu_2022}, and neutral excitation processes within the BSE~\cite{bse_nguyen_2019,bse_dong_2021,bse_yu_2024}, time-dependent density functional theory (TDDFT)~\cite{tddft_jin_2022,tddft_jin_2023}, and quantum defect embedding theory~\cite{qdet_ma_2020,qdet_ma_2021,qdet_sheng_2022}. WEST adopts algorithms that circumvent the explicit summation over empty states, a common bottleneck of conventional implementations of MBPT that hinders the calculation of the Green's function and of response functions for large systems. Starting from a DFT electronic structure computed using the semi-local Perdew-Burke-Ernzerhof (PBE)~\cite{pbe_perdew_1996} exchange-correlation (XC) functional, we demonstrated that the G$_0$W$_0$ implementation in WEST scales up to 524,288 central processing unit (CPU) cores~\cite{west_govoni_2015} and 25,920 graphics processing units (GPUs)~\cite{west_yu_2022,roadmap_gavini_2023}.

The results of G$_0$W$_0$ and BSE are sensitive to the choice of the XC functional employed to compute the DFT single-particle energies and wave functions, which are used as the starting point to compute, for instance, G and W. While the dependence on the starting point can be reduced or eliminated -- although at a much higher computational cost -- by incorporating various levels of self-consistency in Hedin's equations~\cite{hedin_hedin_1965}, reaching self-consistency does not necessarily lead to better accuracy, as several comparisons with experiments or with quantum chemistry methods such as coupled cluster singles, doubles, and perturbative triples [CCSD(T)]~\cite{gw_golze_2019,scgw_schone_1998,exx_fuchs_2007,scgw_shishkin_2007,exx_marom_2012,exx_knight_2016,scgw_grumet_2018} have shown. XC functionals provide parameters that can be tuned to provide an optimal starting point for one-shot MBPT calculations~\cite{exx_gerosa_2017}. In the generalized KS framework, hybrid functionals mix a fraction of the exact Fock exchange (EXX) with the local exchange~\cite{gks_becke_1993,gks_seidl_1996}, leading to an improved description of the electronic structure for both isolated and extended systems. The pursuit of non-empirical, universally applicable hybrid functionals remains an active research area~\cite{hybrid_marques_2011,ddh_skone_2014,ddh_skone_2016,hybrid_brawand_2017,hybrid_zheng_2019,hybrid_wing_2021,hybrid_shukla_2022,hybrid_gomez_2023,hybrid_zhan_2023}.

The adoption of hybrid DFT as the starting point in MBPT calculations has been increasingly reported for GW~\cite{gw_ma_2019,exx_fuchs_2007,exx_marom_2012,exx_knight_2016,exx_rinke_2005,exx_marom_2011,exx_korzdorfer_2012,exx_bruneval_2013,exx_atalla_2013,exx_luftner_2014,exx_kang_2014,exx_chen_2014,exx_gallandi_2015,exx_gallandi_2016,exx_caruso_2016,exx_dauth_2016,exx_bois_2017,exx_chen_2017,exx_leppert_2019,exx_gant_2022} and BSE~\cite{exx_mckeon_2022,exx_ohad_2023}. In global hybrid functionals of the $\alpha$-PBE0 form, the parameter $\alpha \in [0,1]$ determines the fraction of EXX introduced by the functional. For molecular systems, K\"{o}rzd\"{o}rfer and Marom~\cite{exx_korzdorfer_2012} iteratively tuned $\alpha$ to find a KS eigenspectrum as close as possible to the G$_0$W$_0$ quasiparticle (QP) spectrum. The computed photoemission spectra agreed well with experiments for several organic molecules. Dauth et al.~\cite{exx_dauth_2016} and Caruso et al.~\cite{exx_caruso_2016} chose $\alpha$ that minimizes the deviation from the straight line error, i.e., the spurious nonlinearity of the total energy as a function of fractional particle number. Ionization potentials (IPs) computed in this way were found to agree with the CCSD(T) reference within 0.2 eV. Atalla et al.~\cite{exx_atalla_2013} proposed to minimize the G$_0$W$_0$ QP correction for the highest occupied molecular orbital (HOMO) as a function of $\alpha$, obtaining IPs deviating from CCSD(T) reference by $\sim$3\%. For aqueous solutions, Gaiduk et al.~\cite{exx_gaiduk_2016} and Pham et al.~\cite{exx_pham_2017} used $\alpha$ equal to the inverse of the macroscopic dielectric constant, obtaining G$_0$W$_0$ electronic structure and photoelectron spectrum in good agreement with experiments.

In range-separated hybrid functionals, the exchange is partitioned into short-range and long-range components. Two parameters $\alpha$ and $\beta$ control the fraction of EXX in the long-range and short-range parts, respectively, while the $\gamma$ parameter controls the separation range. For molecular systems, Bruneval and Marques~\cite{exx_bruneval_2013} benchmarked Hartree-Fock (HF), local density approximation (LDA), generalized gradient approximation (GGA), and hybrid starting points for G$_0$W$_0$ and showed that global and range-separated hybrid starting points outperform HF, LDA, and GGA starting points in predicting IPs.
Gallandi et al.~\cite{exx_gallandi_2015,exx_gallandi_2016} tuned the range separation parameter to minimize the G$_0$W$_0$ QP correction for the HOMO, obtaining mean absolute errors below 0.1 eV for IPs and electron affinities. For solids, Chen and Pasquarello~\cite{exx_chen_2017} used G$_0$W$_0$@HSE06, with $\alpha$ tuned to reproduce experimental bandgaps, to compute defect levels originating from point defects in semiconductors. They obtained defect levels with an accuracy of $\sim$0.1 eV with respect to experimental data. Fuchs et al.~\cite{exx_fuchs_2007} found that for materials with shallow d states, bandgaps predicted by G$_0$W$_0$@HSE03 were in good agreement with experiment, whereas G$_0$W$_0$@PBE0 overestimated bandgaps possibly due to underscreening. Gant et al.~\cite{exx_gant_2022} reported the use of a Wannier-localized optimally tuned screened range-separated hybrid functional as the starting point for G$_0$W$_0$, which, for a range of semiconductors and insulators, predicted bandgaps with a level of accuracy comparable to that of more involved methods such as QP self-consistent GW. The same hybrid functional was later used in G$_0$W$_0$-BSE~\cite{exx_mckeon_2022,exx_ohad_2023}, yielding IPs and low-lying neutral excitations of molecules as well as optical absorption spectra of metal oxides in reasonable agreement with experimental or computational reference.

Compared to semi-local functionals, hybrid functionals typically entail a significantly higher computational cost, primarily due to the long-range nature of the EXX operator, i.e., $V_{\mathrm{EXX}}$. To address this challenge, various techniques have been proposed to reduce the computational cost associated with hybrid functionals in plane wave pseudopotential implementations of DFT~\cite{exx_wu_2009,exx_gygi_2013,exx_damle_2015,ace_lin_2016,exx_dawson_2015,exx_hu_2017,exx_ko_2020}. One particularly efficient and easy-to-implement method is the adaptively compressed exchange (ACE) developed by Lin~\cite{ace_lin_2016}. The computational gain introduced by the ACE method stems from a low-rank approximation, $V_{\mathrm{ACE}}$, of the EXX operator $V_{\mathrm{EXX}}$. The low-rank approximation is exact when applied to the occupied KS wave functions. Unlike techniques that accelerate the evaluation of $V_{\mathrm{EXX}}$ by exploiting the near-sightedness principle, which is only applicable to gapped systems, the ACE method does not rely on any localization properties, thus it can be applied to insulating, semiconducting, and metallic systems. By implementing the ACE method, hybrid DFT computations become only marginally more demanding than those employing semi-local functionals. Originally proposed for ground-state hybrid DFT and HF calculations, the ACE method has been extended to excited-state calculations using hybrid TDDFT~\cite{tddft_jin_2023,ace_wu_2022,ace_liu_2022,ace_li_2023,ace_chen_2023}.

Separately from the development of hybrid functionals, MBPT has been extensively used to obtain accurate electronic structure or absorption spectra. Explicit summations over empty states in MBPT, appearing, for instance, in the calculation of G and W, hinder the applicability of MBPT to large heterogeneous systems. Iterative solvers within the framework of density functional perturbation theory (DFPT) for G$_0$W$_0$~\cite{dielectric_wilson_2008,dielectric_wilson_2009,gw_umari_2009,gw_umari_2010,pdep_nguyen_2012,pdep_pham_2013,west_govoni_2015,west_yu_2022} or density matrix perturbation theory (DMPT) for BSE (and TDDFT)~\cite{bse_rocca_2010,bse_rocca_2012,bse_ping_2012,bse_nguyen_2019,tddft_jin_2023,turbotddft_malcioglu_2011} were used to avoid the problem caused by explicit summations over empty states. In such iterative solvers, the KS (or generalized KS) Hamiltonian $H_{\mathrm{KS}}$ is never represented as a dense matrix, but, thanks to fast Fourier transform (FFT) operations that scale as $\mathcal{O}(N\mathrm{log}N)$, $H_{\mathrm{KS}}$ is instead applied to trial functions using the dual space technique, i.e., the kinetic part of $H_{\mathrm{KS}}$ is applied in Fourier space, while the local potential is applied in real space. When $H_{\mathrm{KS}}$ contains the EXX term $V_{\mathrm{EXX}}$, i.e., when a hybrid starting point is used to carry out MBPT, the evaluation of $V_{\mathrm{EXX}}$ in $H_{\mathrm{KS}}$ stands out as a major computational bottleneck.

In this work, we combine the ACE approximation and recently developed methods for MBPT that allow us to apply the full-frequency G$_0$W$_0$ or the BSE method to heterogeneous complex systems using hybrid functional starting points. We demonstrate a significant speedup of over an order of magnitude by approximating $V_{\mathrm{EXX}}$ with $V_{\mathrm{ACE}}$, while maintaining systematically controllable errors in the QP energies computed with G$_0$W$_0$ and the optical absorption spectra and vertical excitation energies (VEEs) computed by solving the BSE, as implemented in the WEST code. Our approach showcases a promising avenue for efficiently performing MBPT calculations with hybrid functional starting points.

\section{Method}
\label{sec:method}

We briefly introduce the G$_0$W$_0$ and BSE methods and the ACE approximation to the EXX operator, then we discuss how the ACE is combined with MBPT to accelerate G$_0$W$_0$ and BSE with hybrid starting points. We focus the discussion on algorithms for MBPT that circumvent the explicit summation over empty states and the storage and inversion of large dielectric matrices, as implemented in the WEST code. While WEST supports spin polarization and $k$-point sampling for G$_0$W$_0$ and spin polarization for BSE, the spin and $k$-point indices are omitted for simplicity.

In KS-DFT~\cite{dft_hohenberg_1964,dft_kohn_1965}, the ground state of a system of interacting electrons in the external field of the ions may be obtained by solving a set of single-particle equations,
\begin{equation}
\label{eq:ks}
H_{\mathrm{KS}} \psi_v = \varepsilon_v \psi_v \,,
\end{equation}
where $\psi_v$ and $\varepsilon_v$ correspond to the wave function and energy of the $v$-th KS state, respectively. The KS Hamiltonian, $H_{\mathrm{KS}}$, includes the single-particle kinetic energy operator and the Hartree, external (ionic), and XC potential operators. QP states may be obtained by solving a similar set of equations,
\begin{equation}
H_{\mathrm{QP}} \psi_v^{\mathrm{QP}} = \varepsilon_v^{\mathrm{QP}} \psi_v^{\mathrm{QP}} \,,
\end{equation}
where the QP Hamiltonian, $H_{\mathrm{QP}}$, is obtained from $H_{\mathrm{KS}}$ by replacing the XC potential $v_{\mathrm{xc}}$ with the electron self-energy $\Sigma$. The latter is a frequency-dependent and nonlocal operator that may be expressed as
\begin{equation}
\Sigma = i G W \Gamma \,,
\end{equation}
with $G$, $W$, and $\Gamma$ being the Green's function, the screened Coulomb interaction, and the vertex operator, respectively. Within the G$_0$W$_0$ approximation~\cite{gw_hybertsen_1985,gw_hybertsen_1986,gw_aryasetiawan_1998}, $\Gamma$ is treated as the identity and the self-energy is evaluated as
\begin{equation}
\Sigma (\mathbf{r}, \mathbf{r}'; \omega) = i \int_{- \infty}^{+ \infty} \frac{\mathrm{d} \omega'}{2 \pi} G_0 (\mathbf{r}, \mathbf{r}'; \omega + \omega') W_0 (\mathbf{r}, \mathbf{r}'; \omega') \,.
\label{eq:sigmag0w0}
\end{equation}
Within the full-frequency G$_0$W$_0$ method, the frequency integration in equation~\eqref{eq:sigmag0w0} is evaluated numerically with the contour deformation technique~\cite{contour_godby_1988,contour_lebegue_2003}. The non-self-consistent $G_0$ and $W_0$ are computed from the KS wave functions and energies obtained by solving equation~\eqref{eq:ks} with the chosen XC functional. QP energies, corresponding to charged excitations measured in photoelectron spectroscopy, are found using perturbation theory starting from the solution of equation~\ref{eq:ks},
\begin{align}
\varepsilon_v^{\mathrm{QP}} & = \varepsilon_v + \braket{\psi_v | \Delta_{\mathrm{QP}} | \psi_v} \nonumber \\
& = \varepsilon_v + \braket{\psi_v | \Sigma (\varepsilon_v^{\mathrm{QP}}) - v_{\mathrm{xc}} | \psi_v} \,,
\label{eq:qp}
\end{align}
where $\Delta_{\mathrm{QP}} = H_{\mathrm{QP}} - H_{\mathrm{KS}}$.

Neutral excitations measured in UV/VIS spectroscopy can be obtained from the imaginary part of the frequency-dependent macroscopic inverse dielectric function in the long wavelength limit~\cite{bse_albrecht_1998,bse_onida_2002}, i.e., $\mathrm{Abs}(\omega) = (\mathrm{Im} \{ \epsilon^{-1}_{\mathbf{G}=\mathbf{0}, \mathbf{G'}=\mathbf{0}}(\mathbf{q}\to 0, \omega)\} )^{-1}$, where $\mathbf{G}$ and $\mathbf{G'}$ are reciprocal lattice vectors and
\begin{equation}
\epsilon^{-1}(\omega) = 1 + v\chi(\omega) \,.
\label{eq:epsm1}
\end{equation}
In equation~\ref{eq:epsm1}, $v$ is the bare Coulomb potential, $\chi(1,2) = L(1,1^+,2,2^+)$ is the density-density response function of the material, and $L(1,2,3,4)$ is the two-particle propagator. The latter satisfies the BSE~\cite{bse_albrecht_1998,bse_onida_2002},
\begin{equation}
L = L_0 + L_0 \Xi L \,,
\end{equation}
where $L_0(1,2,3,4) = -i G(1,3) G(4,2)$ is the free two-particle propagator and the BSE kernel is
\begin{equation}
\Xi(1,2,3,4) = v(1,3) \delta(1,2) \delta(3,4) + i \frac{\delta \Sigma(1,2)}{\delta G(3,4)} \,.
\label{eq:BSE}
\end{equation}
The BSE is numerically solved by inserting equation~\ref{eq:sigmag0w0} into equation~\ref{eq:BSE}, and neglecting retardation effects in $W$ and the derivative of $W$ with respect to $G$.

When a hybrid functional is used, the generalized KS Hamiltonian $H_{\mathrm{KS}}$ in equation~\ref{eq:ks} contains the EXX operator $V_{\mathrm{EXX}}$, defined as
\begin{equation}
V_{\mathrm{EXX}} (\mathbf{r}, \mathbf{r}') = - \sum_{v=1}^{N_{\mathrm{occ}}} \frac{\psi_v (\mathbf{r}) \psi^*_v (\mathbf{r}')}{|\mathbf{r} - \mathbf{r}'|} \,,
\end{equation}
where $N_{\mathrm{occ}}$ is the number of occupied KS states.
Lin~\cite{ace_lin_2016} proposed a low-rank approximation to $V_{\mathrm{EXX}}$, known as the ACE operator $V_{\mathrm{ACE}}$ and defined as
\begin{equation}
V_{\mathrm{ACE}} (\mathbf{r},\mathbf{r}') = - \sum_{k=1}^{N_{\mathrm{ACE}}} \xi_k (\mathbf{r}) \xi^*_k (\mathbf{r}') \,,
\label{eq:ace}
\end{equation}
where $N_{\mathrm{ACE}}$ is the number of KS wave functions included in the summation that defines the operator. The ACE orbitals $\{\xi_k\}$ are obtained as
\begin{equation}
\xi_k (\mathbf{r}) = \sum_{i=1}^{N_{\mathrm{ACE}}} P_i (\mathbf{r}) \left( L^{-T} \right)_{ik} \,,
\end{equation}
\begin{equation}
P_i (\mathbf{r}) = \int V_{\mathrm{EXX}}(\mathbf{r}, \mathbf{r}') \psi_i (\mathbf{r}') \mathrm{d} \mathbf{r}' \,,
\end{equation}
and the lower triangular matrix $L$ is obtained from the Cholesky decomposition of $-M = L L^T$, with matrix elements
\begin{equation}
M_{ij} = \int \psi^*_i (\mathbf{r}) P_j (\mathbf{r}) \mathrm{d} \mathbf{r} \,.
\end{equation}
The ACE method was originally proposed in the context of ground-state hybrid DFT and HF, where $V_{\mathrm{ACE}}$ exactly reproduces the full $V_{\mathrm{EXX}}$ when applied to the lowest $N_{\mathrm{ACE}}$ KS wave functions. After its construction, the application of $V_{\mathrm{ACE}}$ is computationally inexpensive at a cost on par with the application of the non-local part of the pseudopotential. In particular, the computational cost of applying $V_{\mathrm{ACE}}$ to an arbitrary function scales as $\mathcal{O}(N_{\mathrm{ACE}} N_{\mathrm{pw}})$, which is more favorable than the $\mathcal{O}(N_{\mathrm{occ}} N_{\mathrm{pw}} \mathrm{log} N_{\mathrm{pw}})$ scaling of directly applying $V_{\mathrm{EXX}}$, where $N_{\mathrm{pw}}$ is the number of plane waves to represent the wave functions as dictated by the chosen kinetic energy cutoff. Adopting the ACE method in hybrid DFT leads to only a marginal increase in computational cost compared to semi-local DFT.

As $N_{\mathrm{ACE}} \rightarrow N_{\mathrm{pw}}$, $V_{\mathrm{ACE}}$ converges to $V_{\mathrm{EXX}}$. Therefore, even in excited-state calculations, there exists a minimum value of $N_{\mathrm{ACE}}$ that ensures the desired accuracy. Recent studies~\cite{tddft_jin_2023,ace_wu_2022,ace_liu_2022,ace_li_2023,ace_chen_2023} have demonstrated that $V_{\mathrm{ACE}}$, when constructed with $N_{\mathrm{ACE}} > N_{\mathrm{occ}}$, can indeed be applied to speed up excited-state calculations within the framework of hybrid TDDFT for computing VEEs and analytical forces.

In the next sections, we show that the ACE method can be used to speed up full-frequency G$_0$W$_0$ (equation~\ref{eq:qp}) and BSE (equation~\ref{eq:epsm1}) calculations for large systems whenever hybrid functionals are used in the DFT starting point. The computational cost and accuracy of the ACE approximation can be systematically controlled through a single parameter, $N_{\mathrm{ACE}}$. A significant speedup can be achieved with $N_{\mathrm{ACE}}$ being just two to four times $N_{\mathrm{occ}}$, while introducing negligible errors in QP energies, absorption spectra, and excitation energies.

\subsection{ACE-accelerated low-rank decomposition of $W$}
\label{sec:pdep}

The screened coulomb potential, $W$, is a quantity that needs to be computed from first principles in both G$_0$W$_0$ and BSE. We partition $W$ into two parts, $W = v + W_p$, where $v$ is the bare Coulomb potential, and the frequency-dependent part $W_p$ is related to the density-density response function $\chi$ as
\begin{equation}
W_p = v \chi v = v^{1/2} \bar{\chi} v^{1/2} \,,
\label{eq:wp}
\end{equation}
with $\bar{\chi}$ being the symmetrized counterpart of $\chi$. To compute $\bar{\chi}$ at zero frequency without explicitly computing many empty states or inverting large dielectric matrices, we use the projective dielectric eigenpotentials (PDEP) technique~\cite{dielectric_wilson_2008,dielectric_wilson_2009,pdep_nguyen_2012,pdep_pham_2013}, which iteratively diagonalizes the symmetrized irreducible density-density response function $\bar{\chi}^0$ to obtain its leading eigenfunctions $\{\bar{\varphi}_i\}$. The density response is evaluated within DFPT~\cite{dfpt_baroni_1987,dfpt_baroni_2001}. Starting from a set of trial functions $\{\bar{\varphi}_i\}$, we take each function and use it as a perturbation, i.e., $v_i^{\mathrm{pert}} = \bar{\varphi}_i$. For each perturbation, we compute the linear change $\Delta \psi_v^i$ of each occupied state $\psi_v$ of the unperturbed system using the Sternheimer equation~\cite{sternheimer_sternheimer_1954}
\begin{equation}
\label{eq:sternheimer}
(H_{\mathrm{KS}} - \varepsilon_v) \Delta \psi_v^i = -P_{\mathrm{c}} v_i^{\mathrm{pert}} \psi_v \,,
\end{equation}
where $P_{\mathrm{c}} = 1 - \sum_{v=1}^{N_{\mathrm{occ}}} \ket{\psi_v} \bra{\psi_v}$ is the projector onto the unoccupied KS subspace. $\Delta \psi_v^i$ is determined by solving a sparse linear system with the conjugated gradient method. The ACE method of equation~\ref{eq:ace} is used to approximate the computationally demanding $V_{\mathrm{EXX}}$ term in $H_{\mathrm{KS}}$ of equation~\ref{eq:sternheimer}, reducing the computational cost of applying $H_{\mathrm{KS}}$ to $\Delta \psi_v^i$ and, hence, accelerating the computation of $\Delta \psi_v^i$.

The linear change of the density caused by the $i$-th perturbation is
\begin{equation}
\Delta n_i = \sum_{v=1}^{N_{\mathrm{occ}}} \ket{\Delta \psi_v^i} \bra{\psi_v} + \mathrm{c.c.} \,,
\end{equation}
and the matrix elements of $\bar{\chi}^0$ are, by definition,
\begin{equation}
\bar{\chi}^0_{ij} = \braket{\bar{\varphi}_i | \Delta n_j} \,.
\end{equation}
Using the leading eigenfunctions of $\bar{\chi}^0$ at zero frequency, i.e., $\{\bar{\varphi}_i\}$ referred to as the PDEP basis set, $W_p$ can be conveniently expressed with the following low-rank representation:
\begin{equation}
\label{eq:pdep}
W_p (\omega) = \Phi(\omega) + \frac{1}{\Omega} \sum_{ij}^{N_{\mathrm{PDEP}}} \ket{\bar{\varphi}_i} \Lambda_{ij} (\omega) \bra{\bar{\varphi}_j} \,,
\end{equation}
where the $N_{\mathrm{PDEP}} \times N_{\mathrm{PDEP}}$ matrix $\Lambda_{ij} (\omega)$ represents $W_p (\omega)$ on the PDEP basis set, $N_{\mathrm{PDEP}}$ denotes the size of the PDEP basis set, the first term on the right-hand side of equation~\ref{eq:pdep}, $\Phi(\omega) = v^{1/2} \bar{\chi}_{\mathbf{G}=0,\mathbf{G'}=0}(\omega) v^{1/2}$, takes into account the long-range dielectric response, and $\Omega$ is the volume of the simulation cell. We note that within the random phase approximation, we have $\Lambda_{ij}(\omega=0) = \frac{\lambda_i}{1 - \lambda_i} \delta_{ij}$, where $\lambda_i$ is the $i$-th eigenvalue of $\bar{\chi}^0$. The parameter $N_{\mathrm{PDEP}}$ controls the accuracy of the low-rank decomposition in equation~\ref{eq:pdep}. Benchmarks~\cite{west_govoni_2015,soc_scherpelz_2016,gw100_govoni_2018} suggest that converged results can be obtained with $N_{\mathrm{PDEP}}$ being just a few times the number of electrons. The computational complexity of the PDEP method scales as $N_{\mathrm{occ}}^2 \times N_{\mathrm{PDEP}} \times N_{\mathrm{pw}}$, which is more favorable than that of the conventional Adler-Wiser formula~\cite{dielectric_adler_1962,dielectric_wiser_1963} that scales as $N_{\mathrm{occ}} \times N_{\mathrm{empty}} \times N_{\mathrm{pw}}^2$, where $N_{\mathrm{empty}}$ is the number of empty states, and $N_{\mathrm{pw}} \gg N_{\mathrm{PDEP}}$ in practice.

\subsection{ACE-accelerated full-frequency G$_0$W$_0$ self-energy}
\label{sec:g0w0}

Using equation~\eqref{eq:wp}, the self-energy operator, $\Sigma$, is computed as the sum of the exchange self-energy operator,
\begin{equation}
\Sigma_{\mathrm{X}} (\mathbf{r}, \mathbf{r}') = i \int_{- \infty}^{+ \infty} \frac{\mathrm{d} \omega'}{2 \pi} G_0 (\mathbf{r}, \mathbf{r}'; \omega + \omega') v (\mathbf{r}, \mathbf{r}') \,,
\label{eq:sigmax}
\end{equation}
and the correlation self-energy operator,
\begin{equation}
\Sigma_{\mathrm{C}} (\mathbf{r}, \mathbf{r}'; \omega) = i \int_{- \infty}^{+ \infty} \frac{\mathrm{d} \omega'}{2 \pi} G_0 (\mathbf{r}, \mathbf{r}'; \omega + \omega') W_p (\mathbf{r}, \mathbf{r}'; \omega') \,.
\end{equation}
We write the Lehmann representation of $G_0$ in terms of projector operators,
\begin{equation}
\label{eq:g0}
G_0 (\mathbf{r}, \mathbf{r}'; \omega) = P_{\mathrm{v}} O_{\mathrm{KS}} (\omega - i \eta) P_{\mathrm{v}} + P_{\mathrm{c}} O_{\mathrm{KS}}(\omega + i \eta) P_{\mathrm{c}} \,,
\end{equation}
where $P_{\mathrm{v}} = \sum_{v=1}^{N_{\mathrm{occ}}} \ket{\psi_v} \bra{\psi_v}$ is the projector onto the occupied KS subspace, $O_{\mathrm{KS}} (\omega) = (\omega - H_{\mathrm{KS}})^{-1}$, and $\eta$ is a small positive number. Combining equations~\ref{eq:sigmax} and \ref{eq:g0}, it can be shown that the exchange part of the self-energy yields the EXX potential. Combining equations~\ref{eq:pdep} and \ref{eq:g0}, the correlation self-energy becomes
\begin{equation}
\braket{\psi_v | \Sigma_{\mathrm{C}} (\omega) | \psi_v} = A_v (\omega) + B_v (\omega) + C_v (\omega) + D_v (\omega) \,,
\end{equation}
with
\begin{widetext}
\begin{equation}
A_v (\omega) = i \int_{- \infty}^{+ \infty} \frac{\mathrm{d} \omega'}{2 \pi} \Phi (\omega') \braket{\psi_v | P_{\mathrm{v}} O_{\mathrm{KS}} (\omega + \omega' - i \eta) P_{\mathrm{v}} | \psi_v} \,,
\end{equation}
\begin{equation}
B_v (\omega) = i \int_{- \infty}^{+ \infty} \frac{\mathrm{d} \omega'}{2 \pi} \Phi (\omega') \braket{\psi_v | P_{\mathrm{c}} O_{\mathrm{KS}} (\omega + \omega' + i \eta) P_{\mathrm{c}} | \psi_v} \,,
\end{equation}
\begin{equation}
C_v (\omega) = \frac{i}{\Omega} \int_{- \infty}^{+ \infty} \frac{\mathrm{d} \omega'}{2 \pi} \sum_{i, j}^{N_{\mathrm{PDEP}}} \Lambda_{ij} (\omega') \braket{\psi_v \tilde{\varphi}_i | P_{\mathrm{v}} O_{\mathrm{KS}} (\omega + \omega' - i \eta) P_{\mathrm{v}} | \psi_v \tilde{\varphi}_j} \,,
\end{equation}
\begin{equation}
D_v (\omega) = \frac{i}{\Omega} \int_{- \infty}^{+ \infty} \frac{\mathrm{d} \omega'}{2 \pi} \sum_{i, j}^{N_{\mathrm{PDEP}}} \Lambda_{ij} (\omega') \braket{\psi_v \tilde{\varphi}_i | P_{\mathrm{c}} O_{\mathrm{KS}} (\omega + \omega' + i \eta) P_{\mathrm{c}} | \psi_v \tilde{\varphi}_j} \,.
\end{equation}
\end{widetext}
The contributions from the occupied KS states, $A_v$ and $C_v$, can be written in a generic form as
\begin{align}
M_{v; LR} (\omega) & = \braket{L | P_{\mathrm{v}} O_{\mathrm{KS}} (\omega) P_{\mathrm{v}} | R} \nonumber \\
& = \sum_{v=1}^{N_{\mathrm{occ}}} \frac{\braket{L | \psi_v} \braket{\psi_v | R}}{\omega - \varepsilon_v} \,,
\end{align}
where $\ket{L}$ and $\ket{R}$ are two generic vectors, and the evaluation of $M_{v; LR} (\omega)$ only requires the occupied KS states. The contributions from the unoccupied KS states, $B_v$ and $D_v$, can be written in a generic form as
\begin{align}
M_{c; LR} (\omega) & = \braket{L | P_{\mathrm{c}} O_{\mathrm{KS}} (\omega) P_{\mathrm{c}} | R} \nonumber \\
& = \braket{L | P_{\mathrm{c}} (\omega - H_{\mathrm{KS}})^{-1} P_{\mathrm{c}} | R} \,.
\end{align}
To evaluate $M_{c; LR}$ with only the occupied KS states, we construct a so-called Lanczos chain of vectors~\cite{lanczos_walker_2006,lanczos_rocca_2008},
\begin{equation}
\label{eq:q}
\ket{q_{n+1}} = \frac{1}{\beta_{n+1}} [(H_{\mathrm{KS}} - \alpha_n) \ket{q_n} - \beta_n \ket{q_{n-1}}] \,,
\end{equation}
which is initialized with $\ket{q_0} = 0$ and $\ket{q_1} = P_{\mathrm{c}} \ket{R}$, and
\begin{equation}
\label{eq:alpha}
\alpha_n = \braket{q_n | H_{\mathrm{KS}} | q_n} \,,
\end{equation}
\begin{equation}
\label{eq:beta}
\beta_{n+1} = || (H_{\mathrm{KS}} - \alpha_n) \ket{q_n} - \beta_n \ket{q_{n-1}} || \,.
\end{equation}
The ACE method of equation~\ref{eq:ace} is used to approximate the computationally demanding $V_{\mathrm{EXX}}$ term in $H_{\mathrm{KS}}$ of equation~\ref{eq:q}, reducing the computational cost of applying $H_{\mathrm{KS}}$ to $q_n$ and, hence, accelerating the construction of the Lanczos chains.

The Lanczos vectors transform $H_{\mathrm{KS}}$ into a tridiagonal matrix $T$, with $\alpha_n$ along the diagonal and $\beta_n$ along the sub- and super-diagonal. Using the eigenvalues $d_n$ and eigenvectors $U_{n n'}$ of $T$, we have
\begin{equation}
M_{c; LR} (\omega) = \sum_{n, n'}^{N_{\mathrm{Lanczos}}} \braket{L | q_n} U_{n n'} (\omega - d_{n'})^{-1} U_{1 n'} \,.
\end{equation}
Therefore, once the Lanczos chain is constructed with $N_{\mathrm{Lanczos}}$ elements and the tridiagonal matrix $T$ is diagonalized, the frequency dependence of $M_{c; LR}$ becomes analytically known, enabling the evaluation of the correlation self-energy at multiple frequencies without any explicit summation over empty states.

\subsection{ACE-accelerated Bethe-Salpeter equation solver}
\label{sec:bse}

The BSE within the Tamm-Dancoff approximation~\cite{tda_hirata_1999} can be conveniently solved using DMPT, i.e., by evaluating the response of the density matrix to a perturbation~\cite{bse_rocca_2010,bse_rocca_2012,bse_ping_2012,bse_nguyen_2019}. Within such framework, the VEE $\omega_s$ from the ground to the $s$-th excited state can be computed by solving the eigenvalue problem of the Liouville operator,
\begin{equation}
\label{eq:evp}
(D + K^{1e} - K^{1d}) A_s = \omega_s A_s \,,
\end{equation}
where $A_s = \{ \ket{a_{s,v}}: v = 1, \dots, N_{\mathrm{occ}} \}$ denotes a set of vectors that enter the definition of the linear change of the density with respect to the ground state due to the $s$-th neutral excitation as
\begin{equation}
\Delta n_s = \sum_{v=1}^{N_{\mathrm{occ}}} \ket{a_{s,v}} \bra{\psi_v} \,.
\end{equation}
The $D$, $K^{1e}$, and $K^{1d}$ terms in equation~\ref{eq:evp} are defined as follows:
\begin{widetext}
\begin{equation}
\label{eq:d}
D A_s = \left\{ P_{\mathrm{c}} ( H_{\mathrm{KS}} + \Delta_{\mathrm{QP}} - \varepsilon_v^{\mathrm{QP}} ) \ket{a_{s,v}} : v = 1, \dots, N_{\mathrm{occ}} \right\} \,,
\end{equation}
\begin{equation}
K^{1e} A_s = \left\{ 2 \int \mathrm{d} \mathbf{r}' P_{\mathrm{c}} (\mathbf{r}, \mathbf{r}') \psi_v (\mathbf{r}') \sum_{v'=1}^{N_{\mathrm{occ}}} \int \mathrm{d} \mathbf{r}'' v (\mathbf{r}', \mathbf{r}'') \psi_{v'}^* (\mathbf{r}'') a_{s,v'} (\mathbf{r}'') : v = 1, \dots, N_{\mathrm{occ}} \right\} \,,
\end{equation}
\begin{equation}
K^{1d} A_s = \left\{ \int \mathrm{d} \mathbf{r}' P_{\mathrm{c}} (\mathbf{r}, \mathbf{r}') \sum_{v'=1}^{N_{\mathrm{occ}}} \tau_{vv'} (\mathbf{r'}) a_{s,v'} (\mathbf{r'}) : v = 1, \dots, N_{\mathrm{occ}} \right\} \,,
\end{equation}
\end{widetext}
where $\tau_{vv'} (\mathbf{r})$ is the statically screened Coulomb integral
\begin{equation}
\tau_{vv'}(\mathbf{r}) = \int \mathrm{d}\mathbf{r}' W_0 (\mathbf{r}, \mathbf{r}', \omega=0) \psi_{v} (\mathbf{r}') \psi_{v'}^* (\mathbf{r}') \,.
\label{eq:tauvv}
\end{equation}
The ACE method of equation~\ref{eq:ace} is used to approximate the computationally demanding $V_{\mathrm{EXX}}$ term in $H_{\mathrm{KS}}$ of equation~\ref{eq:d}, reducing the computational cost of applying $H_{\mathrm{KS}}$ to $a_{s,v}$ and, hence, accelerating the iterative diagonalization of the Liouville operator using, for instance, the Davidson method.

In equation~\ref{eq:tauvv}, the evaluation of $\tau_{vv'}$ for all pairs of occupied states constitutes a severe computational burden when solving the BSE, which can be reduced by introducing the low-rank decomposition of $W$ of equation~\ref{eq:pdep} and by reducing the number of integrals $\tau_{vv'}$ to be evaluated by localizing the KS wave functions with the recursive subspace method~\cite{bisection_gygi_2009} or by transforming them to maximally localized Wannier orbitals. We approximate $\tau_{vv'} = 0$ when a pair of non-overlapping localized wave functions are considered~\cite{bse_nguyen_2019,tddft_jin_2023}.

\section{Computational accuracy}
\label{sec:accuracy}

We assess the computational accuracy of employing the ACE method in combination with the implementation of MBPT presented in section~\ref{sec:method}, focusing on full-frequency G$_0$W$_0$ and BSE calculations starting from hybrid DFT. We consider five test cases, namely the bulk silicon $4 \times 4 \times 4$ supercell (denoted Si), a box of 64 water molecules extracted from a first principles molecular dynamics trajectory~\cite{bse_dong_2021} (H$_2$O), the cadmium selenide nanoparticle (CdSe), the kk divacancy in 3C SiC $3 \times 3 \times 3$ supercell (VV$^0$), and the nitrogen-vacancy in diamond $3 \times 3 \times 3$ supercell (NV$^-$). The VV$^0$ and NV$^-$ are representative spin defects for solid-state quantum information technologies. Details of the test systems are summarized in table~\ref{tab:tests}.

\begin{table}[ht!]
\centering
\caption{Test cases considered in this work. $N_{\mathrm{atom}}$, $N_{\mathrm{electron}}$, and $N_{\mathrm{spin}}$ denote the numbers of atoms, electrons, and spin channels, respectively. $\alpha$ is the fraction of the exact exchange in the hybrid functional.}
\begin{tabular}{c|cccc}
\hline
System & $N_{\mathrm{atom}}$ & $N_{\mathrm{electron}}$ & $N_{\mathrm{spin}}$ & $\alpha$ \\
\hline
Si     & 128 & 512 & 1 & 0.084 \\
H$_2$O & 192 & 512 & 1 & 0.610 \\
CdSe   &  68 & 884 & 1 & 0.250 \\
VV$^0$ & 214 & 859 & 2 & 0.153 \\
NV$^-$ & 215 & 862 & 2 & 0.178 \\
\hline
\end{tabular}
\label{tab:tests}
\end{table}

The \texttt{pwscf} code in the Quantum ESPRESSO software suite~\cite{qe_giannozzi_2020} was employed for all ground-state DFT calculations. We used a kinetic energy cutoff of 60 Ry for the plane wave basis set, the SG15 optimized norm-conserving Vanderbilt pseudopotentials~\cite{oncv_hamann_2013,oncv_schlipf_2015}, and the dielectric dependent hybrid (DDH) functional, which includes a fraction of the exact exchange with the mixing parameter $\alpha$ equal to the inverse of the macroscopic dielectric constant of the material~\cite{ddh_skone_2014,ddh_skone_2016,exx_gaiduk_2016,exx_pham_2017}. For the CdSe we used the PBE0 hybrid functional with $\alpha = 0.25$. The Brillouin zone was sampled at the $\Gamma$-point. The \texttt{wstat} code of WEST was employed to iteratively diagonalize the static dielectric matrix to obtain the PDEP basis set following the method described in section~\ref{sec:pdep}, with $N_{\mathrm{PDEP}}$ set equal to twice the number of electrons in the system. The \texttt{wfreq} code of WEST was employed to compute the QP energies using the full-frequency G$_0$W$_0$ method as described in section~\ref{sec:g0w0}. The \texttt{wbse} code of WEST was employed to solve the BSE as described in section~\ref{sec:bse}, hence to compute the absorption spectra of Si, H$_2$O, and CdSe, and the VEEs of VV$^0$ and NV$^-$. We note that the G$_0$W$_0$ and BSE benchmarks presented here are intended to establish the applicability of the ACE method in MBPT calculations. We, therefore, focus the discussion on comparing the numerical results and computational costs obtained with and without using the ACE method to approximate the EXX operator.

\subsection{G$_0$W$_0$ quasiparticle energies}

\begin{figure}[ht!]
\centering
\includegraphics[width=0.4\textwidth]{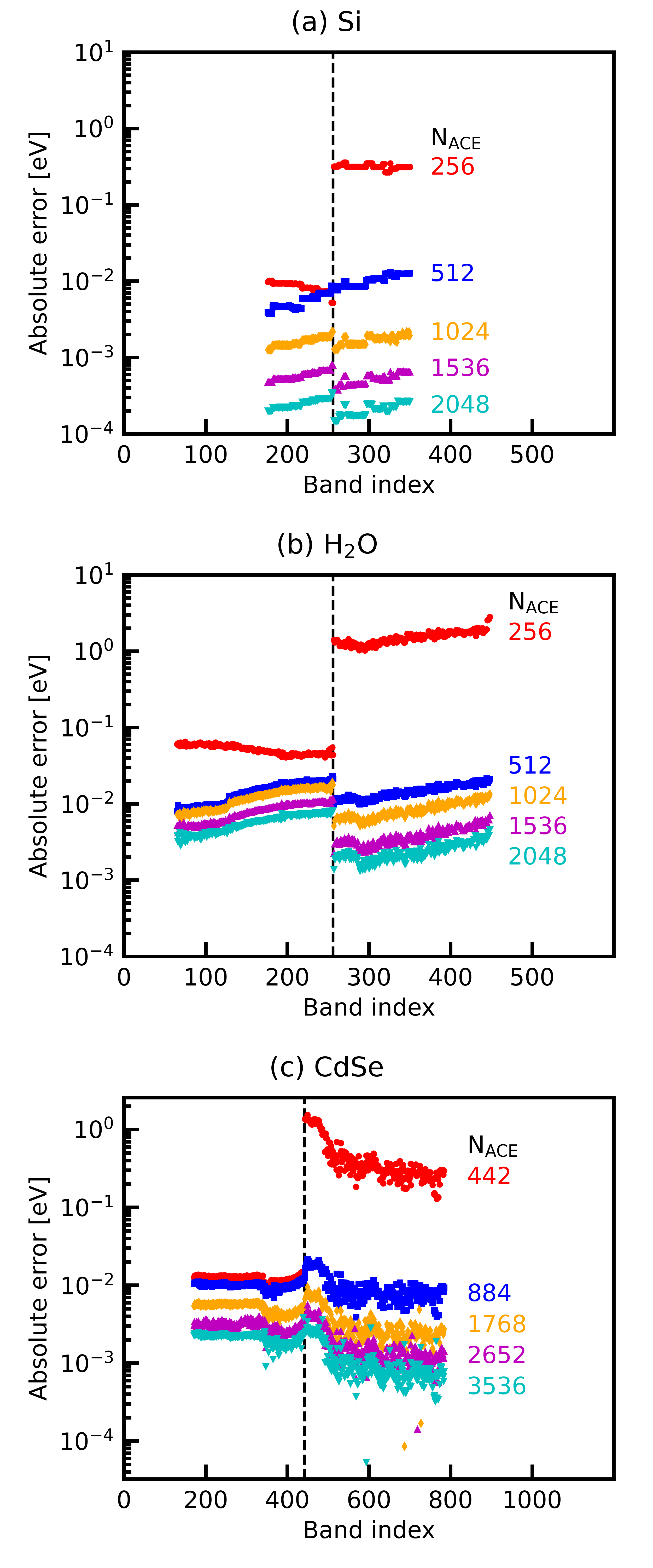}
\caption{Absolute errors in QP energies as a function of the band index, measured as the difference between QP energies computed with and without the ACE approximation in the full-frequency G$_0$W$_0$ calculations for (a) bulk silicon (Si), (b) 64 water molecules (H$_2$O), and (c) cadmium selenide nanoparticle (CdSe). The DDH functional is used in the DFT starting point for Si and H$_2$O, while the PBE0 functional is used for CdSe. Bands on the left (right) of the dashed line are occupied (unoccupied). $N_{\mathrm{ACE}}$ identifies the number of KS wave functions used to build the ACE operator.}
\label{fig:qp_err}
\end{figure}

Figure~\ref{fig:qp_err} shows the errors introduced by using $V_{\mathrm{ACE}}$ instead of $V_{\mathrm{EXX}}$ in full-frequency G$_0$W$_0$ calculations employing the DDH functional for the Si and H$_2$O and the PBE0 functional for the CdSe. We present the absolute errors in QP energies, measured as the difference between energies computed with and without the ACE approximation, for values of $N_{\mathrm{ACE}}$ ranging from $N_{\mathrm{occ}}$ to $8 \times N_{\mathrm{occ}}$. For the Si, when $N_{\mathrm{ACE}} = N_{\mathrm{occ}} = 256$, the ACE approximation introduces errors of at most 0.01 eV for the occupied states and errors of $\sim$0.3 eV for the unoccupied states. In this case, $V_{\mathrm{ACE}}$ is constructed solely from the occupied KS wave functions. Therefore, it is expected to be an inaccurate approximation of $V_{\mathrm{EXX}}$ in the unoccupied subspace. With $N_{\mathrm{ACE}} = 2 \times N_{\mathrm{occ}}$, the errors in both occupied and unoccupied states are reduced to below 0.02 eV. As we further increase $N_{\mathrm{ACE}}$, the errors continue to decrease as expected. Similar to the Si, for the H$_2$O and CdSe, when $N_{\mathrm{ACE}} = N_{\mathrm{occ}}$, the ACE approximation introduces moderate errors of up to 0.1 eV for the occupied states, and large errors of over 1 eV for the unoccupied states. The errors in both occupied and unoccupied states are consistently reduced as we increase $N_{\mathrm{ACE}}$.

\begin{figure*}[ht!]
\centering
\includegraphics[width=0.8\textwidth]{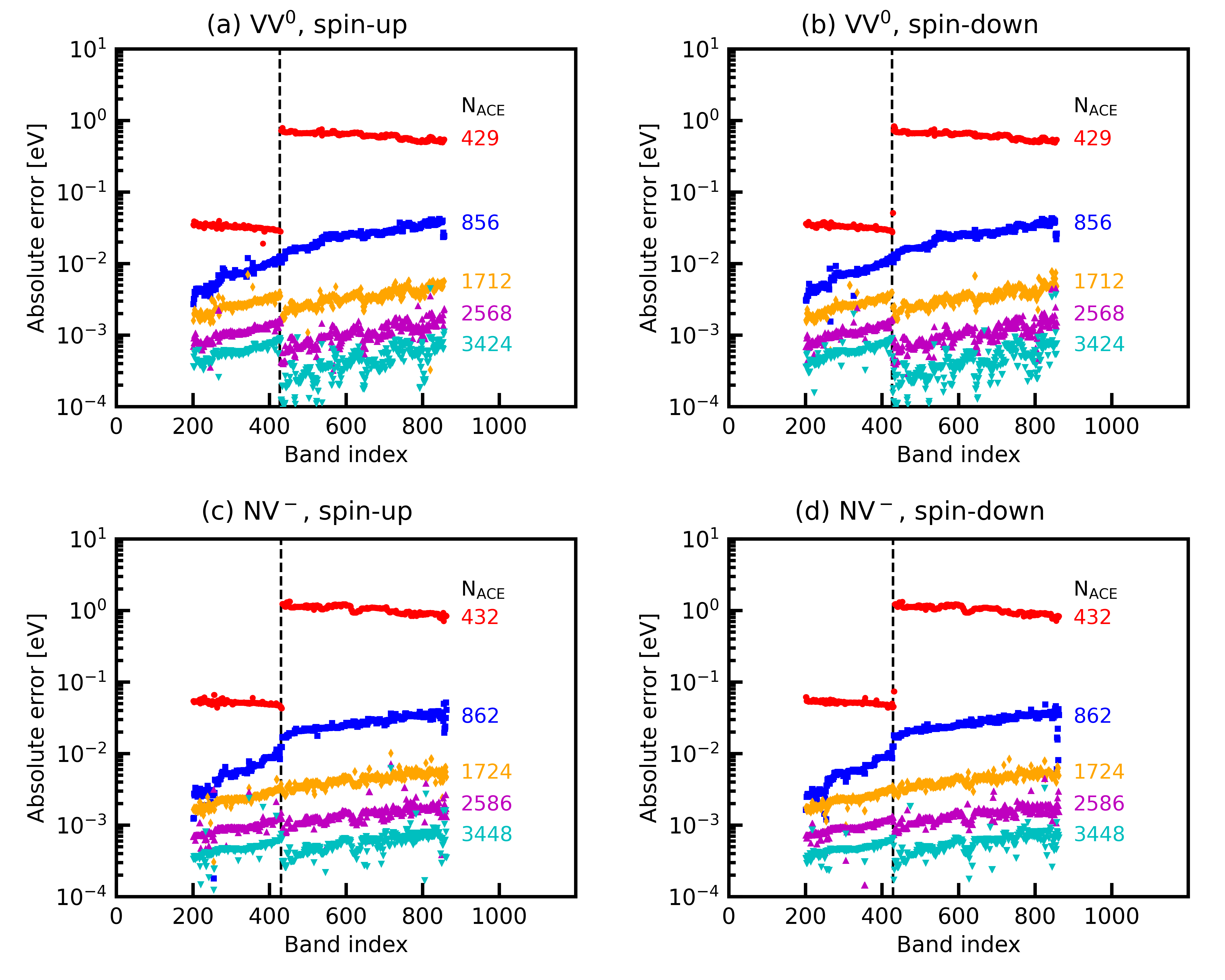}
\caption{Absolute errors in QP energies as a function of the band index, measured as the difference between QP energies computed with and without the ACE approximation in the full-frequency G$_0$W$_0$ calculations for (a) kk divacancy center in 3C SiC (VV$^0$), spin-up channel, (b) VV$^0$, spin-down channel, (c) nitrogen-vacancy center in diamond (NV$^-$), spin-up channel, and (d) NV$^-$, spin-down channel. The DDH functional is used in the DFT starting point. Bands on the left (right) of the dashed line are occupied (unoccupied). $N_{\mathrm{ACE}}$ identifies the number of KS wave functions used to build the ACE operator.}
\label{fig:qp_err_spin}
\end{figure*}

Figure~\ref{fig:qp_err_spin} shows the errors introduced by using $V_{\mathrm{ACE}}$ instead of $V_{\mathrm{EXX}}$ in full-frequency G$_0$W$_0$ calculations employing the DDH functional for the spin-polarized VV$^0$ and NV$^-$. We present the absolute errors in QP energies, measured as the difference between energies computed with and without the ACE approximation, for values of $N_{\mathrm{ACE}}$ ranging from $N_{\mathrm{occ}}$ to $8 \times N_{\mathrm{occ}}$. For the VV$^0$, when $N_{\mathrm{ACE}} = N_{\mathrm{occ}} = 429$, the ACE approximation introduces errors of 0.02 to 0.04 eV for the occupied states and errors of $\sim$1 eV for the unoccupied states. With $N_{\mathrm{ACE}} = 2 \times N_{\mathrm{occ}}$, the errors in both occupied and unoccupied states are reduced to below 0.01 eV. As we further increase $N_{\mathrm{ACE}}$, the errors continue to decrease as expected. For the same value of $N_{\mathrm{ACE}}$, the spin-up and spin-down channels have errors of the same magnitude. Similar to the VV$^0$, for the NV$^-$, when $N_{\mathrm{ACE}} = N_{\mathrm{occ}}$, the ACE approximation introduces moderate errors of 0.04 to 0.07 eV for the occupied states, and large errors of over 1 eV for the unoccupied states. The errors in both occupied and unoccupied states are consistently reduced as we increase $N_{\mathrm{ACE}}$.

\subsection{BSE absorption spectra}

\begin{figure}[ht!]
\centering
\includegraphics[width=0.4\textwidth]{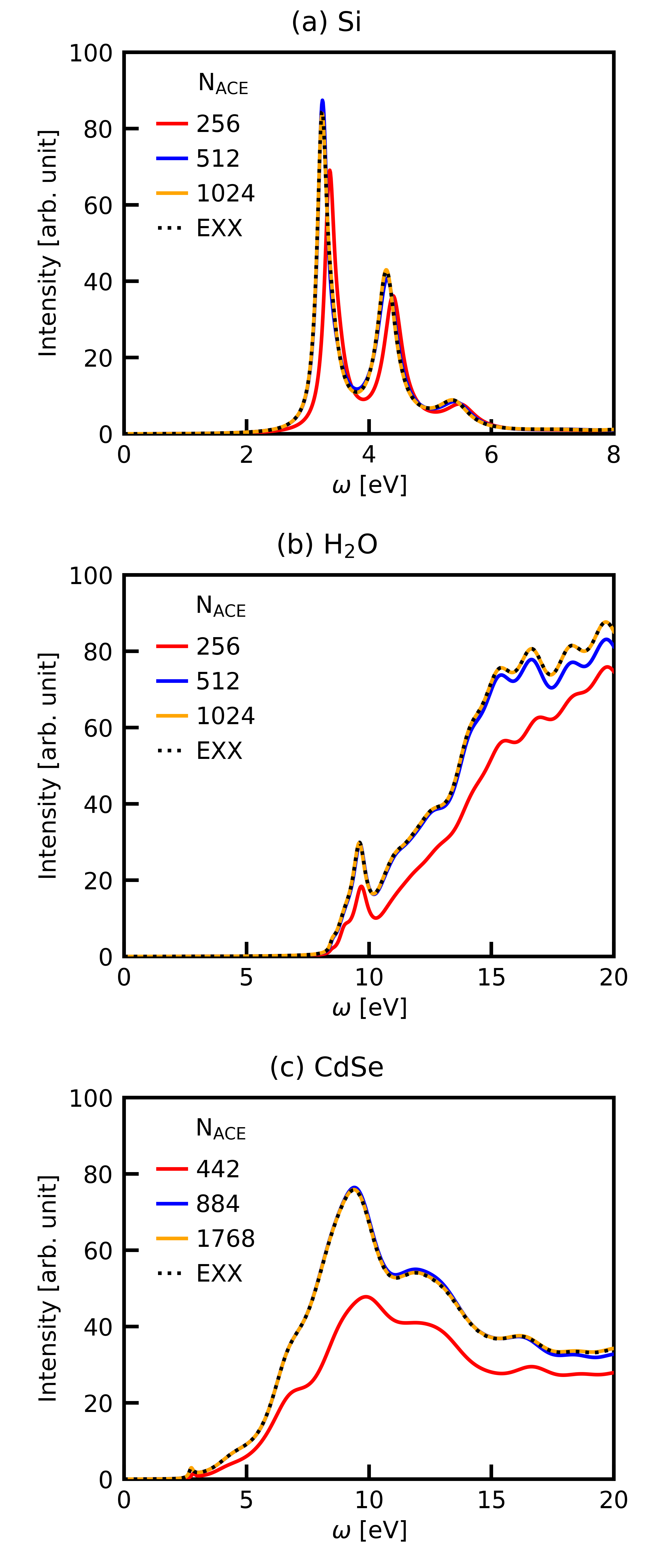}
\caption{Optical absorption spectra obtained by solving the BSE with or without the ACE approximation for (a) bulk silicon (Si), (b) 64 water molecules (H$_2$O), and (c) cadmium selenide nanoparticle (CdSe). The DDH functional is used in the DFT starting point for Si and H$_2$O, while the PBE0 functional is used for CdSe. The black dotted line corresponds to the reference result computed with the EXX operator without the ACE approximation. $N_{\mathrm{ACE}}$ identifies the number of KS wave functions used to build the ACE operator.}
\label{fig:spectrum_err}
\end{figure}

Figure~\ref{fig:spectrum_err} shows the errors introduced by using $V_{\mathrm{ACE}}$ instead of $V_{\mathrm{EXX}}$ in G$_0$W$_0$-BSE calculations employing the DDH functional for the Si and H$_2$O and the PBE0 functional for the CdSe. We present the absorption spectra for values of $N_{\mathrm{ACE}}$ ranging from $N_{\mathrm{occ}}$ to $4 \times N_{\mathrm{occ}}$. The spectra computed directly using $V_{\mathrm{EXX}}$ are shown as the reference. For all three test systems, the spectra obtained with $N_{\mathrm{ACE}} = N_{\mathrm{occ}}$ appear to capture the main features of the reference while blue-shifting the peaks by $\sim$0.1 eV and underestimating the intensity. The spectra computed with $N_{\mathrm{ACE}} = 2 \times N_{\mathrm{occ}}$ exhibit slight discrepancies to the reference, in particular at > 14 eV for the H$_2$O. The spectra computed with $N_{\mathrm{ACE}} = 4 \times N_{\mathrm{occ}}$ are visually indistinguishable from the reference.

In general, for any fixed value of $N_{\mathrm{ACE}}$, the errors are larger in the high-energy range than in the low-energy range. This is because $V_{\mathrm{ACE}}$, by construction, exactly reproduces $V_{\mathrm{EXX}}$ within the subspace spanned by the lowest $N_{\mathrm{ACE}}$ wave functions. Therefore, when an excitation takes place between wave functions within the range $[1, N_{\mathrm{ACE}}]$, using $V_{\mathrm{ACE}}$ should theoretically introduce no error to this excitation, although minor numerical discrepancies are unavoidable in practice, due to factors such as the finite convergence precision of iterative diagonalization and the use of finite arithmetic. For higher-lying excitations that involve transitions to wave functions beyond $N_{\mathrm{ACE}}$, $V_{\mathrm{ACE}}$ is inherently less accurate, resulting in larger errors in these excitations compared to the lower-lying ones. The errors in the high-energy range are reduced by increasing $N_{\mathrm{ACE}}$, as shown in figure~\ref{fig:spectrum_err}.

\subsection{BSE vertical excitation energies}

\begin{figure*}[ht!]
\centering
\includegraphics[width=0.8\textwidth]{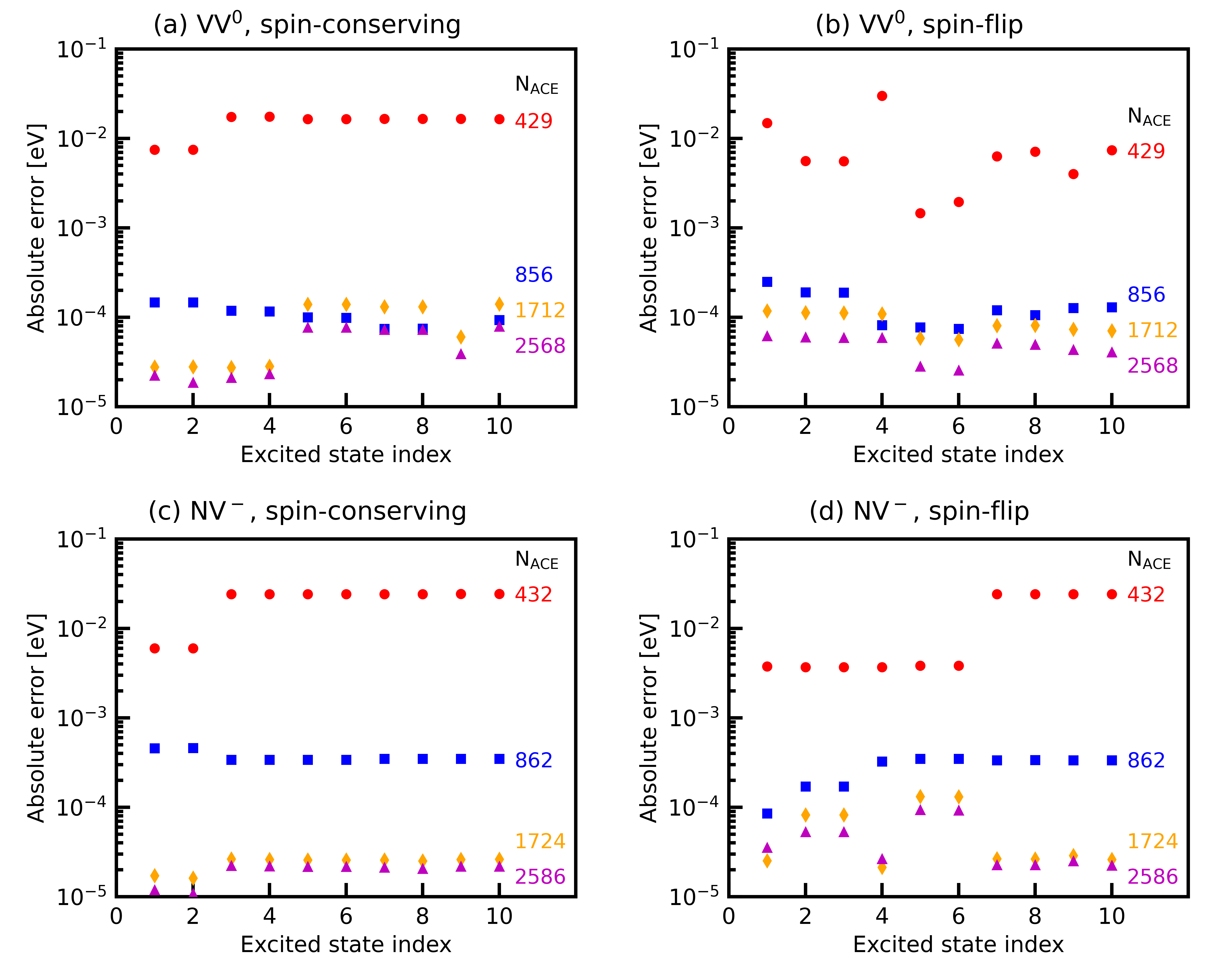}
\caption{Absolute errors in vertical excitation energies for the lowest ten excited states, measured as the difference between excitation energies computed using the ACE operator and the EXX operator without the ACE approximation in BSE calculations for (a) kk divacancy center in 3C SiC (VV$^0$), spin-conserving excitations, (b) VV$^0$, spin-flip excitations, (c) nitrogen-vacancy center in diamond (NV$^-$), spin-conserving excitations, and (d) NV$^-$, spin-flip excitations. The DDH functional is used in the DFT starting point. $N_{\mathrm{ACE}}$ identifies the number of KS wave functions used to build the ACE operator.}
\label{fig:vee_err}
\end{figure*}

Figure~\ref{fig:vee_err} shows the errors introduced by using $V_{\mathrm{ACE}}$ instead of $V_{\mathrm{EXX}}$ in G$_0$W$_0$-BSE calculations employing the DDH functional for the spin-polarized VV$^0$ and NV$^-$. We present the absolute errors in VEEs of the lowest spin-conserving and spin-flip excited states, measured as the difference between energies computed with and without the ACE approximation, for values of $N_{\mathrm{ACE}}$ ranging from $N_{\mathrm{occ}}$ to $6 \times N_{\mathrm{occ}}$. For both test systems, when $N_{\mathrm{ACE}} = N_{\mathrm{occ}}$, the ACE approximation introduces errors of at most 0.03 eV. The errors are reduced to well below 1 meV with $N_{\mathrm{ACE}} \ge 2 \times N_{\mathrm{occ}}$. For the same value of $N_{\mathrm{ACE}}$, spin-conserving and spin-flip excitations have errors of the same magnitude.

\section{Computational cost}
\label{sec:cost}

Using the H$_2$O and NV$^-$ test cases (see table~\ref{tab:tests}), we benchmark the computational cost of employing the ACE method in combination with the implementation of G$_0$W$_0$ and BSE presented in section~\ref{sec:method}.

\begin{figure}[ht!]
\centering
\includegraphics[width=0.4\textwidth]{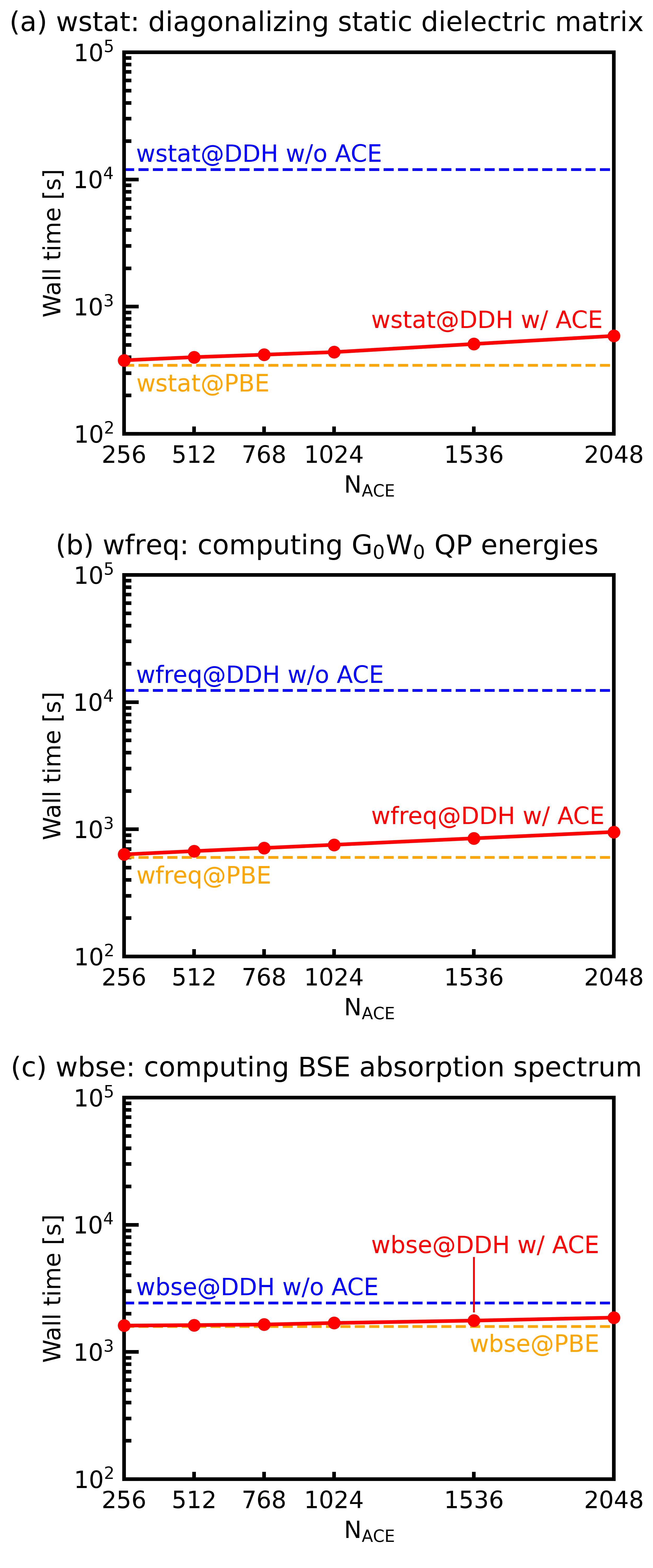}
\caption{Computational time of full-frequency G$_0$W$_0$ and BSE calculations for a sample of 64 water molecules (H$_2$O). The DDH functional is used in the DFT starting point. The calculations were performed on 32 CPU nodes of Perlmutter. (a) Wall clock time of the \texttt{wstat} code that diagonalizes the static dielectric matrix to obtain the PDEP basis set. (b) Wall clock time of the \texttt{wfreq} code that computes the QP energies within the full-frequency G$_0$W$_0$ method. (c) Wall clock time of the \texttt{wbse} code that computes the optical absorption spectrum using the DMPT solver of BSE. The red line indicates the computational cost of calculations using the DDH functional within the ACE approximation. The blue dashed line indicates the computational cost of calculations using the DDH functional and the EXX operator without the ACE approximation. The orange dashed line indicates the computational cost of calculations using the PBE functional. $N_{\mathrm{ACE}}$ identifies the number of KS wave functions used to build the ACE operator.}
\label{fig:h2o_time}
\end{figure}

Figure~\ref{fig:h2o_time} shows the computational time saved for the H$_2$O by introducing the ACE approximation in full-frequency G$_0$W$_0$ and BSE calculations starting from hybrid DFT. The benchmarks were conducted on 32 CPU nodes of Perlmutter, the supercomputer of the National Energy Research Scientific Computing Center. Each CPU node of Perlmutter is equipped with two AMD EPYC 7763 CPUs. In figure~\ref{fig:h2o_time}(a), (b), and (c), we compare the wall clock time of the \texttt{wstat}, \texttt{wfreq}, and \texttt{wbse} codes, respectively, using $V_{\mathrm{EXX}}$ (blue dashed line) versus $V_{\mathrm{ACE}}$ with various values of $N_{\mathrm{ACE}}$ (red solid line). The use of $V_{\mathrm{ACE}}$ leads to a significant speedup of over one order of magnitude for both \texttt{wstat}, which computes the PDEP basis set, and \texttt{wfreq}, which computes the full-frequency G$_0$W$_0$ QP energies. For $N_{\mathrm{ACE}} = 1024$, which results in negligible errors as demonstrated in figures~\ref{fig:qp_err} and \ref{fig:spectrum_err}, speedup factors of 27.3x and 16.4x are achieved for \texttt{wstat} and \texttt{wfreq}, respectively. In the case of \texttt{wbse}, which solves the BSE to compute the absorption spectrum, the speedup achieved with $V_{\mathrm{ACE}}$ ranges from 1.3x to 1.5x. The speedup for \texttt{wbse} is less pronounced compared to that of \texttt{wstat} or \texttt{wfreq}, as only the $D$ term in equation~\ref{eq:evp} is accelerated by the ACE approximation, while the other terms are not. The $D$ term amounts to approximately a third of the computational cost of \texttt{wbse}. Therefore, the maximum speedup factor for \texttt{wbse} is limited to 1.5x even if the computational cost of the $D$ term was completely eliminated. Overall, the use of $V_{\mathrm{ACE}}$ reduces the computational cost of G$_0$W$_0$ and BSE calculations with the hybrid DDH functional to only slightly higher than that of G$_0$W$_0$ and BSE calculations with the semi-local PBE functional (orange dashed line).

\begin{figure}[ht!]
\centering
\includegraphics[width=0.4\textwidth]{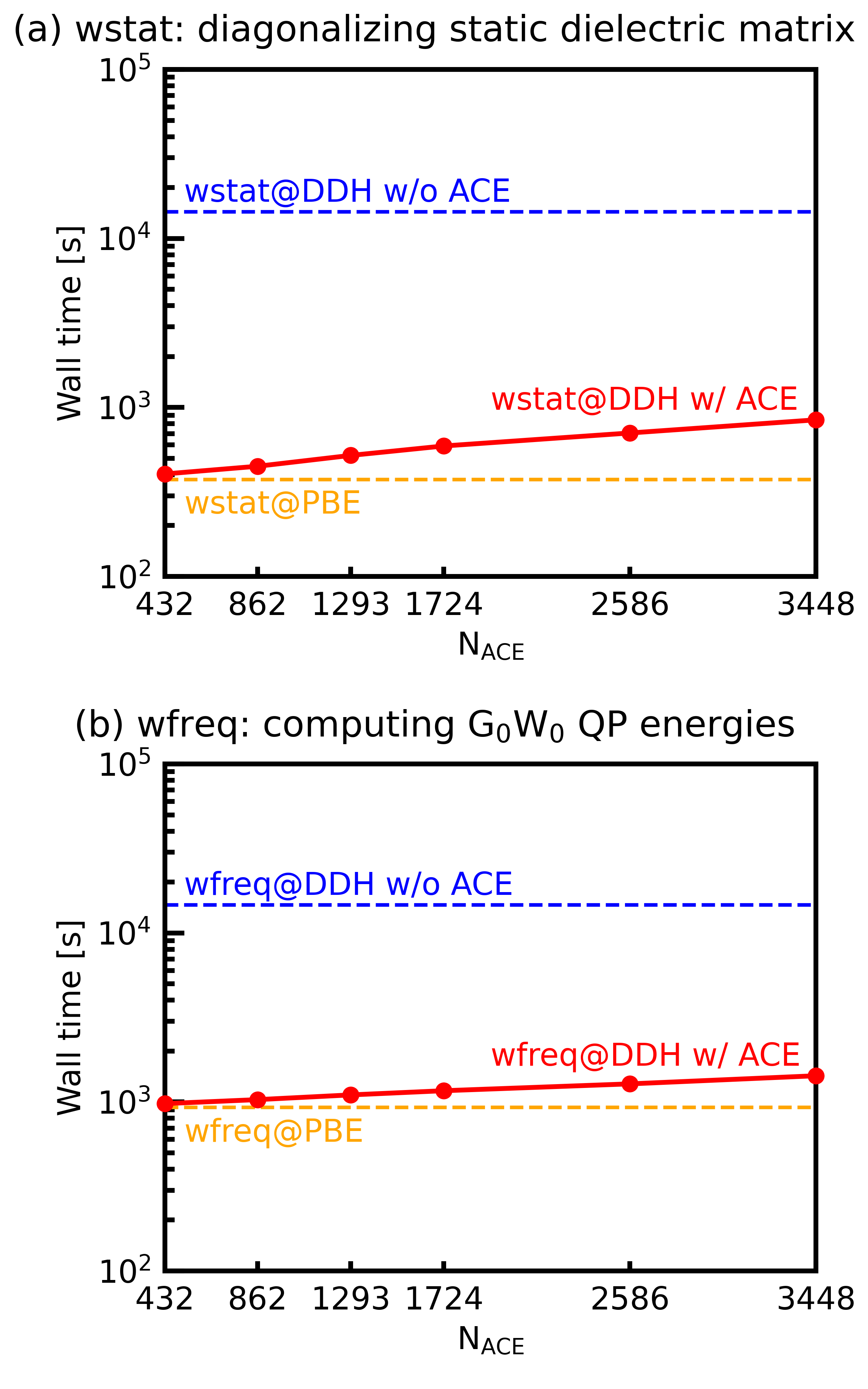}
\caption{Computational time of full-frequency G$_0$W$_0$ and BSE calculations for the nitrogen-vacancy center in diamond (NV$^-$). The DDH functional is used in the DFT starting point. The calculations were performed on 256 CPU nodes of Perlmutter. (a) Wall clock time of the \texttt{wstat} code that diagonalizes the static dielectric matrix to obtain the PDEP basis set. (b) Wall clock time of the \texttt{wfreq} code that computes the QP energies within the full-frequency G$_0$W$_0$ method. The red line indicates the computational cost of calculations using the DDH functional and the ACE approximation. The blue dashed line indicates the computational cost of calculations using the DDH functional and the EXX operator without the ACE approximation. The orange dashed line indicates the computational cost of calculations using the PBE functional. $N_{\mathrm{ACE}}$ identifies the number of KS wave functions used to build the ACE operator.}
\label{fig:nv_time}
\end{figure}

Figure~\ref{fig:nv_time} shows the computational time saved for the NV$^-$ by using the ACE approximation in G$_0$W$_0$ and BSE calculations carried out starting from an electronic structure obtained with the DDH functional. The benchmarks were conducted on 256 CPU nodes of Perlmutter. In figure~\ref{fig:nv_time}(a) and (b), we compare the wall clock time of the \texttt{wstat} and \texttt{wfreq} codes, respectively, using $V_{\mathrm{EXX}}$ (blue dashed line) versus $V_{\mathrm{ACE}}$ with various values of $N_{\mathrm{ACE}}$ (red solid line). The use of the ACE approximation leads to a significant speedup of over an order of magnitude for both \texttt{wstat} and \texttt{wfreq}. For $N_{\mathrm{ACE}} = 1724$, which results in negligible errors as demonstrated in figures~\ref{fig:qp_err_spin} and \ref{fig:vee_err}, speedup factors of 24.3x and 12.6x are achieved for \texttt{wstat} and \texttt{wfreq}, respectively. The use of the ACE approximation reduces the computational cost of full-frequency G$_0$W$_0$ calculations with the hybrid functional to only slightly higher than that of G$_0$W$_0$ calculations with the semi-local PBE functional (orange dashed line). The time spent by \texttt{wbse}, which solves the BSE to compute the VEEs, was negligible (less than 1 minute) and therefore it is not discussed here.

\begin{figure}[ht!]
\centering
\includegraphics[width=0.4\textwidth]{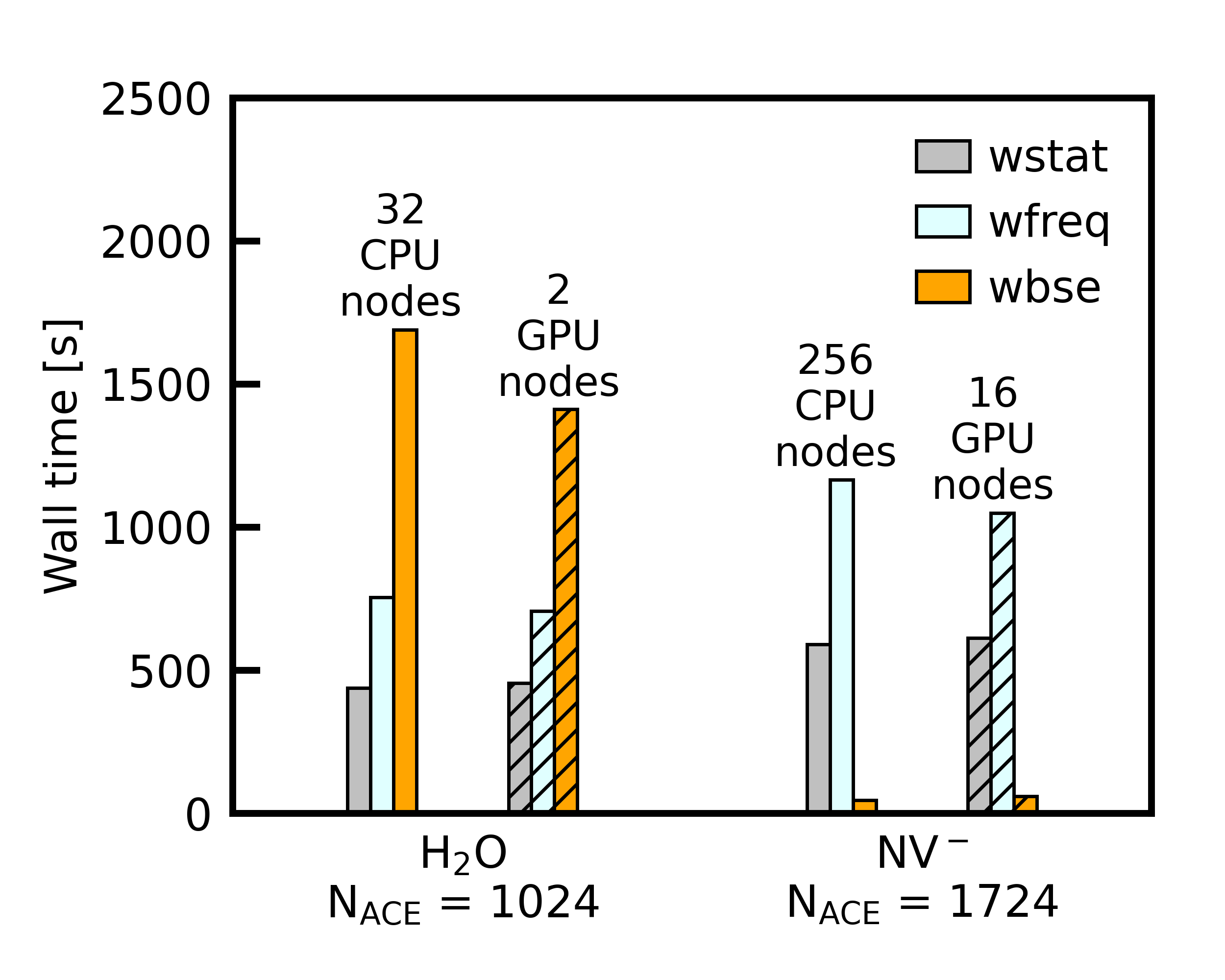}
\caption{Computational time of full-frequency G$_0$W$_0$ and BSE calculations for 64 water molecules (H$_2$O) and the nitrogen-vacancy center in diamond (NV$^-$). We have used the DDH functional and the ACE operator. $N_{\mathrm{ACE}}$ identifies the number of KS wave functions used to build the ACE operator. The calculations of the H$_2$O sample were performed on 32 CPU nodes or two GPU nodes of Perlmutter. The calculations of the NV$^-$ were performed on 256 CPU nodes or 16 GPU nodes of Perlmutter. The solid (striped) bars represent the wall clock time of calculations carried out on CPUs (GPUs).}
\label{fig:gpu_time}
\end{figure}

Finally, in figure~\ref{fig:gpu_time}, we present the performance of the GPU-accelerated \texttt{wstat}, \texttt{wfreq}, and \texttt{wbse} codes in G$_0$W$_0$ and BSE calculations, utilizing the DDH functional and the ACE operator. The benchmarks were conducted using the GPU nodes of Perlmutter, each of which is equipped with one AMD EPYC 7763 CPU and four NVIDIA A100 GPUs. The theoretical peak performance of one GPU node is approximately equivalent to that of 16 CPU nodes. For the H$_2$O, we compare the performance using two GPU nodes versus 32 CPU nodes. The ACE operator with $N_{\mathrm{ACE}} = 1024$ was used. The results indicate that the performance of \texttt{wstat}, \texttt{wfreq}, and \texttt{wbse} using two GPU nodes is comparable to or better than that using 32 CPU nodes. For the NV$^-$, we compare the performance using 16 GPU nodes versus 256 CPU nodes. The ACE operator with $N_{\mathrm{ACE}} = 1724$ was used. Again, the performance using 16 GPU nodes is comparable to or better than that using 256 CPU nodes.

\section{Conclusions}
\label{sec:conclusions}

We have implemented and benchmarked the use of the ACE operator in MBPT calculations with hybrid DFT starting points. The formulation we used for G$_0$W$_0$ and BSE sidesteps any explicit summation over empty states, relying instead on the application of the ground-state KS Hamiltonian in various iterative solvers. By approximating the EXX operator in the Hamiltonian with the ACE operator, a remarkable speedup of over one order of magnitude is achieved, greatly reducing the computational cost of G$_0$W$_0$ and BSE calculations with hybrid starting points. The computational precision of the ACE approximation can be systematically controlled by adjusting $N_{\mathrm{ACE}}$, the number of KS wave functions entering the construction of the ACE operator. We have ported the method described here to GPUs, thereby enabling large-scale simulations on GPU-accelerated computers. We envision that our work will facilitate the exploration and evaluation of fine-tuned hybrid starting points aimed at enhancing the accuracy of one-shot GW and BSE calculations without the need for computationally demanding self-consistency in Hedin's equations.

\begin{acknowledgments}
This research was supported by the Midwest Integrated Center for Computational Materials (MICCoM), as part of the Computational Materials Sciences Program funded by the U.S. Department of Energy, Office of Science, Basic Energy Sciences, Materials Sciences, and Engineering Division through Argonne National Laboratory. This research used resources of the National Energy Research Scientific Computing Center, a U.S. Department of Energy Office of Science User Facility operated under Contract No. DE-AC02-05CH11231. We acknowledge helpful discussions with Dr. Yu Jin and Prof. Giulia Galli.
\end{acknowledgments}

\section*{Data Availability}

The data that support the findings of this study are organized using the Qresp software~\cite{qresp_govoni_2019} and are available online at \url{https://paperstack.uchicago.edu}.

\nocite{*}
\bibliographystyle{apsrev4-1}
\bibliography{ace}

\begin{thebibliography}{129}%
\makeatletter
\providecommand \@ifxundefined [1]{%
 \@ifx{#1\undefined}
}%
\providecommand \@ifnum [1]{%
 \ifnum #1\expandafter \@firstoftwo
 \else \expandafter \@secondoftwo
 \fi
}%
\providecommand \@ifx [1]{%
 \ifx #1\expandafter \@firstoftwo
 \else \expandafter \@secondoftwo
 \fi
}%
\providecommand \natexlab [1]{#1}%
\providecommand \enquote  [1]{``#1''}%
\providecommand \bibnamefont  [1]{#1}%
\providecommand \bibfnamefont [1]{#1}%
\providecommand \citenamefont [1]{#1}%
\providecommand \href@noop [0]{\@secondoftwo}%
\providecommand \href [0]{\begingroup \@sanitize@url \@href}%
\providecommand \@href[1]{\@@startlink{#1}\@@href}%
\providecommand \@@href[1]{\endgroup#1\@@endlink}%
\providecommand \@sanitize@url [0]{\catcode `\\12\catcode `\$12\catcode `\&12\catcode `\#12\catcode `\^12\catcode `\_12\catcode `\%12\relax}%
\providecommand \@@startlink[1]{}%
\providecommand \@@endlink[0]{}%
\providecommand \url  [0]{\begingroup\@sanitize@url \@url }%
\providecommand \@url [1]{\endgroup\@href {#1}{\urlprefix }}%
\providecommand \urlprefix  [0]{URL }%
\providecommand \Eprint [0]{\href }%
\providecommand \doibase [0]{http://dx.doi.org/}%
\providecommand \selectlanguage [0]{\@gobble}%
\providecommand \bibinfo  [0]{\@secondoftwo}%
\providecommand \bibfield  [0]{\@secondoftwo}%
\providecommand \translation [1]{[#1]}%
\providecommand \BibitemOpen [0]{}%
\providecommand \bibitemStop [0]{}%
\providecommand \bibitemNoStop [0]{.\EOS\space}%
\providecommand \EOS [0]{\spacefactor3000\relax}%
\providecommand \BibitemShut  [1]{\csname bibitem#1\endcsname}%
\let\auto@bib@innerbib\@empty
\bibitem [{\citenamefont {Hohenberg}\ and\ \citenamefont {Kohn}(1964)}]{dft_hohenberg_1964}%
  \BibitemOpen
  \bibfield  {author} {\bibinfo {author} {\bibfnamefont {P.}~\bibnamefont {Hohenberg}}\ and\ \bibinfo {author} {\bibfnamefont {W.}~\bibnamefont {Kohn}},\ }\href@noop {} {\bibfield  {journal} {\bibinfo  {journal} {Phys. Rev.}\ }\textbf {\bibinfo {volume} {136}},\ \bibinfo {pages} {B864} (\bibinfo {year} {1964})}\BibitemShut {NoStop}%
\bibitem [{\citenamefont {Kohn}\ and\ \citenamefont {Sham}(1965)}]{dft_kohn_1965}%
  \BibitemOpen
  \bibfield  {author} {\bibinfo {author} {\bibfnamefont {W.}~\bibnamefont {Kohn}}\ and\ \bibinfo {author} {\bibfnamefont {L.~J.}\ \bibnamefont {Sham}},\ }\href@noop {} {\bibfield  {journal} {\bibinfo  {journal} {Phys. Rev.}\ }\textbf {\bibinfo {volume} {140}},\ \bibinfo {pages} {A1133} (\bibinfo {year} {1965})}\BibitemShut {NoStop}%
\bibitem [{\citenamefont {Hybertsen}\ and\ \citenamefont {Louie}(1985)}]{gw_hybertsen_1985}%
  \BibitemOpen
  \bibfield  {author} {\bibinfo {author} {\bibfnamefont {M.~S.}\ \bibnamefont {Hybertsen}}\ and\ \bibinfo {author} {\bibfnamefont {S.~G.}\ \bibnamefont {Louie}},\ }\href@noop {} {\bibfield  {journal} {\bibinfo  {journal} {Phys. Rev. Lett.}\ }\textbf {\bibinfo {volume} {55}},\ \bibinfo {pages} {1418} (\bibinfo {year} {1985})}\BibitemShut {NoStop}%
\bibitem [{\citenamefont {Hybertsen}\ and\ \citenamefont {Louie}(1986)}]{gw_hybertsen_1986}%
  \BibitemOpen
  \bibfield  {author} {\bibinfo {author} {\bibfnamefont {M.~S.}\ \bibnamefont {Hybertsen}}\ and\ \bibinfo {author} {\bibfnamefont {S.~G.}\ \bibnamefont {Louie}},\ }\href@noop {} {\bibfield  {journal} {\bibinfo  {journal} {Phys. Rev. B}\ }\textbf {\bibinfo {volume} {34}},\ \bibinfo {pages} {5390} (\bibinfo {year} {1986})}\BibitemShut {NoStop}%
\bibitem [{\citenamefont {Strinati}(1988)}]{gw_strinati_1988}%
  \BibitemOpen
  \bibfield  {author} {\bibinfo {author} {\bibfnamefont {G.}~\bibnamefont {Strinati}},\ }\href@noop {} {\bibfield  {journal} {\bibinfo  {journal} {Riv. Nuovo Cim.}\ }\textbf {\bibinfo {volume} {11}},\ \bibinfo {pages} {1} (\bibinfo {year} {1988})}\BibitemShut {NoStop}%
\bibitem [{\citenamefont {Aryasetiawan}\ and\ \citenamefont {Gunnarsson}(1998)}]{gw_aryasetiawan_1998}%
  \BibitemOpen
  \bibfield  {author} {\bibinfo {author} {\bibfnamefont {F.}~\bibnamefont {Aryasetiawan}}\ and\ \bibinfo {author} {\bibfnamefont {O.}~\bibnamefont {Gunnarsson}},\ }\href@noop {} {\bibfield  {journal} {\bibinfo  {journal} {Rep. Prog. Phys.}\ }\textbf {\bibinfo {volume} {61}},\ \bibinfo {pages} {237} (\bibinfo {year} {1998})}\BibitemShut {NoStop}%
\bibitem [{\citenamefont {Golze}\ \emph {et~al.}(2019)\citenamefont {Golze}, \citenamefont {Dvorak},\ and\ \citenamefont {Rinke}}]{gw_golze_2019}%
  \BibitemOpen
  \bibfield  {author} {\bibinfo {author} {\bibfnamefont {D.}~\bibnamefont {Golze}}, \bibinfo {author} {\bibfnamefont {M.}~\bibnamefont {Dvorak}}, \ and\ \bibinfo {author} {\bibfnamefont {P.}~\bibnamefont {Rinke}},\ }\href@noop {} {\bibfield  {journal} {\bibinfo  {journal} {Front. Chem.}\ }\textbf {\bibinfo {volume} {7}},\ \bibinfo {pages} {377} (\bibinfo {year} {2019})}\BibitemShut {NoStop}%
\bibitem [{\citenamefont {Salpeter}\ and\ \citenamefont {Bethe}(1951)}]{bse_salpeter_1951}%
  \BibitemOpen
  \bibfield  {author} {\bibinfo {author} {\bibfnamefont {E.~E.}\ \bibnamefont {Salpeter}}\ and\ \bibinfo {author} {\bibfnamefont {H.~A.}\ \bibnamefont {Bethe}},\ }\href@noop {} {\bibfield  {journal} {\bibinfo  {journal} {Phys. Rev.}\ }\textbf {\bibinfo {volume} {84}},\ \bibinfo {pages} {1232} (\bibinfo {year} {1951})}\BibitemShut {NoStop}%
\bibitem [{\citenamefont {Albrecht}\ \emph {et~al.}(1998)\citenamefont {Albrecht}, \citenamefont {Reining}, \citenamefont {Del~Sole},\ and\ \citenamefont {Onida}}]{bse_albrecht_1998}%
  \BibitemOpen
  \bibfield  {author} {\bibinfo {author} {\bibfnamefont {S.}~\bibnamefont {Albrecht}}, \bibinfo {author} {\bibfnamefont {L.}~\bibnamefont {Reining}}, \bibinfo {author} {\bibfnamefont {R.}~\bibnamefont {Del~Sole}}, \ and\ \bibinfo {author} {\bibfnamefont {G.}~\bibnamefont {Onida}},\ }\href@noop {} {\bibfield  {journal} {\bibinfo  {journal} {Phys. Rev. Lett.}\ }\textbf {\bibinfo {volume} {80}},\ \bibinfo {pages} {4510} (\bibinfo {year} {1998})}\BibitemShut {NoStop}%
\bibitem [{\citenamefont {Onida}\ \emph {et~al.}(2002)\citenamefont {Onida}, \citenamefont {Reining},\ and\ \citenamefont {Rubio}}]{bse_onida_2002}%
  \BibitemOpen
  \bibfield  {author} {\bibinfo {author} {\bibfnamefont {G.}~\bibnamefont {Onida}}, \bibinfo {author} {\bibfnamefont {L.}~\bibnamefont {Reining}}, \ and\ \bibinfo {author} {\bibfnamefont {A.}~\bibnamefont {Rubio}},\ }\href@noop {} {\bibfield  {journal} {\bibinfo  {journal} {Rev. Mod. Phys.}\ }\textbf {\bibinfo {volume} {74}},\ \bibinfo {pages} {601} (\bibinfo {year} {2002})}\BibitemShut {NoStop}%
\bibitem [{\citenamefont {Bruneval}\ and\ \citenamefont {Gonze}(2008)}]{gw_bruneval_2008}%
  \BibitemOpen
  \bibfield  {author} {\bibinfo {author} {\bibfnamefont {F.}~\bibnamefont {Bruneval}}\ and\ \bibinfo {author} {\bibfnamefont {X.}~\bibnamefont {Gonze}},\ }\href@noop {} {\bibfield  {journal} {\bibinfo  {journal} {Phys. Rev. B}\ }\textbf {\bibinfo {volume} {78}},\ \bibinfo {pages} {085125} (\bibinfo {year} {2008})}\BibitemShut {NoStop}%
\bibitem [{\citenamefont {Umari}\ \emph {et~al.}(2009)\citenamefont {Umari}, \citenamefont {Stenuit},\ and\ \citenamefont {Baroni}}]{gw_umari_2009}%
  \BibitemOpen
  \bibfield  {author} {\bibinfo {author} {\bibfnamefont {P.}~\bibnamefont {Umari}}, \bibinfo {author} {\bibfnamefont {G.}~\bibnamefont {Stenuit}}, \ and\ \bibinfo {author} {\bibfnamefont {S.}~\bibnamefont {Baroni}},\ }\href@noop {} {\bibfield  {journal} {\bibinfo  {journal} {Phys. Rev. B}\ }\textbf {\bibinfo {volume} {79}},\ \bibinfo {pages} {201104} (\bibinfo {year} {2009})}\BibitemShut {NoStop}%
\bibitem [{\citenamefont {Umari}\ \emph {et~al.}(2010)\citenamefont {Umari}, \citenamefont {Stenuit},\ and\ \citenamefont {Baroni}}]{gw_umari_2010}%
  \BibitemOpen
  \bibfield  {author} {\bibinfo {author} {\bibfnamefont {P.}~\bibnamefont {Umari}}, \bibinfo {author} {\bibfnamefont {G.}~\bibnamefont {Stenuit}}, \ and\ \bibinfo {author} {\bibfnamefont {S.}~\bibnamefont {Baroni}},\ }\href@noop {} {\bibfield  {journal} {\bibinfo  {journal} {Phys. Rev. B}\ }\textbf {\bibinfo {volume} {81}},\ \bibinfo {pages} {115104} (\bibinfo {year} {2010})}\BibitemShut {NoStop}%
\bibitem [{\citenamefont {Giustino}\ \emph {et~al.}(2010)\citenamefont {Giustino}, \citenamefont {Cohen},\ and\ \citenamefont {Louie}}]{gw_giustino_2010}%
  \BibitemOpen
  \bibfield  {author} {\bibinfo {author} {\bibfnamefont {F.}~\bibnamefont {Giustino}}, \bibinfo {author} {\bibfnamefont {M.~L.}\ \bibnamefont {Cohen}}, \ and\ \bibinfo {author} {\bibfnamefont {S.~G.}\ \bibnamefont {Louie}},\ }\href@noop {} {\bibfield  {journal} {\bibinfo  {journal} {Phys. Rev. B}\ }\textbf {\bibinfo {volume} {81}},\ \bibinfo {pages} {115105} (\bibinfo {year} {2010})}\BibitemShut {NoStop}%
\bibitem [{\citenamefont {Berger}\ \emph {et~al.}(2010)\citenamefont {Berger}, \citenamefont {Reining},\ and\ \citenamefont {Sottile}}]{gw_berger_2010}%
  \BibitemOpen
  \bibfield  {author} {\bibinfo {author} {\bibfnamefont {J.~A.}\ \bibnamefont {Berger}}, \bibinfo {author} {\bibfnamefont {L.}~\bibnamefont {Reining}}, \ and\ \bibinfo {author} {\bibfnamefont {F.}~\bibnamefont {Sottile}},\ }\href@noop {} {\bibfield  {journal} {\bibinfo  {journal} {Phys. Rev. B}\ }\textbf {\bibinfo {volume} {82}},\ \bibinfo {pages} {041103} (\bibinfo {year} {2010})}\BibitemShut {NoStop}%
\bibitem [{\citenamefont {Kang}\ and\ \citenamefont {Hybertsen}(2010)}]{gw_kang_2010}%
  \BibitemOpen
  \bibfield  {author} {\bibinfo {author} {\bibfnamefont {W.}~\bibnamefont {Kang}}\ and\ \bibinfo {author} {\bibfnamefont {M.~S.}\ \bibnamefont {Hybertsen}},\ }\href@noop {} {\bibfield  {journal} {\bibinfo  {journal} {Phys. Rev. B}\ }\textbf {\bibinfo {volume} {82}},\ \bibinfo {pages} {195108} (\bibinfo {year} {2010})}\BibitemShut {NoStop}%
\bibitem [{\citenamefont {Foerster}\ \emph {et~al.}(2011)\citenamefont {Foerster}, \citenamefont {Koval},\ and\ \citenamefont {S\'{a}nchez-Portal}}]{gw_foerster_2011}%
  \BibitemOpen
  \bibfield  {author} {\bibinfo {author} {\bibfnamefont {D.}~\bibnamefont {Foerster}}, \bibinfo {author} {\bibfnamefont {P.}~\bibnamefont {Koval}}, \ and\ \bibinfo {author} {\bibfnamefont {D.}~\bibnamefont {S\'{a}nchez-Portal}},\ }\href@noop {} {\bibfield  {journal} {\bibinfo  {journal} {J. Chem. Phys.}\ }\textbf {\bibinfo {volume} {135}},\ \bibinfo {pages} {074105} (\bibinfo {year} {2011})}\BibitemShut {NoStop}%
\bibitem [{\citenamefont {Berger}\ \emph {et~al.}(2012)\citenamefont {Berger}, \citenamefont {Reining},\ and\ \citenamefont {Sottile}}]{gw_berger_2012}%
  \BibitemOpen
  \bibfield  {author} {\bibinfo {author} {\bibfnamefont {J.~A.}\ \bibnamefont {Berger}}, \bibinfo {author} {\bibfnamefont {L.}~\bibnamefont {Reining}}, \ and\ \bibinfo {author} {\bibfnamefont {F.}~\bibnamefont {Sottile}},\ }\href@noop {} {\bibfield  {journal} {\bibinfo  {journal} {Phys. Rev. B}\ }\textbf {\bibinfo {volume} {85}},\ \bibinfo {pages} {085126} (\bibinfo {year} {2012})}\BibitemShut {NoStop}%
\bibitem [{\citenamefont {Nguyen}\ \emph {et~al.}(2012)\citenamefont {Nguyen}, \citenamefont {Pham}, \citenamefont {Rocca},\ and\ \citenamefont {Galli}}]{pdep_nguyen_2012}%
  \BibitemOpen
  \bibfield  {author} {\bibinfo {author} {\bibfnamefont {H.-V.}\ \bibnamefont {Nguyen}}, \bibinfo {author} {\bibfnamefont {T.~A.}\ \bibnamefont {Pham}}, \bibinfo {author} {\bibfnamefont {D.}~\bibnamefont {Rocca}}, \ and\ \bibinfo {author} {\bibfnamefont {G.}~\bibnamefont {Galli}},\ }\href@noop {} {\bibfield  {journal} {\bibinfo  {journal} {Phys. Rev. B}\ }\textbf {\bibinfo {volume} {85}},\ \bibinfo {pages} {081101} (\bibinfo {year} {2012})}\BibitemShut {NoStop}%
\bibitem [{\citenamefont {Pham}\ \emph {et~al.}(2013)\citenamefont {Pham}, \citenamefont {Nguyen}, \citenamefont {Rocca},\ and\ \citenamefont {Galli}}]{pdep_pham_2013}%
  \BibitemOpen
  \bibfield  {author} {\bibinfo {author} {\bibfnamefont {T.~A.}\ \bibnamefont {Pham}}, \bibinfo {author} {\bibfnamefont {H.-V.}\ \bibnamefont {Nguyen}}, \bibinfo {author} {\bibfnamefont {D.}~\bibnamefont {Rocca}}, \ and\ \bibinfo {author} {\bibfnamefont {G.}~\bibnamefont {Galli}},\ }\href@noop {} {\bibfield  {journal} {\bibinfo  {journal} {Phys. Rev. B}\ }\textbf {\bibinfo {volume} {87}},\ \bibinfo {pages} {155148} (\bibinfo {year} {2013})}\BibitemShut {NoStop}%
\bibitem [{\citenamefont {Lambert}\ and\ \citenamefont {Giustino}(2013)}]{gw_lambert_2013}%
  \BibitemOpen
  \bibfield  {author} {\bibinfo {author} {\bibfnamefont {H.}~\bibnamefont {Lambert}}\ and\ \bibinfo {author} {\bibfnamefont {F.}~\bibnamefont {Giustino}},\ }\href@noop {} {\bibfield  {journal} {\bibinfo  {journal} {Phys. Rev. B}\ }\textbf {\bibinfo {volume} {88}},\ \bibinfo {pages} {075117} (\bibinfo {year} {2013})}\BibitemShut {NoStop}%
\bibitem [{\citenamefont {Neuhauser}\ \emph {et~al.}(2014)\citenamefont {Neuhauser}, \citenamefont {Gao}, \citenamefont {Arntsen}, \citenamefont {Karshenas}, \citenamefont {Rabani},\ and\ \citenamefont {Baer}}]{stochastic_neuhauser_2014}%
  \BibitemOpen
  \bibfield  {author} {\bibinfo {author} {\bibfnamefont {D.}~\bibnamefont {Neuhauser}}, \bibinfo {author} {\bibfnamefont {Y.}~\bibnamefont {Gao}}, \bibinfo {author} {\bibfnamefont {C.}~\bibnamefont {Arntsen}}, \bibinfo {author} {\bibfnamefont {C.}~\bibnamefont {Karshenas}}, \bibinfo {author} {\bibfnamefont {E.}~\bibnamefont {Rabani}}, \ and\ \bibinfo {author} {\bibfnamefont {R.}~\bibnamefont {Baer}},\ }\href@noop {} {\bibfield  {journal} {\bibinfo  {journal} {Phys. Rev. Lett.}\ }\textbf {\bibinfo {volume} {113}},\ \bibinfo {pages} {076402} (\bibinfo {year} {2014})}\BibitemShut {NoStop}%
\bibitem [{\citenamefont {Janssen}\ \emph {et~al.}(2015)\citenamefont {Janssen}, \citenamefont {Rousseau},\ and\ \citenamefont {Côté}}]{gw_janssen_2015}%
  \BibitemOpen
  \bibfield  {author} {\bibinfo {author} {\bibfnamefont {J.~L.}\ \bibnamefont {Janssen}}, \bibinfo {author} {\bibfnamefont {B.}~\bibnamefont {Rousseau}}, \ and\ \bibinfo {author} {\bibfnamefont {M.}~\bibnamefont {Côté}},\ }\href@noop {} {\bibfield  {journal} {\bibinfo  {journal} {Phys. Rev. B}\ }\textbf {\bibinfo {volume} {91}},\ \bibinfo {pages} {125120} (\bibinfo {year} {2015})}\BibitemShut {NoStop}%
\bibitem [{\citenamefont {Bruneval}(2016)}]{gw_bruneval_2016}%
  \BibitemOpen
  \bibfield  {author} {\bibinfo {author} {\bibfnamefont {F.}~\bibnamefont {Bruneval}},\ }\href@noop {} {\bibfield  {journal} {\bibinfo  {journal} {J. Chem. Phys.}\ }\textbf {\bibinfo {volume} {145}},\ \bibinfo {pages} {234110} (\bibinfo {year} {2016})}\BibitemShut {NoStop}%
\bibitem [{\citenamefont {Liu}\ \emph {et~al.}(2016)\citenamefont {Liu}, \citenamefont {Kaltak}, \citenamefont {Klime\v{s}},\ and\ \citenamefont {Kresse}}]{gw_liu_2016}%
  \BibitemOpen
  \bibfield  {author} {\bibinfo {author} {\bibfnamefont {P.}~\bibnamefont {Liu}}, \bibinfo {author} {\bibfnamefont {M.}~\bibnamefont {Kaltak}}, \bibinfo {author} {\bibfnamefont {J.}~\bibnamefont {Klime\v{s}}}, \ and\ \bibinfo {author} {\bibfnamefont {G.}~\bibnamefont {Kresse}},\ }\href@noop {} {\bibfield  {journal} {\bibinfo  {journal} {Phys. Rev. B}\ }\textbf {\bibinfo {volume} {94}},\ \bibinfo {pages} {165109} (\bibinfo {year} {2016})}\BibitemShut {NoStop}%
\bibitem [{\citenamefont {Vl\v{c}ek}\ \emph {et~al.}(2017)\citenamefont {Vl\v{c}ek}, \citenamefont {Rabani}, \citenamefont {Neuhauser},\ and\ \citenamefont {Baer}}]{stochastic_vlcek_2017}%
  \BibitemOpen
  \bibfield  {author} {\bibinfo {author} {\bibfnamefont {V.}~\bibnamefont {Vl\v{c}ek}}, \bibinfo {author} {\bibfnamefont {E.}~\bibnamefont {Rabani}}, \bibinfo {author} {\bibfnamefont {D.}~\bibnamefont {Neuhauser}}, \ and\ \bibinfo {author} {\bibfnamefont {R.}~\bibnamefont {Baer}},\ }\href@noop {} {\bibfield  {journal} {\bibinfo  {journal} {J. Chem. Theory Comput.}\ }\textbf {\bibinfo {volume} {13}},\ \bibinfo {pages} {4997} (\bibinfo {year} {2017})}\BibitemShut {NoStop}%
\bibitem [{\citenamefont {Vl\v{c}ek}\ \emph {et~al.}(2018)\citenamefont {Vl\v{c}ek}, \citenamefont {Li}, \citenamefont {Baer}, \citenamefont {Rabani},\ and\ \citenamefont {Neuhauser}}]{stochastic_vlcek_2018}%
  \BibitemOpen
  \bibfield  {author} {\bibinfo {author} {\bibfnamefont {V.}~\bibnamefont {Vl\v{c}ek}}, \bibinfo {author} {\bibfnamefont {W.}~\bibnamefont {Li}}, \bibinfo {author} {\bibfnamefont {R.}~\bibnamefont {Baer}}, \bibinfo {author} {\bibfnamefont {E.}~\bibnamefont {Rabani}}, \ and\ \bibinfo {author} {\bibfnamefont {D.}~\bibnamefont {Neuhauser}},\ }\href@noop {} {\bibfield  {journal} {\bibinfo  {journal} {Phys. Rev. B}\ }\textbf {\bibinfo {volume} {98}},\ \bibinfo {pages} {075107} (\bibinfo {year} {2018})}\BibitemShut {NoStop}%
\bibitem [{\citenamefont {Wilhelm}\ \emph {et~al.}(2018)\citenamefont {Wilhelm}, \citenamefont {Golze}, \citenamefont {Talirz}, \citenamefont {Hutter},\ and\ \citenamefont {Pignedoli}}]{gw_wilhelm_2018}%
  \BibitemOpen
  \bibfield  {author} {\bibinfo {author} {\bibfnamefont {J.}~\bibnamefont {Wilhelm}}, \bibinfo {author} {\bibfnamefont {D.}~\bibnamefont {Golze}}, \bibinfo {author} {\bibfnamefont {L.}~\bibnamefont {Talirz}}, \bibinfo {author} {\bibfnamefont {J.}~\bibnamefont {Hutter}}, \ and\ \bibinfo {author} {\bibfnamefont {C.~A.}\ \bibnamefont {Pignedoli}},\ }\href@noop {} {\bibfield  {journal} {\bibinfo  {journal} {J. Phys. Chem. Lett.}\ }\textbf {\bibinfo {volume} {9}},\ \bibinfo {pages} {306} (\bibinfo {year} {2018})}\BibitemShut {NoStop}%
\bibitem [{\citenamefont {Del~Ben}\ \emph {et~al.}(2020)\citenamefont {Del~Ben}, \citenamefont {Yang}, \citenamefont {Li}, \citenamefont {Jornada}, \citenamefont {Louie},\ and\ \citenamefont {Deslippe}}]{berkelygw_delben_2020}%
  \BibitemOpen
  \bibfield  {author} {\bibinfo {author} {\bibfnamefont {M.}~\bibnamefont {Del~Ben}}, \bibinfo {author} {\bibfnamefont {C.}~\bibnamefont {Yang}}, \bibinfo {author} {\bibfnamefont {Z.}~\bibnamefont {Li}}, \bibinfo {author} {\bibfnamefont {F.~H.~d.}\ \bibnamefont {Jornada}}, \bibinfo {author} {\bibfnamefont {S.~G.}\ \bibnamefont {Louie}}, \ and\ \bibinfo {author} {\bibfnamefont {J.}~\bibnamefont {Deslippe}},\ }in\ \href@noop {} {\emph {\bibinfo {booktitle} {SC20: International Conference for High Performance Computing, Networking, Storage and Analysis}}}\ (\bibinfo  {publisher} {IEEE},\ \bibinfo {year} {2020})\ pp.\ \bibinfo {pages} {1--11}\BibitemShut {NoStop}%
\bibitem [{\citenamefont {F\"{o}rster}\ and\ \citenamefont {Visscher}(2020)}]{gw_forster_2020}%
  \BibitemOpen
  \bibfield  {author} {\bibinfo {author} {\bibfnamefont {A.}~\bibnamefont {F\"{o}rster}}\ and\ \bibinfo {author} {\bibfnamefont {L.}~\bibnamefont {Visscher}},\ }\href@noop {} {\bibfield  {journal} {\bibinfo  {journal} {J. Chem. Theory Comput.}\ }\textbf {\bibinfo {volume} {16}},\ \bibinfo {pages} {7381} (\bibinfo {year} {2020})}\BibitemShut {NoStop}%
\bibitem [{\citenamefont {F\"{o}rster}\ and\ \citenamefont {Visscher}(2021)}]{gw_forster_2021}%
  \BibitemOpen
  \bibfield  {author} {\bibinfo {author} {\bibfnamefont {A.}~\bibnamefont {F\"{o}rster}}\ and\ \bibinfo {author} {\bibfnamefont {L.}~\bibnamefont {Visscher}},\ }\href@noop {} {\bibfield  {journal} {\bibinfo  {journal} {Front. Chem.}\ }\textbf {\bibinfo {volume} {9}},\ \bibinfo {pages} {736591} (\bibinfo {year} {2021})}\BibitemShut {NoStop}%
\bibitem [{\citenamefont {Duchemin}\ and\ \citenamefont {Blase}(2021)}]{gw_duchemin_2021}%
  \BibitemOpen
  \bibfield  {author} {\bibinfo {author} {\bibfnamefont {I.}~\bibnamefont {Duchemin}}\ and\ \bibinfo {author} {\bibfnamefont {X.}~\bibnamefont {Blase}},\ }\href@noop {} {\bibfield  {journal} {\bibinfo  {journal} {J. Chem. Theory Comput.}\ }\textbf {\bibinfo {volume} {17}},\ \bibinfo {pages} {2383} (\bibinfo {year} {2021})}\BibitemShut {NoStop}%
\bibitem [{\citenamefont {Wilhelm}\ \emph {et~al.}(2021)\citenamefont {Wilhelm}, \citenamefont {Seewald},\ and\ \citenamefont {Golze}}]{gw_wilhelm_2021}%
  \BibitemOpen
  \bibfield  {author} {\bibinfo {author} {\bibfnamefont {J.}~\bibnamefont {Wilhelm}}, \bibinfo {author} {\bibfnamefont {P.}~\bibnamefont {Seewald}}, \ and\ \bibinfo {author} {\bibfnamefont {D.}~\bibnamefont {Golze}},\ }\href@noop {} {\bibfield  {journal} {\bibinfo  {journal} {J. Chem. Theory Comput.}\ }\textbf {\bibinfo {volume} {17}},\ \bibinfo {pages} {1662} (\bibinfo {year} {2021})}\BibitemShut {NoStop}%
\bibitem [{\citenamefont {Ma}\ \emph {et~al.}(2021{\natexlab{a}})\citenamefont {Ma}, \citenamefont {Wang}, \citenamefont {Wan}, \citenamefont {Li}, \citenamefont {Qin}, \citenamefont {Liu}, \citenamefont {Hu}, \citenamefont {Lin}, \citenamefont {Yang},\ and\ \citenamefont {Yang}}]{gw_ma_2021}%
  \BibitemOpen
  \bibfield  {author} {\bibinfo {author} {\bibfnamefont {H.}~\bibnamefont {Ma}}, \bibinfo {author} {\bibfnamefont {L.}~\bibnamefont {Wang}}, \bibinfo {author} {\bibfnamefont {L.}~\bibnamefont {Wan}}, \bibinfo {author} {\bibfnamefont {J.}~\bibnamefont {Li}}, \bibinfo {author} {\bibfnamefont {X.}~\bibnamefont {Qin}}, \bibinfo {author} {\bibfnamefont {J.}~\bibnamefont {Liu}}, \bibinfo {author} {\bibfnamefont {W.}~\bibnamefont {Hu}}, \bibinfo {author} {\bibfnamefont {L.}~\bibnamefont {Lin}}, \bibinfo {author} {\bibfnamefont {C.}~\bibnamefont {Yang}}, \ and\ \bibinfo {author} {\bibfnamefont {J.}~\bibnamefont {Yang}},\ }\href@noop {} {\bibfield  {journal} {\bibinfo  {journal} {J. Phys. Chem. A}\ }\textbf {\bibinfo {volume} {125}},\ \bibinfo {pages} {7545} (\bibinfo {year} {2021}{\natexlab{a}})}\BibitemShut {NoStop}%
\bibitem [{\citenamefont {Leon}\ \emph {et~al.}(2021)\citenamefont {Leon}, \citenamefont {Cardoso}, \citenamefont {Chiarotti}, \citenamefont {Varsano}, \citenamefont {Molinari},\ and\ \citenamefont {Ferretti}}]{gw_leon_2021}%
  \BibitemOpen
  \bibfield  {author} {\bibinfo {author} {\bibfnamefont {D.~A.}\ \bibnamefont {Leon}}, \bibinfo {author} {\bibfnamefont {C.}~\bibnamefont {Cardoso}}, \bibinfo {author} {\bibfnamefont {T.}~\bibnamefont {Chiarotti}}, \bibinfo {author} {\bibfnamefont {D.}~\bibnamefont {Varsano}}, \bibinfo {author} {\bibfnamefont {E.}~\bibnamefont {Molinari}}, \ and\ \bibinfo {author} {\bibfnamefont {A.}~\bibnamefont {Ferretti}},\ }\href@noop {} {\bibfield  {journal} {\bibinfo  {journal} {Phys. Rev. B}\ }\textbf {\bibinfo {volume} {104}},\ \bibinfo {pages} {115157} (\bibinfo {year} {2021})}\BibitemShut {NoStop}%
\bibitem [{\citenamefont {Leon}\ \emph {et~al.}(2023)\citenamefont {Leon}, \citenamefont {Ferretti}, \citenamefont {Varsano}, \citenamefont {Molinari},\ and\ \citenamefont {Cardoso}}]{gw_leon_2023}%
  \BibitemOpen
  \bibfield  {author} {\bibinfo {author} {\bibfnamefont {D.~A.}\ \bibnamefont {Leon}}, \bibinfo {author} {\bibfnamefont {A.}~\bibnamefont {Ferretti}}, \bibinfo {author} {\bibfnamefont {D.}~\bibnamefont {Varsano}}, \bibinfo {author} {\bibfnamefont {E.}~\bibnamefont {Molinari}}, \ and\ \bibinfo {author} {\bibfnamefont {C.}~\bibnamefont {Cardoso}},\ }\href@noop {} {\bibfield  {journal} {\bibinfo  {journal} {Phys. Rev. B}\ }\textbf {\bibinfo {volume} {107}},\ \bibinfo {pages} {155130} (\bibinfo {year} {2023})}\BibitemShut {NoStop}%
\bibitem [{\citenamefont {Rocca}\ \emph {et~al.}(2010)\citenamefont {Rocca}, \citenamefont {Lu},\ and\ \citenamefont {Galli}}]{bse_rocca_2010}%
  \BibitemOpen
  \bibfield  {author} {\bibinfo {author} {\bibfnamefont {D.}~\bibnamefont {Rocca}}, \bibinfo {author} {\bibfnamefont {D.}~\bibnamefont {Lu}}, \ and\ \bibinfo {author} {\bibfnamefont {G.}~\bibnamefont {Galli}},\ }\href@noop {} {\bibfield  {journal} {\bibinfo  {journal} {J. Chem. Phys.}\ }\textbf {\bibinfo {volume} {133}},\ \bibinfo {pages} {164109} (\bibinfo {year} {2010})}\BibitemShut {NoStop}%
\bibitem [{\citenamefont {Rocca}\ \emph {et~al.}(2012)\citenamefont {Rocca}, \citenamefont {Ping}, \citenamefont {Gebauer},\ and\ \citenamefont {Galli}}]{bse_rocca_2012}%
  \BibitemOpen
  \bibfield  {author} {\bibinfo {author} {\bibfnamefont {D.}~\bibnamefont {Rocca}}, \bibinfo {author} {\bibfnamefont {Y.}~\bibnamefont {Ping}}, \bibinfo {author} {\bibfnamefont {R.}~\bibnamefont {Gebauer}}, \ and\ \bibinfo {author} {\bibfnamefont {G.}~\bibnamefont {Galli}},\ }\href@noop {} {\bibfield  {journal} {\bibinfo  {journal} {Phys. Rev. B}\ }\textbf {\bibinfo {volume} {85}},\ \bibinfo {pages} {045116} (\bibinfo {year} {2012})}\BibitemShut {NoStop}%
\bibitem [{\citenamefont {Ping}\ \emph {et~al.}(2012)\citenamefont {Ping}, \citenamefont {Rocca}, \citenamefont {Lu},\ and\ \citenamefont {Galli}}]{bse_ping_2012}%
  \BibitemOpen
  \bibfield  {author} {\bibinfo {author} {\bibfnamefont {Y.}~\bibnamefont {Ping}}, \bibinfo {author} {\bibfnamefont {D.}~\bibnamefont {Rocca}}, \bibinfo {author} {\bibfnamefont {D.}~\bibnamefont {Lu}}, \ and\ \bibinfo {author} {\bibfnamefont {G.}~\bibnamefont {Galli}},\ }\href@noop {} {\bibfield  {journal} {\bibinfo  {journal} {Phys. Rev. B}\ }\textbf {\bibinfo {volume} {85}},\ \bibinfo {pages} {035316} (\bibinfo {year} {2012})}\BibitemShut {NoStop}%
\bibitem [{\citenamefont {Marsili}\ \emph {et~al.}(2017)\citenamefont {Marsili}, \citenamefont {Mosconi}, \citenamefont {De~Angelis},\ and\ \citenamefont {Umari}}]{bse_marsili_2017}%
  \BibitemOpen
  \bibfield  {author} {\bibinfo {author} {\bibfnamefont {M.}~\bibnamefont {Marsili}}, \bibinfo {author} {\bibfnamefont {E.}~\bibnamefont {Mosconi}}, \bibinfo {author} {\bibfnamefont {F.}~\bibnamefont {De~Angelis}}, \ and\ \bibinfo {author} {\bibfnamefont {P.}~\bibnamefont {Umari}},\ }\href@noop {} {\bibfield  {journal} {\bibinfo  {journal} {Phys. Rev. B}\ }\textbf {\bibinfo {volume} {95}},\ \bibinfo {pages} {075415} (\bibinfo {year} {2017})}\BibitemShut {NoStop}%
\bibitem [{\citenamefont {Krause}\ and\ \citenamefont {Klopper}(2017)}]{bse_krause_2017}%
  \BibitemOpen
  \bibfield  {author} {\bibinfo {author} {\bibfnamefont {K.}~\bibnamefont {Krause}}\ and\ \bibinfo {author} {\bibfnamefont {W.}~\bibnamefont {Klopper}},\ }\href@noop {} {\bibfield  {journal} {\bibinfo  {journal} {J. Comput. Chem.}\ }\textbf {\bibinfo {volume} {38}},\ \bibinfo {pages} {383} (\bibinfo {year} {2017})}\BibitemShut {NoStop}%
\bibitem [{\citenamefont {Bradbury}\ \emph {et~al.}(2022)\citenamefont {Bradbury}, \citenamefont {Nguyen}, \citenamefont {Caram},\ and\ \citenamefont {Neuhauser}}]{bse_bradbury_2022}%
  \BibitemOpen
  \bibfield  {author} {\bibinfo {author} {\bibfnamefont {N.~C.}\ \bibnamefont {Bradbury}}, \bibinfo {author} {\bibfnamefont {M.}~\bibnamefont {Nguyen}}, \bibinfo {author} {\bibfnamefont {J.~R.}\ \bibnamefont {Caram}}, \ and\ \bibinfo {author} {\bibfnamefont {D.}~\bibnamefont {Neuhauser}},\ }\href@noop {} {\bibfield  {journal} {\bibinfo  {journal} {J. Chem. Phys.}\ }\textbf {\bibinfo {volume} {157}},\ \bibinfo {pages} {031104} (\bibinfo {year} {2022})}\BibitemShut {NoStop}%
\bibitem [{\citenamefont {Franzke}\ \emph {et~al.}(2022)\citenamefont {Franzke}, \citenamefont {Holzer},\ and\ \citenamefont {Mack}}]{bse_franzke_2022}%
  \BibitemOpen
  \bibfield  {author} {\bibinfo {author} {\bibfnamefont {Y.~J.}\ \bibnamefont {Franzke}}, \bibinfo {author} {\bibfnamefont {C.}~\bibnamefont {Holzer}}, \ and\ \bibinfo {author} {\bibfnamefont {F.}~\bibnamefont {Mack}},\ }\href@noop {} {\bibfield  {journal} {\bibinfo  {journal} {J. Chem. Theory Comput.}\ }\textbf {\bibinfo {volume} {18}},\ \bibinfo {pages} {1030} (\bibinfo {year} {2022})}\BibitemShut {NoStop}%
\bibitem [{\citenamefont {Merkel}\ and\ \citenamefont {Ortmann}(2024)}]{bse_merkel_2024}%
  \BibitemOpen
  \bibfield  {author} {\bibinfo {author} {\bibfnamefont {K.}~\bibnamefont {Merkel}}\ and\ \bibinfo {author} {\bibfnamefont {F.}~\bibnamefont {Ortmann}},\ }\href@noop {} {\bibfield  {journal} {\bibinfo  {journal} {J. Phys. Mater.}\ }\textbf {\bibinfo {volume} {7}},\ \bibinfo {pages} {015001} (\bibinfo {year} {2024})}\BibitemShut {NoStop}%
\bibitem [{wes(2024)}]{west_website}%
  \BibitemOpen
  \href@noop {} {\enquote {\bibinfo {title} {{WEST}},}\ }\bibinfo {howpublished} {https://west-code.org} (\bibinfo {year} {(accessed October 20, 2024)})\BibitemShut {NoStop}%
\bibitem [{\citenamefont {Govoni}\ and\ \citenamefont {Galli}(2015)}]{west_govoni_2015}%
  \BibitemOpen
  \bibfield  {author} {\bibinfo {author} {\bibfnamefont {M.}~\bibnamefont {Govoni}}\ and\ \bibinfo {author} {\bibfnamefont {G.}~\bibnamefont {Galli}},\ }\href@noop {} {\bibfield  {journal} {\bibinfo  {journal} {J. Chem. Theory Comput.}\ }\textbf {\bibinfo {volume} {11}},\ \bibinfo {pages} {2680} (\bibinfo {year} {2015})}\BibitemShut {NoStop}%
\bibitem [{\citenamefont {Scherpelz}\ \emph {et~al.}(2016)\citenamefont {Scherpelz}, \citenamefont {Govoni}, \citenamefont {Hamada},\ and\ \citenamefont {Galli}}]{soc_scherpelz_2016}%
  \BibitemOpen
  \bibfield  {author} {\bibinfo {author} {\bibfnamefont {P.}~\bibnamefont {Scherpelz}}, \bibinfo {author} {\bibfnamefont {M.}~\bibnamefont {Govoni}}, \bibinfo {author} {\bibfnamefont {I.}~\bibnamefont {Hamada}}, \ and\ \bibinfo {author} {\bibfnamefont {G.}~\bibnamefont {Galli}},\ }\href@noop {} {\bibfield  {journal} {\bibinfo  {journal} {J. Chem. Theory Comput.}\ }\textbf {\bibinfo {volume} {12}},\ \bibinfo {pages} {3523} (\bibinfo {year} {2016})}\BibitemShut {NoStop}%
\bibitem [{\citenamefont {Govoni}\ and\ \citenamefont {Galli}(2018)}]{gw100_govoni_2018}%
  \BibitemOpen
  \bibfield  {author} {\bibinfo {author} {\bibfnamefont {M.}~\bibnamefont {Govoni}}\ and\ \bibinfo {author} {\bibfnamefont {G.}~\bibnamefont {Galli}},\ }\href@noop {} {\bibfield  {journal} {\bibinfo  {journal} {J. Chem. Theory Comput.}\ }\textbf {\bibinfo {volume} {14}},\ \bibinfo {pages} {1895} (\bibinfo {year} {2018})}\BibitemShut {NoStop}%
\bibitem [{\citenamefont {Ma}\ \emph {et~al.}(2019)\citenamefont {Ma}, \citenamefont {Govoni}, \citenamefont {Gygi},\ and\ \citenamefont {Galli}}]{gw_ma_2019}%
  \BibitemOpen
  \bibfield  {author} {\bibinfo {author} {\bibfnamefont {H.}~\bibnamefont {Ma}}, \bibinfo {author} {\bibfnamefont {M.}~\bibnamefont {Govoni}}, \bibinfo {author} {\bibfnamefont {F.}~\bibnamefont {Gygi}}, \ and\ \bibinfo {author} {\bibfnamefont {G.}~\bibnamefont {Galli}},\ }\href@noop {} {\bibfield  {journal} {\bibinfo  {journal} {J. Chem. Theory Comput.}\ }\textbf {\bibinfo {volume} {15}},\ \bibinfo {pages} {154} (\bibinfo {year} {2019})}\BibitemShut {NoStop}%
\bibitem [{\citenamefont {Yang}\ \emph {et~al.}(2019)\citenamefont {Yang}, \citenamefont {Govoni},\ and\ \citenamefont {Galli}}]{gw_yang_2019}%
  \BibitemOpen
  \bibfield  {author} {\bibinfo {author} {\bibfnamefont {H.}~\bibnamefont {Yang}}, \bibinfo {author} {\bibfnamefont {M.}~\bibnamefont {Govoni}}, \ and\ \bibinfo {author} {\bibfnamefont {G.}~\bibnamefont {Galli}},\ }\href@noop {} {\bibfield  {journal} {\bibinfo  {journal} {J. Chem. Phys.}\ }\textbf {\bibinfo {volume} {151}},\ \bibinfo {pages} {224102} (\bibinfo {year} {2019})}\BibitemShut {NoStop}%
\bibitem [{\citenamefont {Yu}\ and\ \citenamefont {Govoni}(2022)}]{west_yu_2022}%
  \BibitemOpen
  \bibfield  {author} {\bibinfo {author} {\bibfnamefont {V.~W.-z.}\ \bibnamefont {Yu}}\ and\ \bibinfo {author} {\bibfnamefont {M.}~\bibnamefont {Govoni}},\ }\href@noop {} {\bibfield  {journal} {\bibinfo  {journal} {J. Chem. Theory Comput.}\ }\textbf {\bibinfo {volume} {18}},\ \bibinfo {pages} {4690} (\bibinfo {year} {2022})}\BibitemShut {NoStop}%
\bibitem [{\citenamefont {Nguyen}\ \emph {et~al.}(2019)\citenamefont {Nguyen}, \citenamefont {Ma}, \citenamefont {Govoni}, \citenamefont {Gygi},\ and\ \citenamefont {Galli}}]{bse_nguyen_2019}%
  \BibitemOpen
  \bibfield  {author} {\bibinfo {author} {\bibfnamefont {N.~L.}\ \bibnamefont {Nguyen}}, \bibinfo {author} {\bibfnamefont {H.}~\bibnamefont {Ma}}, \bibinfo {author} {\bibfnamefont {M.}~\bibnamefont {Govoni}}, \bibinfo {author} {\bibfnamefont {F.}~\bibnamefont {Gygi}}, \ and\ \bibinfo {author} {\bibfnamefont {G.}~\bibnamefont {Galli}},\ }\href@noop {} {\bibfield  {journal} {\bibinfo  {journal} {Phys. Rev. Lett.}\ }\textbf {\bibinfo {volume} {122}},\ \bibinfo {pages} {237402} (\bibinfo {year} {2019})}\BibitemShut {NoStop}%
\bibitem [{\citenamefont {Dong}\ \emph {et~al.}(2021)\citenamefont {Dong}, \citenamefont {Govoni},\ and\ \citenamefont {Galli}}]{bse_dong_2021}%
  \BibitemOpen
  \bibfield  {author} {\bibinfo {author} {\bibfnamefont {S.~S.}\ \bibnamefont {Dong}}, \bibinfo {author} {\bibfnamefont {M.}~\bibnamefont {Govoni}}, \ and\ \bibinfo {author} {\bibfnamefont {G.}~\bibnamefont {Galli}},\ }\href@noop {} {\bibfield  {journal} {\bibinfo  {journal} {Chem. Sci.}\ }\textbf {\bibinfo {volume} {12}},\ \bibinfo {pages} {4970} (\bibinfo {year} {2021})}\BibitemShut {NoStop}%
\bibitem [{\citenamefont {Yu}\ \emph {et~al.}(2024)\citenamefont {Yu}, \citenamefont {Jin}, \citenamefont {Galli},\ and\ \citenamefont {Govoni}}]{bse_yu_2024}%
  \BibitemOpen
  \bibfield  {author} {\bibinfo {author} {\bibfnamefont {V.~W.-z.}\ \bibnamefont {Yu}}, \bibinfo {author} {\bibfnamefont {Y.}~\bibnamefont {Jin}}, \bibinfo {author} {\bibfnamefont {G.}~\bibnamefont {Galli}}, \ and\ \bibinfo {author} {\bibfnamefont {M.}~\bibnamefont {Govoni}},\ }\href@noop {} {\bibfield  {journal} {\bibinfo  {journal} {J. Chem. Theory Comput.}\ }\textbf {\bibinfo {volume} {20}},\ \bibinfo {pages} {10899} (\bibinfo {year} {2024})}\BibitemShut {NoStop}%
\bibitem [{\citenamefont {Jin}\ \emph {et~al.}(2022)\citenamefont {Jin}, \citenamefont {Govoni},\ and\ \citenamefont {Galli}}]{tddft_jin_2022}%
  \BibitemOpen
  \bibfield  {author} {\bibinfo {author} {\bibfnamefont {Y.}~\bibnamefont {Jin}}, \bibinfo {author} {\bibfnamefont {M.}~\bibnamefont {Govoni}}, \ and\ \bibinfo {author} {\bibfnamefont {G.}~\bibnamefont {Galli}},\ }\href@noop {} {\bibfield  {journal} {\bibinfo  {journal} {npj Comput. Mater.}\ }\textbf {\bibinfo {volume} {8}},\ \bibinfo {pages} {238} (\bibinfo {year} {2022})}\BibitemShut {NoStop}%
\bibitem [{\citenamefont {Jin}\ \emph {et~al.}(2023)\citenamefont {Jin}, \citenamefont {Yu}, \citenamefont {Govoni}, \citenamefont {Xu},\ and\ \citenamefont {Galli}}]{tddft_jin_2023}%
  \BibitemOpen
  \bibfield  {author} {\bibinfo {author} {\bibfnamefont {Y.}~\bibnamefont {Jin}}, \bibinfo {author} {\bibfnamefont {V.~W.-z.}\ \bibnamefont {Yu}}, \bibinfo {author} {\bibfnamefont {M.}~\bibnamefont {Govoni}}, \bibinfo {author} {\bibfnamefont {A.~C.}\ \bibnamefont {Xu}}, \ and\ \bibinfo {author} {\bibfnamefont {G.}~\bibnamefont {Galli}},\ }\href@noop {} {\bibfield  {journal} {\bibinfo  {journal} {J. Chem. Theory Comput.}\ }\textbf {\bibinfo {volume} {19}},\ \bibinfo {pages} {8689} (\bibinfo {year} {2023})}\BibitemShut {NoStop}%
\bibitem [{\citenamefont {Ma}\ \emph {et~al.}(2020)\citenamefont {Ma}, \citenamefont {Sheng}, \citenamefont {Govoni},\ and\ \citenamefont {Galli}}]{qdet_ma_2020}%
  \BibitemOpen
  \bibfield  {author} {\bibinfo {author} {\bibfnamefont {H.}~\bibnamefont {Ma}}, \bibinfo {author} {\bibfnamefont {N.}~\bibnamefont {Sheng}}, \bibinfo {author} {\bibfnamefont {M.}~\bibnamefont {Govoni}}, \ and\ \bibinfo {author} {\bibfnamefont {G.}~\bibnamefont {Galli}},\ }\href@noop {} {\bibfield  {journal} {\bibinfo  {journal} {Phys. Chem. Chem. Phys.}\ }\textbf {\bibinfo {volume} {22}},\ \bibinfo {pages} {25522} (\bibinfo {year} {2020})}\BibitemShut {NoStop}%
\bibitem [{\citenamefont {Ma}\ \emph {et~al.}(2021{\natexlab{b}})\citenamefont {Ma}, \citenamefont {Sheng}, \citenamefont {Govoni},\ and\ \citenamefont {Galli}}]{qdet_ma_2021}%
  \BibitemOpen
  \bibfield  {author} {\bibinfo {author} {\bibfnamefont {H.}~\bibnamefont {Ma}}, \bibinfo {author} {\bibfnamefont {N.}~\bibnamefont {Sheng}}, \bibinfo {author} {\bibfnamefont {M.}~\bibnamefont {Govoni}}, \ and\ \bibinfo {author} {\bibfnamefont {G.}~\bibnamefont {Galli}},\ }\href@noop {} {\bibfield  {journal} {\bibinfo  {journal} {J. Chem. Theory Comput.}\ }\textbf {\bibinfo {volume} {17}},\ \bibinfo {pages} {2116} (\bibinfo {year} {2021}{\natexlab{b}})}\BibitemShut {NoStop}%
\bibitem [{\citenamefont {Sheng}\ \emph {et~al.}(2022)\citenamefont {Sheng}, \citenamefont {Vorwerk}, \citenamefont {Govoni},\ and\ \citenamefont {Galli}}]{qdet_sheng_2022}%
  \BibitemOpen
  \bibfield  {author} {\bibinfo {author} {\bibfnamefont {N.}~\bibnamefont {Sheng}}, \bibinfo {author} {\bibfnamefont {C.}~\bibnamefont {Vorwerk}}, \bibinfo {author} {\bibfnamefont {M.}~\bibnamefont {Govoni}}, \ and\ \bibinfo {author} {\bibfnamefont {G.}~\bibnamefont {Galli}},\ }\href@noop {} {\bibfield  {journal} {\bibinfo  {journal} {J. Chem. Theory Comput.}\ }\textbf {\bibinfo {volume} {18}},\ \bibinfo {pages} {3512} (\bibinfo {year} {2022})}\BibitemShut {NoStop}%
\bibitem [{\citenamefont {Perdew}\ \emph {et~al.}(1996)\citenamefont {Perdew}, \citenamefont {Burke},\ and\ \citenamefont {Ernzerhof}}]{pbe_perdew_1996}%
  \BibitemOpen
  \bibfield  {author} {\bibinfo {author} {\bibfnamefont {J.~P.}\ \bibnamefont {Perdew}}, \bibinfo {author} {\bibfnamefont {K.}~\bibnamefont {Burke}}, \ and\ \bibinfo {author} {\bibfnamefont {M.}~\bibnamefont {Ernzerhof}},\ }\href@noop {} {\bibfield  {journal} {\bibinfo  {journal} {Phys. Rev. Lett.}\ }\textbf {\bibinfo {volume} {77}},\ \bibinfo {pages} {3865} (\bibinfo {year} {1996})}\BibitemShut {NoStop}%
\bibitem [{\citenamefont {Gavini}\ \emph {et~al.}(2023)\citenamefont {Gavini}, \citenamefont {Baroni}, \citenamefont {Blum}, \citenamefont {Bowler}, \citenamefont {Buccheri}, \citenamefont {Chelikowsky}, \citenamefont {Das}, \citenamefont {Dawson}, \citenamefont {Delugas}, \citenamefont {Dogan}, \citenamefont {Draxl}, \citenamefont {Galli}, \citenamefont {Genovese}, \citenamefont {Giannozzi}, \citenamefont {Giantomassi}, \citenamefont {Gonze}, \citenamefont {Govoni}, \citenamefont {Gygi}, \citenamefont {Gulans}, \citenamefont {Herbert}, \citenamefont {Kokott}, \citenamefont {K\"{u}hne}, \citenamefont {Liou}, \citenamefont {Miyazaki}, \citenamefont {Motamarri}, \citenamefont {Nakata}, \citenamefont {Pask}, \citenamefont {Plessl}, \citenamefont {Ratcliff}, \citenamefont {Richard}, \citenamefont {Rossi}, \citenamefont {Schade}, \citenamefont {Scheffler}, \citenamefont {Sch\"{u}tt}, \citenamefont {Suryanarayana}, \citenamefont {Torrent}, \citenamefont {Truflandier}, \citenamefont {Windus}, \citenamefont {Xu},
  \citenamefont {Yu},\ and\ \citenamefont {Perez}}]{roadmap_gavini_2023}%
  \BibitemOpen
  \bibfield  {author} {\bibinfo {author} {\bibfnamefont {V.}~\bibnamefont {Gavini}}, \bibinfo {author} {\bibfnamefont {S.}~\bibnamefont {Baroni}}, \bibinfo {author} {\bibfnamefont {V.}~\bibnamefont {Blum}}, \bibinfo {author} {\bibfnamefont {D.~R.}\ \bibnamefont {Bowler}}, \bibinfo {author} {\bibfnamefont {A.}~\bibnamefont {Buccheri}}, \bibinfo {author} {\bibfnamefont {J.~R.}\ \bibnamefont {Chelikowsky}}, \bibinfo {author} {\bibfnamefont {S.}~\bibnamefont {Das}}, \bibinfo {author} {\bibfnamefont {W.}~\bibnamefont {Dawson}}, \bibinfo {author} {\bibfnamefont {P.}~\bibnamefont {Delugas}}, \bibinfo {author} {\bibfnamefont {M.}~\bibnamefont {Dogan}}, \bibinfo {author} {\bibfnamefont {C.}~\bibnamefont {Draxl}}, \bibinfo {author} {\bibfnamefont {G.}~\bibnamefont {Galli}}, \bibinfo {author} {\bibfnamefont {L.}~\bibnamefont {Genovese}}, \bibinfo {author} {\bibfnamefont {P.}~\bibnamefont {Giannozzi}}, \bibinfo {author} {\bibfnamefont {M.}~\bibnamefont {Giantomassi}}, \bibinfo {author} {\bibfnamefont {X.}~\bibnamefont
  {Gonze}}, \bibinfo {author} {\bibfnamefont {M.}~\bibnamefont {Govoni}}, \bibinfo {author} {\bibfnamefont {F.}~\bibnamefont {Gygi}}, \bibinfo {author} {\bibfnamefont {A.}~\bibnamefont {Gulans}}, \bibinfo {author} {\bibfnamefont {J.~M.}\ \bibnamefont {Herbert}}, \bibinfo {author} {\bibfnamefont {S.}~\bibnamefont {Kokott}}, \bibinfo {author} {\bibfnamefont {T.~D.}\ \bibnamefont {K\"{u}hne}}, \bibinfo {author} {\bibfnamefont {K.-H.}\ \bibnamefont {Liou}}, \bibinfo {author} {\bibfnamefont {T.}~\bibnamefont {Miyazaki}}, \bibinfo {author} {\bibfnamefont {P.}~\bibnamefont {Motamarri}}, \bibinfo {author} {\bibfnamefont {A.}~\bibnamefont {Nakata}}, \bibinfo {author} {\bibfnamefont {J.~E.}\ \bibnamefont {Pask}}, \bibinfo {author} {\bibfnamefont {C.}~\bibnamefont {Plessl}}, \bibinfo {author} {\bibfnamefont {L.~E.}\ \bibnamefont {Ratcliff}}, \bibinfo {author} {\bibfnamefont {R.~M.}\ \bibnamefont {Richard}}, \bibinfo {author} {\bibfnamefont {M.}~\bibnamefont {Rossi}}, \bibinfo {author} {\bibfnamefont {R.}~\bibnamefont
  {Schade}}, \bibinfo {author} {\bibfnamefont {M.}~\bibnamefont {Scheffler}}, \bibinfo {author} {\bibfnamefont {O.}~\bibnamefont {Sch\"{u}tt}}, \bibinfo {author} {\bibfnamefont {P.}~\bibnamefont {Suryanarayana}}, \bibinfo {author} {\bibfnamefont {M.}~\bibnamefont {Torrent}}, \bibinfo {author} {\bibfnamefont {L.}~\bibnamefont {Truflandier}}, \bibinfo {author} {\bibfnamefont {T.~L.}\ \bibnamefont {Windus}}, \bibinfo {author} {\bibfnamefont {Q.}~\bibnamefont {Xu}}, \bibinfo {author} {\bibfnamefont {V.~W.-z.}\ \bibnamefont {Yu}}, \ and\ \bibinfo {author} {\bibfnamefont {D.}~\bibnamefont {Perez}},\ }\href@noop {} {\bibfield  {journal} {\bibinfo  {journal} {Model. Simul. Mat. Sci. Eng.}\ }\textbf {\bibinfo {volume} {31}},\ \bibinfo {pages} {063301} (\bibinfo {year} {2023})}\BibitemShut {NoStop}%
\bibitem [{\citenamefont {Hedin}(1965)}]{hedin_hedin_1965}%
  \BibitemOpen
  \bibfield  {author} {\bibinfo {author} {\bibfnamefont {L.}~\bibnamefont {Hedin}},\ }\href@noop {} {\bibfield  {journal} {\bibinfo  {journal} {Phys. Rev.}\ }\textbf {\bibinfo {volume} {139}},\ \bibinfo {pages} {A796} (\bibinfo {year} {1965})}\BibitemShut {NoStop}%
\bibitem [{\citenamefont {Sch\"{o}ne}\ and\ \citenamefont {Eguiluz}(1998)}]{scgw_schone_1998}%
  \BibitemOpen
  \bibfield  {author} {\bibinfo {author} {\bibfnamefont {W.-D.}\ \bibnamefont {Sch\"{o}ne}}\ and\ \bibinfo {author} {\bibfnamefont {A.~G.}\ \bibnamefont {Eguiluz}},\ }\href@noop {} {\bibfield  {journal} {\bibinfo  {journal} {Phys. Rev. Lett.}\ }\textbf {\bibinfo {volume} {81}},\ \bibinfo {pages} {1662} (\bibinfo {year} {1998})}\BibitemShut {NoStop}%
\bibitem [{\citenamefont {Fuchs}\ \emph {et~al.}(2007)\citenamefont {Fuchs}, \citenamefont {Furthm\"{u}ller}, \citenamefont {Bechstedt}, \citenamefont {Shishkin},\ and\ \citenamefont {Kresse}}]{exx_fuchs_2007}%
  \BibitemOpen
  \bibfield  {author} {\bibinfo {author} {\bibfnamefont {F.}~\bibnamefont {Fuchs}}, \bibinfo {author} {\bibfnamefont {J.}~\bibnamefont {Furthm\"{u}ller}}, \bibinfo {author} {\bibfnamefont {F.}~\bibnamefont {Bechstedt}}, \bibinfo {author} {\bibfnamefont {M.}~\bibnamefont {Shishkin}}, \ and\ \bibinfo {author} {\bibfnamefont {G.}~\bibnamefont {Kresse}},\ }\href@noop {} {\bibfield  {journal} {\bibinfo  {journal} {Phys. Rev. B}\ }\textbf {\bibinfo {volume} {76}},\ \bibinfo {pages} {115109} (\bibinfo {year} {2007})}\BibitemShut {NoStop}%
\bibitem [{\citenamefont {Shishkin}\ and\ \citenamefont {Kresse}(2007)}]{scgw_shishkin_2007}%
  \BibitemOpen
  \bibfield  {author} {\bibinfo {author} {\bibfnamefont {M.}~\bibnamefont {Shishkin}}\ and\ \bibinfo {author} {\bibfnamefont {G.}~\bibnamefont {Kresse}},\ }\href@noop {} {\bibfield  {journal} {\bibinfo  {journal} {Phys. Rev. B}\ }\textbf {\bibinfo {volume} {75}},\ \bibinfo {pages} {235102} (\bibinfo {year} {2007})}\BibitemShut {NoStop}%
\bibitem [{\citenamefont {Marom}\ \emph {et~al.}(2012)\citenamefont {Marom}, \citenamefont {Caruso}, \citenamefont {Ren}, \citenamefont {Hofmann}, \citenamefont {K\"{o}rzd\"{o}rfer}, \citenamefont {Chelikowsky}, \citenamefont {Rubio}, \citenamefont {Scheffler},\ and\ \citenamefont {Rinke}}]{exx_marom_2012}%
  \BibitemOpen
  \bibfield  {author} {\bibinfo {author} {\bibfnamefont {N.}~\bibnamefont {Marom}}, \bibinfo {author} {\bibfnamefont {F.}~\bibnamefont {Caruso}}, \bibinfo {author} {\bibfnamefont {X.}~\bibnamefont {Ren}}, \bibinfo {author} {\bibfnamefont {O.~T.}\ \bibnamefont {Hofmann}}, \bibinfo {author} {\bibfnamefont {T.}~\bibnamefont {K\"{o}rzd\"{o}rfer}}, \bibinfo {author} {\bibfnamefont {J.~R.}\ \bibnamefont {Chelikowsky}}, \bibinfo {author} {\bibfnamefont {A.}~\bibnamefont {Rubio}}, \bibinfo {author} {\bibfnamefont {M.}~\bibnamefont {Scheffler}}, \ and\ \bibinfo {author} {\bibfnamefont {P.}~\bibnamefont {Rinke}},\ }\href@noop {} {\bibfield  {journal} {\bibinfo  {journal} {Phys. Rev. B}\ }\textbf {\bibinfo {volume} {86}},\ \bibinfo {pages} {245127} (\bibinfo {year} {2012})}\BibitemShut {NoStop}%
\bibitem [{\citenamefont {Knight}\ \emph {et~al.}(2016)\citenamefont {Knight}, \citenamefont {Wang}, \citenamefont {Gallandi}, \citenamefont {Dolgounitcheva}, \citenamefont {Ren}, \citenamefont {Ortiz}, \citenamefont {Rinke}, \citenamefont {K\"{o}rzd\"{o}rfer},\ and\ \citenamefont {Marom}}]{exx_knight_2016}%
  \BibitemOpen
  \bibfield  {author} {\bibinfo {author} {\bibfnamefont {J.~W.}\ \bibnamefont {Knight}}, \bibinfo {author} {\bibfnamefont {X.}~\bibnamefont {Wang}}, \bibinfo {author} {\bibfnamefont {L.}~\bibnamefont {Gallandi}}, \bibinfo {author} {\bibfnamefont {O.}~\bibnamefont {Dolgounitcheva}}, \bibinfo {author} {\bibfnamefont {X.}~\bibnamefont {Ren}}, \bibinfo {author} {\bibfnamefont {J.~V.}\ \bibnamefont {Ortiz}}, \bibinfo {author} {\bibfnamefont {P.}~\bibnamefont {Rinke}}, \bibinfo {author} {\bibfnamefont {T.}~\bibnamefont {K\"{o}rzd\"{o}rfer}}, \ and\ \bibinfo {author} {\bibfnamefont {N.}~\bibnamefont {Marom}},\ }\href@noop {} {\bibfield  {journal} {\bibinfo  {journal} {J. Chem. Theory Comput.}\ }\textbf {\bibinfo {volume} {12}},\ \bibinfo {pages} {615} (\bibinfo {year} {2016})}\BibitemShut {NoStop}%
\bibitem [{\citenamefont {Grumet}\ \emph {et~al.}(2018)\citenamefont {Grumet}, \citenamefont {Liu}, \citenamefont {Kaltak}, \citenamefont {Klime\v{s}},\ and\ \citenamefont {Kresse}}]{scgw_grumet_2018}%
  \BibitemOpen
  \bibfield  {author} {\bibinfo {author} {\bibfnamefont {M.}~\bibnamefont {Grumet}}, \bibinfo {author} {\bibfnamefont {P.}~\bibnamefont {Liu}}, \bibinfo {author} {\bibfnamefont {M.}~\bibnamefont {Kaltak}}, \bibinfo {author} {\bibfnamefont {J.}~\bibnamefont {Klime\v{s}}}, \ and\ \bibinfo {author} {\bibfnamefont {G.}~\bibnamefont {Kresse}},\ }\href@noop {} {\bibfield  {journal} {\bibinfo  {journal} {Phys. Rev. B}\ }\textbf {\bibinfo {volume} {98}},\ \bibinfo {pages} {155143} (\bibinfo {year} {2018})}\BibitemShut {NoStop}%
\bibitem [{\citenamefont {Gerosa}\ \emph {et~al.}(2017)\citenamefont {Gerosa}, \citenamefont {Bottani}, \citenamefont {Valentin}, \citenamefont {Onida},\ and\ \citenamefont {Pacchioni}}]{exx_gerosa_2017}%
  \BibitemOpen
  \bibfield  {author} {\bibinfo {author} {\bibfnamefont {M.}~\bibnamefont {Gerosa}}, \bibinfo {author} {\bibfnamefont {C.~E.}\ \bibnamefont {Bottani}}, \bibinfo {author} {\bibfnamefont {C.~D.}\ \bibnamefont {Valentin}}, \bibinfo {author} {\bibfnamefont {G.}~\bibnamefont {Onida}}, \ and\ \bibinfo {author} {\bibfnamefont {G.}~\bibnamefont {Pacchioni}},\ }\href@noop {} {\bibfield  {journal} {\bibinfo  {journal} {J. Phys. Condens. Matter}\ }\textbf {\bibinfo {volume} {30}},\ \bibinfo {pages} {044003} (\bibinfo {year} {2017})}\BibitemShut {NoStop}%
\bibitem [{\citenamefont {Becke}(1993)}]{gks_becke_1993}%
  \BibitemOpen
  \bibfield  {author} {\bibinfo {author} {\bibfnamefont {A.~D.}\ \bibnamefont {Becke}},\ }\href@noop {} {\bibfield  {journal} {\bibinfo  {journal} {J. Chem. Phys.}\ }\textbf {\bibinfo {volume} {98}},\ \bibinfo {pages} {1372} (\bibinfo {year} {1993})}\BibitemShut {NoStop}%
\bibitem [{\citenamefont {Seidl}\ \emph {et~al.}(1996)\citenamefont {Seidl}, \citenamefont {G\"orling}, \citenamefont {Vogl}, \citenamefont {Majewski},\ and\ \citenamefont {Levy}}]{gks_seidl_1996}%
  \BibitemOpen
  \bibfield  {author} {\bibinfo {author} {\bibfnamefont {A.}~\bibnamefont {Seidl}}, \bibinfo {author} {\bibfnamefont {A.}~\bibnamefont {G\"orling}}, \bibinfo {author} {\bibfnamefont {P.}~\bibnamefont {Vogl}}, \bibinfo {author} {\bibfnamefont {J.~A.}\ \bibnamefont {Majewski}}, \ and\ \bibinfo {author} {\bibfnamefont {M.}~\bibnamefont {Levy}},\ }\href@noop {} {\bibfield  {journal} {\bibinfo  {journal} {Phys. Rev. B}\ }\textbf {\bibinfo {volume} {53}},\ \bibinfo {pages} {3764} (\bibinfo {year} {1996})}\BibitemShut {NoStop}%
\bibitem [{\citenamefont {Marques}\ \emph {et~al.}(2011)\citenamefont {Marques}, \citenamefont {Vidal}, \citenamefont {Oliveira}, \citenamefont {Reining},\ and\ \citenamefont {Botti}}]{hybrid_marques_2011}%
  \BibitemOpen
  \bibfield  {author} {\bibinfo {author} {\bibfnamefont {M.~A.~L.}\ \bibnamefont {Marques}}, \bibinfo {author} {\bibfnamefont {J.}~\bibnamefont {Vidal}}, \bibinfo {author} {\bibfnamefont {M.~J.~T.}\ \bibnamefont {Oliveira}}, \bibinfo {author} {\bibfnamefont {L.}~\bibnamefont {Reining}}, \ and\ \bibinfo {author} {\bibfnamefont {S.}~\bibnamefont {Botti}},\ }\href@noop {} {\bibfield  {journal} {\bibinfo  {journal} {Physical Review B}\ }\textbf {\bibinfo {volume} {83}},\ \bibinfo {pages} {035119} (\bibinfo {year} {2011})}\BibitemShut {NoStop}%
\bibitem [{\citenamefont {Skone}\ \emph {et~al.}(2014)\citenamefont {Skone}, \citenamefont {Govoni},\ and\ \citenamefont {Galli}}]{ddh_skone_2014}%
  \BibitemOpen
  \bibfield  {author} {\bibinfo {author} {\bibfnamefont {J.~H.}\ \bibnamefont {Skone}}, \bibinfo {author} {\bibfnamefont {M.}~\bibnamefont {Govoni}}, \ and\ \bibinfo {author} {\bibfnamefont {G.}~\bibnamefont {Galli}},\ }\href@noop {} {\bibfield  {journal} {\bibinfo  {journal} {Phys. Rev. B}\ }\textbf {\bibinfo {volume} {89}},\ \bibinfo {pages} {195112} (\bibinfo {year} {2014})}\BibitemShut {NoStop}%
\bibitem [{\citenamefont {Skone}\ \emph {et~al.}(2016)\citenamefont {Skone}, \citenamefont {Govoni},\ and\ \citenamefont {Galli}}]{ddh_skone_2016}%
  \BibitemOpen
  \bibfield  {author} {\bibinfo {author} {\bibfnamefont {J.~H.}\ \bibnamefont {Skone}}, \bibinfo {author} {\bibfnamefont {M.}~\bibnamefont {Govoni}}, \ and\ \bibinfo {author} {\bibfnamefont {G.}~\bibnamefont {Galli}},\ }\href@noop {} {\bibfield  {journal} {\bibinfo  {journal} {Phys. Rev. B}\ }\textbf {\bibinfo {volume} {93}},\ \bibinfo {pages} {235106} (\bibinfo {year} {2016})}\BibitemShut {NoStop}%
\bibitem [{\citenamefont {Brawand}\ \emph {et~al.}(2017)\citenamefont {Brawand}, \citenamefont {Govoni}, \citenamefont {V\"{o}r\"{o}s},\ and\ \citenamefont {Galli}}]{hybrid_brawand_2017}%
  \BibitemOpen
  \bibfield  {author} {\bibinfo {author} {\bibfnamefont {N.~P.}\ \bibnamefont {Brawand}}, \bibinfo {author} {\bibfnamefont {M.}~\bibnamefont {Govoni}}, \bibinfo {author} {\bibfnamefont {M.}~\bibnamefont {V\"{o}r\"{o}s}}, \ and\ \bibinfo {author} {\bibfnamefont {G.}~\bibnamefont {Galli}},\ }\href@noop {} {\bibfield  {journal} {\bibinfo  {journal} {J. Chem. Theory Comput.}\ }\textbf {\bibinfo {volume} {13}},\ \bibinfo {pages} {3318} (\bibinfo {year} {2017})}\BibitemShut {NoStop}%
\bibitem [{\citenamefont {Zheng}\ \emph {et~al.}(2019)\citenamefont {Zheng}, \citenamefont {Govoni},\ and\ \citenamefont {Galli}}]{hybrid_zheng_2019}%
  \BibitemOpen
  \bibfield  {author} {\bibinfo {author} {\bibfnamefont {H.}~\bibnamefont {Zheng}}, \bibinfo {author} {\bibfnamefont {M.}~\bibnamefont {Govoni}}, \ and\ \bibinfo {author} {\bibfnamefont {G.}~\bibnamefont {Galli}},\ }\href@noop {} {\bibfield  {journal} {\bibinfo  {journal} {Phys. Rev. Mater.}\ }\textbf {\bibinfo {volume} {3}},\ \bibinfo {pages} {073803} (\bibinfo {year} {2019})}\BibitemShut {NoStop}%
\bibitem [{\citenamefont {Wing}\ \emph {et~al.}(2021)\citenamefont {Wing}, \citenamefont {Ohad}, \citenamefont {Haber}, \citenamefont {Filip}, \citenamefont {Gant}, \citenamefont {Neaton},\ and\ \citenamefont {Kronik}}]{hybrid_wing_2021}%
  \BibitemOpen
  \bibfield  {author} {\bibinfo {author} {\bibfnamefont {D.}~\bibnamefont {Wing}}, \bibinfo {author} {\bibfnamefont {G.}~\bibnamefont {Ohad}}, \bibinfo {author} {\bibfnamefont {J.~B.}\ \bibnamefont {Haber}}, \bibinfo {author} {\bibfnamefont {M.~R.}\ \bibnamefont {Filip}}, \bibinfo {author} {\bibfnamefont {S.~E.}\ \bibnamefont {Gant}}, \bibinfo {author} {\bibfnamefont {J.~B.}\ \bibnamefont {Neaton}}, \ and\ \bibinfo {author} {\bibfnamefont {L.}~\bibnamefont {Kronik}},\ }\href@noop {} {\bibfield  {journal} {\bibinfo  {journal} {Proc. Natl. Acad. Sci. U.S.A.}\ }\textbf {\bibinfo {volume} {118}},\ \bibinfo {pages} {e2104556118} (\bibinfo {year} {2021})}\BibitemShut {NoStop}%
\bibitem [{\citenamefont {Shukla}\ \emph {et~al.}(2022)\citenamefont {Shukla}, \citenamefont {Jiao}, \citenamefont {Lee}, \citenamefont {Schr\"{o}der}, \citenamefont {Neaton},\ and\ \citenamefont {Hyldgaard}}]{hybrid_shukla_2022}%
  \BibitemOpen
  \bibfield  {author} {\bibinfo {author} {\bibfnamefont {V.}~\bibnamefont {Shukla}}, \bibinfo {author} {\bibfnamefont {Y.}~\bibnamefont {Jiao}}, \bibinfo {author} {\bibfnamefont {J.-H.}\ \bibnamefont {Lee}}, \bibinfo {author} {\bibfnamefont {E.}~\bibnamefont {Schr\"{o}der}}, \bibinfo {author} {\bibfnamefont {J.~B.}\ \bibnamefont {Neaton}}, \ and\ \bibinfo {author} {\bibfnamefont {P.}~\bibnamefont {Hyldgaard}},\ }\href@noop {} {\bibfield  {journal} {\bibinfo  {journal} {Phys. Rev. X}\ }\textbf {\bibinfo {volume} {12}},\ \bibinfo {pages} {041003} (\bibinfo {year} {2022})}\BibitemShut {NoStop}%
\bibitem [{\citenamefont {Camarasa-G\'{o}mez}\ \emph {et~al.}(2023)\citenamefont {Camarasa-G\'{o}mez}, \citenamefont {Ramasubramaniam}, \citenamefont {Neaton},\ and\ \citenamefont {Kronik}}]{hybrid_gomez_2023}%
  \BibitemOpen
  \bibfield  {author} {\bibinfo {author} {\bibfnamefont {M.}~\bibnamefont {Camarasa-G\'{o}mez}}, \bibinfo {author} {\bibfnamefont {A.}~\bibnamefont {Ramasubramaniam}}, \bibinfo {author} {\bibfnamefont {J.~B.}\ \bibnamefont {Neaton}}, \ and\ \bibinfo {author} {\bibfnamefont {L.}~\bibnamefont {Kronik}},\ }\href@noop {} {\bibfield  {journal} {\bibinfo  {journal} {Phys. Rev. Mater.}\ }\textbf {\bibinfo {volume} {7}},\ \bibinfo {pages} {104001} (\bibinfo {year} {2023})}\BibitemShut {NoStop}%
\bibitem [{\citenamefont {Zhan}\ \emph {et~al.}(2023)\citenamefont {Zhan}, \citenamefont {Govoni},\ and\ \citenamefont {Galli}}]{hybrid_zhan_2023}%
  \BibitemOpen
  \bibfield  {author} {\bibinfo {author} {\bibfnamefont {J.}~\bibnamefont {Zhan}}, \bibinfo {author} {\bibfnamefont {M.}~\bibnamefont {Govoni}}, \ and\ \bibinfo {author} {\bibfnamefont {G.}~\bibnamefont {Galli}},\ }\href@noop {} {\bibfield  {journal} {\bibinfo  {journal} {J. Chem. Theory Comput.}\ }\textbf {\bibinfo {volume} {19}},\ \bibinfo {pages} {5851} (\bibinfo {year} {2023})}\BibitemShut {NoStop}%
\bibitem [{\citenamefont {Rinke}\ \emph {et~al.}(2005)\citenamefont {Rinke}, \citenamefont {Qteish}, \citenamefont {Neugebauer}, \citenamefont {Freysoldt},\ and\ \citenamefont {Scheffler}}]{exx_rinke_2005}%
  \BibitemOpen
  \bibfield  {author} {\bibinfo {author} {\bibfnamefont {P.}~\bibnamefont {Rinke}}, \bibinfo {author} {\bibfnamefont {A.}~\bibnamefont {Qteish}}, \bibinfo {author} {\bibfnamefont {J.}~\bibnamefont {Neugebauer}}, \bibinfo {author} {\bibfnamefont {C.}~\bibnamefont {Freysoldt}}, \ and\ \bibinfo {author} {\bibfnamefont {M.}~\bibnamefont {Scheffler}},\ }\href@noop {} {\bibfield  {journal} {\bibinfo  {journal} {New J. Phys.}\ }\textbf {\bibinfo {volume} {7}},\ \bibinfo {pages} {126} (\bibinfo {year} {2005})}\BibitemShut {NoStop}%
\bibitem [{\citenamefont {Marom}\ \emph {et~al.}(2011)\citenamefont {Marom}, \citenamefont {Ren}, \citenamefont {Moussa}, \citenamefont {Chelikowsky},\ and\ \citenamefont {Kronik}}]{exx_marom_2011}%
  \BibitemOpen
  \bibfield  {author} {\bibinfo {author} {\bibfnamefont {N.}~\bibnamefont {Marom}}, \bibinfo {author} {\bibfnamefont {X.}~\bibnamefont {Ren}}, \bibinfo {author} {\bibfnamefont {J.~E.}\ \bibnamefont {Moussa}}, \bibinfo {author} {\bibfnamefont {J.~R.}\ \bibnamefont {Chelikowsky}}, \ and\ \bibinfo {author} {\bibfnamefont {L.}~\bibnamefont {Kronik}},\ }\href@noop {} {\bibfield  {journal} {\bibinfo  {journal} {Phys. Rev. B}\ }\textbf {\bibinfo {volume} {84}},\ \bibinfo {pages} {195143} (\bibinfo {year} {2011})}\BibitemShut {NoStop}%
\bibitem [{\citenamefont {K\"{o}rzd\"{o}rfer}\ and\ \citenamefont {Marom}(2012)}]{exx_korzdorfer_2012}%
  \BibitemOpen
  \bibfield  {author} {\bibinfo {author} {\bibfnamefont {T.}~\bibnamefont {K\"{o}rzd\"{o}rfer}}\ and\ \bibinfo {author} {\bibfnamefont {N.}~\bibnamefont {Marom}},\ }\href@noop {} {\bibfield  {journal} {\bibinfo  {journal} {Phys. Rev. B}\ }\textbf {\bibinfo {volume} {86}},\ \bibinfo {pages} {041110} (\bibinfo {year} {2012})}\BibitemShut {NoStop}%
\bibitem [{\citenamefont {Bruneval}\ and\ \citenamefont {Marques}(2013)}]{exx_bruneval_2013}%
  \BibitemOpen
  \bibfield  {author} {\bibinfo {author} {\bibfnamefont {F.}~\bibnamefont {Bruneval}}\ and\ \bibinfo {author} {\bibfnamefont {M.~A.~L.}\ \bibnamefont {Marques}},\ }\href@noop {} {\bibfield  {journal} {\bibinfo  {journal} {J. Chem. Theory Comput.}\ }\textbf {\bibinfo {volume} {9}},\ \bibinfo {pages} {324} (\bibinfo {year} {2013})}\BibitemShut {NoStop}%
\bibitem [{\citenamefont {Atalla}\ \emph {et~al.}(2013)\citenamefont {Atalla}, \citenamefont {Yoon}, \citenamefont {Caruso}, \citenamefont {Rinke},\ and\ \citenamefont {Scheffler}}]{exx_atalla_2013}%
  \BibitemOpen
  \bibfield  {author} {\bibinfo {author} {\bibfnamefont {V.}~\bibnamefont {Atalla}}, \bibinfo {author} {\bibfnamefont {M.}~\bibnamefont {Yoon}}, \bibinfo {author} {\bibfnamefont {F.}~\bibnamefont {Caruso}}, \bibinfo {author} {\bibfnamefont {P.}~\bibnamefont {Rinke}}, \ and\ \bibinfo {author} {\bibfnamefont {M.}~\bibnamefont {Scheffler}},\ }\href@noop {} {\bibfield  {journal} {\bibinfo  {journal} {Phys. Rev. B}\ }\textbf {\bibinfo {volume} {88}},\ \bibinfo {pages} {165122} (\bibinfo {year} {2013})}\BibitemShut {NoStop}%
\bibitem [{\citenamefont {L\"{u}ftner}\ \emph {et~al.}(2014)\citenamefont {L\"{u}ftner}, \citenamefont {Refaely-Abramson}, \citenamefont {Pachler}, \citenamefont {Resel}, \citenamefont {Ramsey}, \citenamefont {Kronik},\ and\ \citenamefont {Puschnig}}]{exx_luftner_2014}%
  \BibitemOpen
  \bibfield  {author} {\bibinfo {author} {\bibfnamefont {D.}~\bibnamefont {L\"{u}ftner}}, \bibinfo {author} {\bibfnamefont {S.}~\bibnamefont {Refaely-Abramson}}, \bibinfo {author} {\bibfnamefont {M.}~\bibnamefont {Pachler}}, \bibinfo {author} {\bibfnamefont {R.}~\bibnamefont {Resel}}, \bibinfo {author} {\bibfnamefont {M.~G.}\ \bibnamefont {Ramsey}}, \bibinfo {author} {\bibfnamefont {L.}~\bibnamefont {Kronik}}, \ and\ \bibinfo {author} {\bibfnamefont {P.}~\bibnamefont {Puschnig}},\ }\href@noop {} {\bibfield  {journal} {\bibinfo  {journal} {Phys. Rev. B}\ }\textbf {\bibinfo {volume} {90}},\ \bibinfo {pages} {075204} (\bibinfo {year} {2014})}\BibitemShut {NoStop}%
\bibitem [{\citenamefont {Kang}\ \emph {et~al.}(2014)\citenamefont {Kang}, \citenamefont {Kang}, \citenamefont {Nahm}, \citenamefont {Cho}, \citenamefont {Park},\ and\ \citenamefont {Han}}]{exx_kang_2014}%
  \BibitemOpen
  \bibfield  {author} {\bibinfo {author} {\bibfnamefont {Y.}~\bibnamefont {Kang}}, \bibinfo {author} {\bibfnamefont {G.}~\bibnamefont {Kang}}, \bibinfo {author} {\bibfnamefont {H.-H.}\ \bibnamefont {Nahm}}, \bibinfo {author} {\bibfnamefont {S.-H.}\ \bibnamefont {Cho}}, \bibinfo {author} {\bibfnamefont {Y.~S.}\ \bibnamefont {Park}}, \ and\ \bibinfo {author} {\bibfnamefont {S.}~\bibnamefont {Han}},\ }\href@noop {} {\bibfield  {journal} {\bibinfo  {journal} {Phys. Rev. B}\ }\textbf {\bibinfo {volume} {89}},\ \bibinfo {pages} {165130} (\bibinfo {year} {2014})}\BibitemShut {NoStop}%
\bibitem [{\citenamefont {Chen}\ and\ \citenamefont {Pasquarello}(2014)}]{exx_chen_2014}%
  \BibitemOpen
  \bibfield  {author} {\bibinfo {author} {\bibfnamefont {W.}~\bibnamefont {Chen}}\ and\ \bibinfo {author} {\bibfnamefont {A.}~\bibnamefont {Pasquarello}},\ }\href@noop {} {\bibfield  {journal} {\bibinfo  {journal} {Phys. Rev. B}\ }\textbf {\bibinfo {volume} {90}},\ \bibinfo {pages} {165133} (\bibinfo {year} {2014})}\BibitemShut {NoStop}%
\bibitem [{\citenamefont {Gallandi}\ and\ \citenamefont {K\"{o}rzd\"{o}rfer}(2015)}]{exx_gallandi_2015}%
  \BibitemOpen
  \bibfield  {author} {\bibinfo {author} {\bibfnamefont {L.}~\bibnamefont {Gallandi}}\ and\ \bibinfo {author} {\bibfnamefont {T.}~\bibnamefont {K\"{o}rzd\"{o}rfer}},\ }\href@noop {} {\bibfield  {journal} {\bibinfo  {journal} {J. Chem. Theory Comput.}\ }\textbf {\bibinfo {volume} {11}},\ \bibinfo {pages} {5391} (\bibinfo {year} {2015})}\BibitemShut {NoStop}%
\bibitem [{\citenamefont {Gallandi}\ \emph {et~al.}(2016)\citenamefont {Gallandi}, \citenamefont {Marom}, \citenamefont {Rinke},\ and\ \citenamefont {K\"{o}rzd\"{o}rfer}}]{exx_gallandi_2016}%
  \BibitemOpen
  \bibfield  {author} {\bibinfo {author} {\bibfnamefont {L.}~\bibnamefont {Gallandi}}, \bibinfo {author} {\bibfnamefont {N.}~\bibnamefont {Marom}}, \bibinfo {author} {\bibfnamefont {P.}~\bibnamefont {Rinke}}, \ and\ \bibinfo {author} {\bibfnamefont {T.}~\bibnamefont {K\"{o}rzd\"{o}rfer}},\ }\href@noop {} {\bibfield  {journal} {\bibinfo  {journal} {J. Chem. Theory Comput.}\ }\textbf {\bibinfo {volume} {12}},\ \bibinfo {pages} {605} (\bibinfo {year} {2016})}\BibitemShut {NoStop}%
\bibitem [{\citenamefont {Caruso}\ \emph {et~al.}(2016)\citenamefont {Caruso}, \citenamefont {Dauth}, \citenamefont {Van~Setten},\ and\ \citenamefont {Rinke}}]{exx_caruso_2016}%
  \BibitemOpen
  \bibfield  {author} {\bibinfo {author} {\bibfnamefont {F.}~\bibnamefont {Caruso}}, \bibinfo {author} {\bibfnamefont {M.}~\bibnamefont {Dauth}}, \bibinfo {author} {\bibfnamefont {M.~J.}\ \bibnamefont {Van~Setten}}, \ and\ \bibinfo {author} {\bibfnamefont {P.}~\bibnamefont {Rinke}},\ }\href@noop {} {\bibfield  {journal} {\bibinfo  {journal} {J. Chem. Theory Comput.}\ }\textbf {\bibinfo {volume} {12}},\ \bibinfo {pages} {5076} (\bibinfo {year} {2016})}\BibitemShut {NoStop}%
\bibitem [{\citenamefont {Dauth}\ \emph {et~al.}(2016)\citenamefont {Dauth}, \citenamefont {Caruso}, \citenamefont {K\"{u}mmel},\ and\ \citenamefont {Rinke}}]{exx_dauth_2016}%
  \BibitemOpen
  \bibfield  {author} {\bibinfo {author} {\bibfnamefont {M.}~\bibnamefont {Dauth}}, \bibinfo {author} {\bibfnamefont {F.}~\bibnamefont {Caruso}}, \bibinfo {author} {\bibfnamefont {S.}~\bibnamefont {K\"{u}mmel}}, \ and\ \bibinfo {author} {\bibfnamefont {P.}~\bibnamefont {Rinke}},\ }\href@noop {} {\bibfield  {journal} {\bibinfo  {journal} {Phys. Rev. B}\ }\textbf {\bibinfo {volume} {93}},\ \bibinfo {pages} {121115} (\bibinfo {year} {2016})}\BibitemShut {NoStop}%
\bibitem [{\citenamefont {Bois}\ and\ \citenamefont {K\"{o}rzd\"{o}rfer}(2017)}]{exx_bois_2017}%
  \BibitemOpen
  \bibfield  {author} {\bibinfo {author} {\bibfnamefont {J.}~\bibnamefont {Bois}}\ and\ \bibinfo {author} {\bibfnamefont {T.}~\bibnamefont {K\"{o}rzd\"{o}rfer}},\ }\href@noop {} {\bibfield  {journal} {\bibinfo  {journal} {J. Chem. Theory Comput.}\ }\textbf {\bibinfo {volume} {13}},\ \bibinfo {pages} {4962} (\bibinfo {year} {2017})}\BibitemShut {NoStop}%
\bibitem [{\citenamefont {Chen}\ and\ \citenamefont {Pasquarello}(2017)}]{exx_chen_2017}%
  \BibitemOpen
  \bibfield  {author} {\bibinfo {author} {\bibfnamefont {W.}~\bibnamefont {Chen}}\ and\ \bibinfo {author} {\bibfnamefont {A.}~\bibnamefont {Pasquarello}},\ }\href@noop {} {\bibfield  {journal} {\bibinfo  {journal} {Phys. Rev. B}\ }\textbf {\bibinfo {volume} {96}},\ \bibinfo {pages} {020101} (\bibinfo {year} {2017})}\BibitemShut {NoStop}%
\bibitem [{\citenamefont {Leppert}\ \emph {et~al.}(2019)\citenamefont {Leppert}, \citenamefont {Rangel},\ and\ \citenamefont {Neaton}}]{exx_leppert_2019}%
  \BibitemOpen
  \bibfield  {author} {\bibinfo {author} {\bibfnamefont {L.}~\bibnamefont {Leppert}}, \bibinfo {author} {\bibfnamefont {T.}~\bibnamefont {Rangel}}, \ and\ \bibinfo {author} {\bibfnamefont {J.~B.}\ \bibnamefont {Neaton}},\ }\href@noop {} {\bibfield  {journal} {\bibinfo  {journal} {Phys. Rev. Mater.}\ }\textbf {\bibinfo {volume} {3}},\ \bibinfo {pages} {103803} (\bibinfo {year} {2019})}\BibitemShut {NoStop}%
\bibitem [{\citenamefont {Gant}\ \emph {et~al.}(2022)\citenamefont {Gant}, \citenamefont {Haber}, \citenamefont {Filip}, \citenamefont {Sagredo}, \citenamefont {Wing}, \citenamefont {Ohad}, \citenamefont {Kronik},\ and\ \citenamefont {Neaton}}]{exx_gant_2022}%
  \BibitemOpen
  \bibfield  {author} {\bibinfo {author} {\bibfnamefont {S.~E.}\ \bibnamefont {Gant}}, \bibinfo {author} {\bibfnamefont {J.~B.}\ \bibnamefont {Haber}}, \bibinfo {author} {\bibfnamefont {M.~R.}\ \bibnamefont {Filip}}, \bibinfo {author} {\bibfnamefont {F.}~\bibnamefont {Sagredo}}, \bibinfo {author} {\bibfnamefont {D.}~\bibnamefont {Wing}}, \bibinfo {author} {\bibfnamefont {G.}~\bibnamefont {Ohad}}, \bibinfo {author} {\bibfnamefont {L.}~\bibnamefont {Kronik}}, \ and\ \bibinfo {author} {\bibfnamefont {J.~B.}\ \bibnamefont {Neaton}},\ }\href@noop {} {\bibfield  {journal} {\bibinfo  {journal} {Phys. Rev. Mater.}\ }\textbf {\bibinfo {volume} {6}},\ \bibinfo {pages} {053802} (\bibinfo {year} {2022})}\BibitemShut {NoStop}%
\bibitem [{\citenamefont {McKeon}\ \emph {et~al.}(2022)\citenamefont {McKeon}, \citenamefont {Hamed}, \citenamefont {Bruneval},\ and\ \citenamefont {Neaton}}]{exx_mckeon_2022}%
  \BibitemOpen
  \bibfield  {author} {\bibinfo {author} {\bibfnamefont {C.~A.}\ \bibnamefont {McKeon}}, \bibinfo {author} {\bibfnamefont {S.~M.}\ \bibnamefont {Hamed}}, \bibinfo {author} {\bibfnamefont {F.}~\bibnamefont {Bruneval}}, \ and\ \bibinfo {author} {\bibfnamefont {J.~B.}\ \bibnamefont {Neaton}},\ }\href@noop {} {\bibfield  {journal} {\bibinfo  {journal} {J. Chem. Phys.}\ }\textbf {\bibinfo {volume} {157}},\ \bibinfo {pages} {074103} (\bibinfo {year} {2022})}\BibitemShut {NoStop}%
\bibitem [{\citenamefont {Ohad}\ \emph {et~al.}(2023)\citenamefont {Ohad}, \citenamefont {Gant}, \citenamefont {Wing}, \citenamefont {Haber}, \citenamefont {Camarasa-G\'{o}mez}, \citenamefont {Sagredo}, \citenamefont {Filip}, \citenamefont {Neaton},\ and\ \citenamefont {Kronik}}]{exx_ohad_2023}%
  \BibitemOpen
  \bibfield  {author} {\bibinfo {author} {\bibfnamefont {G.}~\bibnamefont {Ohad}}, \bibinfo {author} {\bibfnamefont {S.~E.}\ \bibnamefont {Gant}}, \bibinfo {author} {\bibfnamefont {D.}~\bibnamefont {Wing}}, \bibinfo {author} {\bibfnamefont {J.~B.}\ \bibnamefont {Haber}}, \bibinfo {author} {\bibfnamefont {M.}~\bibnamefont {Camarasa-G\'{o}mez}}, \bibinfo {author} {\bibfnamefont {F.}~\bibnamefont {Sagredo}}, \bibinfo {author} {\bibfnamefont {M.~R.}\ \bibnamefont {Filip}}, \bibinfo {author} {\bibfnamefont {J.~B.}\ \bibnamefont {Neaton}}, \ and\ \bibinfo {author} {\bibfnamefont {L.}~\bibnamefont {Kronik}},\ }\href@noop {} {\bibfield  {journal} {\bibinfo  {journal} {Phys. Rev. Mater.}\ }\textbf {\bibinfo {volume} {7}},\ \bibinfo {pages} {123803} (\bibinfo {year} {2023})}\BibitemShut {NoStop}%
\bibitem [{\citenamefont {Gaiduk}\ \emph {et~al.}(2016)\citenamefont {Gaiduk}, \citenamefont {Govoni}, \citenamefont {Seidel}, \citenamefont {Skone}, \citenamefont {Winter},\ and\ \citenamefont {Galli}}]{exx_gaiduk_2016}%
  \BibitemOpen
  \bibfield  {author} {\bibinfo {author} {\bibfnamefont {A.~P.}\ \bibnamefont {Gaiduk}}, \bibinfo {author} {\bibfnamefont {M.}~\bibnamefont {Govoni}}, \bibinfo {author} {\bibfnamefont {R.}~\bibnamefont {Seidel}}, \bibinfo {author} {\bibfnamefont {J.~H.}\ \bibnamefont {Skone}}, \bibinfo {author} {\bibfnamefont {B.}~\bibnamefont {Winter}}, \ and\ \bibinfo {author} {\bibfnamefont {G.}~\bibnamefont {Galli}},\ }\href@noop {} {\bibfield  {journal} {\bibinfo  {journal} {J. Am. Chem. Soc.}\ }\textbf {\bibinfo {volume} {138}},\ \bibinfo {pages} {6912} (\bibinfo {year} {2016})}\BibitemShut {NoStop}%
\bibitem [{\citenamefont {Pham}\ \emph {et~al.}(2017)\citenamefont {Pham}, \citenamefont {Govoni}, \citenamefont {Seidel}, \citenamefont {Bradforth}, \citenamefont {Schwegler},\ and\ \citenamefont {Galli}}]{exx_pham_2017}%
  \BibitemOpen
  \bibfield  {author} {\bibinfo {author} {\bibfnamefont {T.~A.}\ \bibnamefont {Pham}}, \bibinfo {author} {\bibfnamefont {M.}~\bibnamefont {Govoni}}, \bibinfo {author} {\bibfnamefont {R.}~\bibnamefont {Seidel}}, \bibinfo {author} {\bibfnamefont {S.~E.}\ \bibnamefont {Bradforth}}, \bibinfo {author} {\bibfnamefont {E.}~\bibnamefont {Schwegler}}, \ and\ \bibinfo {author} {\bibfnamefont {G.}~\bibnamefont {Galli}},\ }\href@noop {} {\bibfield  {journal} {\bibinfo  {journal} {Sci. Adv.}\ }\textbf {\bibinfo {volume} {3}},\ \bibinfo {pages} {e1603210} (\bibinfo {year} {2017})}\BibitemShut {NoStop}%
\bibitem [{\citenamefont {Wu}\ \emph {et~al.}(2009)\citenamefont {Wu}, \citenamefont {Selloni},\ and\ \citenamefont {Car}}]{exx_wu_2009}%
  \BibitemOpen
  \bibfield  {author} {\bibinfo {author} {\bibfnamefont {X.}~\bibnamefont {Wu}}, \bibinfo {author} {\bibfnamefont {A.}~\bibnamefont {Selloni}}, \ and\ \bibinfo {author} {\bibfnamefont {R.}~\bibnamefont {Car}},\ }\href@noop {} {\bibfield  {journal} {\bibinfo  {journal} {Phys. Rev. B}\ }\textbf {\bibinfo {volume} {79}},\ \bibinfo {pages} {085102} (\bibinfo {year} {2009})}\BibitemShut {NoStop}%
\bibitem [{\citenamefont {Gygi}\ and\ \citenamefont {Duchemin}(2013)}]{exx_gygi_2013}%
  \BibitemOpen
  \bibfield  {author} {\bibinfo {author} {\bibfnamefont {F.}~\bibnamefont {Gygi}}\ and\ \bibinfo {author} {\bibfnamefont {I.}~\bibnamefont {Duchemin}},\ }\href@noop {} {\bibfield  {journal} {\bibinfo  {journal} {J. Chem. Theory Comput.}\ }\textbf {\bibinfo {volume} {9}},\ \bibinfo {pages} {582} (\bibinfo {year} {2013})}\BibitemShut {NoStop}%
\bibitem [{\citenamefont {Damle}\ \emph {et~al.}(2015)\citenamefont {Damle}, \citenamefont {Lin},\ and\ \citenamefont {Ying}}]{exx_damle_2015}%
  \BibitemOpen
  \bibfield  {author} {\bibinfo {author} {\bibfnamefont {A.}~\bibnamefont {Damle}}, \bibinfo {author} {\bibfnamefont {L.}~\bibnamefont {Lin}}, \ and\ \bibinfo {author} {\bibfnamefont {L.}~\bibnamefont {Ying}},\ }\href@noop {} {\bibfield  {journal} {\bibinfo  {journal} {J. Chem. Theory Comput.}\ }\textbf {\bibinfo {volume} {11}},\ \bibinfo {pages} {1463} (\bibinfo {year} {2015})}\BibitemShut {NoStop}%
\bibitem [{\citenamefont {Lin}(2016)}]{ace_lin_2016}%
  \BibitemOpen
  \bibfield  {author} {\bibinfo {author} {\bibfnamefont {L.}~\bibnamefont {Lin}},\ }\href@noop {} {\bibfield  {journal} {\bibinfo  {journal} {J. Chem. Theory Comput.}\ }\textbf {\bibinfo {volume} {12}},\ \bibinfo {pages} {2242} (\bibinfo {year} {2016})}\BibitemShut {NoStop}%
\bibitem [{\citenamefont {Dawson}\ and\ \citenamefont {Gygi}(2015)}]{exx_dawson_2015}%
  \BibitemOpen
  \bibfield  {author} {\bibinfo {author} {\bibfnamefont {W.}~\bibnamefont {Dawson}}\ and\ \bibinfo {author} {\bibfnamefont {F.}~\bibnamefont {Gygi}},\ }\href@noop {} {\bibfield  {journal} {\bibinfo  {journal} {J. Chem. Theory Comput.}\ }\textbf {\bibinfo {volume} {11}},\ \bibinfo {pages} {4655} (\bibinfo {year} {2015})}\BibitemShut {NoStop}%
\bibitem [{\citenamefont {Hu}\ \emph {et~al.}(2017)\citenamefont {Hu}, \citenamefont {Lin},\ and\ \citenamefont {Yang}}]{exx_hu_2017}%
  \BibitemOpen
  \bibfield  {author} {\bibinfo {author} {\bibfnamefont {W.}~\bibnamefont {Hu}}, \bibinfo {author} {\bibfnamefont {L.}~\bibnamefont {Lin}}, \ and\ \bibinfo {author} {\bibfnamefont {C.}~\bibnamefont {Yang}},\ }\href@noop {} {\bibfield  {journal} {\bibinfo  {journal} {J. Chem. Theory Comput.}\ }\textbf {\bibinfo {volume} {13}},\ \bibinfo {pages} {5420} (\bibinfo {year} {2017})}\BibitemShut {NoStop}%
\bibitem [{\citenamefont {Ko}\ \emph {et~al.}(2020)\citenamefont {Ko}, \citenamefont {Jia}, \citenamefont {Santra}, \citenamefont {Wu}, \citenamefont {Car},\ and\ \citenamefont {DiStasio~Jr}}]{exx_ko_2020}%
  \BibitemOpen
  \bibfield  {author} {\bibinfo {author} {\bibfnamefont {H.-Y.}\ \bibnamefont {Ko}}, \bibinfo {author} {\bibfnamefont {J.}~\bibnamefont {Jia}}, \bibinfo {author} {\bibfnamefont {B.}~\bibnamefont {Santra}}, \bibinfo {author} {\bibfnamefont {X.}~\bibnamefont {Wu}}, \bibinfo {author} {\bibfnamefont {R.}~\bibnamefont {Car}}, \ and\ \bibinfo {author} {\bibfnamefont {R.~A.}\ \bibnamefont {DiStasio~Jr}},\ }\href@noop {} {\bibfield  {journal} {\bibinfo  {journal} {J. Chem. Theory Comput.}\ }\textbf {\bibinfo {volume} {16}},\ \bibinfo {pages} {3757} (\bibinfo {year} {2020})}\BibitemShut {NoStop}%
\bibitem [{\citenamefont {Wu}\ \emph {et~al.}(2022)\citenamefont {Wu}, \citenamefont {Qin}, \citenamefont {Hu},\ and\ \citenamefont {Yang}}]{ace_wu_2022}%
  \BibitemOpen
  \bibfield  {author} {\bibinfo {author} {\bibfnamefont {K.}~\bibnamefont {Wu}}, \bibinfo {author} {\bibfnamefont {X.}~\bibnamefont {Qin}}, \bibinfo {author} {\bibfnamefont {W.}~\bibnamefont {Hu}}, \ and\ \bibinfo {author} {\bibfnamefont {J.}~\bibnamefont {Yang}},\ }\href@noop {} {\bibfield  {journal} {\bibinfo  {journal} {J. Chem. Theory Comput.}\ }\textbf {\bibinfo {volume} {18}},\ \bibinfo {pages} {206} (\bibinfo {year} {2022})}\BibitemShut {NoStop}%
\bibitem [{\citenamefont {Liu}\ \emph {et~al.}(2022)\citenamefont {Liu}, \citenamefont {Hu},\ and\ \citenamefont {Yang}}]{ace_liu_2022}%
  \BibitemOpen
  \bibfield  {author} {\bibinfo {author} {\bibfnamefont {J.}~\bibnamefont {Liu}}, \bibinfo {author} {\bibfnamefont {W.}~\bibnamefont {Hu}}, \ and\ \bibinfo {author} {\bibfnamefont {J.}~\bibnamefont {Yang}},\ }\href@noop {} {\bibfield  {journal} {\bibinfo  {journal} {J. Chem. Theory Comput.}\ }\textbf {\bibinfo {volume} {18}},\ \bibinfo {pages} {6713} (\bibinfo {year} {2022})}\BibitemShut {NoStop}%
\bibitem [{\citenamefont {Li}\ \emph {et~al.}(2023)\citenamefont {Li}, \citenamefont {Wan}, \citenamefont {Jiao}, \citenamefont {Hu},\ and\ \citenamefont {Yang}}]{ace_li_2023}%
  \BibitemOpen
  \bibfield  {author} {\bibinfo {author} {\bibfnamefont {J.}~\bibnamefont {Li}}, \bibinfo {author} {\bibfnamefont {L.}~\bibnamefont {Wan}}, \bibinfo {author} {\bibfnamefont {S.}~\bibnamefont {Jiao}}, \bibinfo {author} {\bibfnamefont {W.}~\bibnamefont {Hu}}, \ and\ \bibinfo {author} {\bibfnamefont {J.}~\bibnamefont {Yang}},\ }\href@noop {} {\bibfield  {journal} {\bibinfo  {journal} {Electron. Struct.}\ }\textbf {\bibinfo {volume} {5}},\ \bibinfo {pages} {014008} (\bibinfo {year} {2023})}\BibitemShut {NoStop}%
\bibitem [{\citenamefont {Chen}\ \emph {et~al.}(2023)\citenamefont {Chen}, \citenamefont {Wu}, \citenamefont {Hu},\ and\ \citenamefont {Yang}}]{ace_chen_2023}%
  \BibitemOpen
  \bibfield  {author} {\bibinfo {author} {\bibfnamefont {S.}~\bibnamefont {Chen}}, \bibinfo {author} {\bibfnamefont {K.}~\bibnamefont {Wu}}, \bibinfo {author} {\bibfnamefont {W.}~\bibnamefont {Hu}}, \ and\ \bibinfo {author} {\bibfnamefont {J.}~\bibnamefont {Yang}},\ }\href@noop {} {\bibfield  {journal} {\bibinfo  {journal} {J. Chem. Phys.}\ }\textbf {\bibinfo {volume} {158}},\ \bibinfo {pages} {134106} (\bibinfo {year} {2023})}\BibitemShut {NoStop}%
\bibitem [{\citenamefont {Wilson}\ \emph {et~al.}(2008)\citenamefont {Wilson}, \citenamefont {Gygi},\ and\ \citenamefont {Galli}}]{dielectric_wilson_2008}%
  \BibitemOpen
  \bibfield  {author} {\bibinfo {author} {\bibfnamefont {H.~F.}\ \bibnamefont {Wilson}}, \bibinfo {author} {\bibfnamefont {F.}~\bibnamefont {Gygi}}, \ and\ \bibinfo {author} {\bibfnamefont {G.}~\bibnamefont {Galli}},\ }\href@noop {} {\bibfield  {journal} {\bibinfo  {journal} {Phys. Rev. B}\ }\textbf {\bibinfo {volume} {78}},\ \bibinfo {pages} {113303} (\bibinfo {year} {2008})}\BibitemShut {NoStop}%
\bibitem [{\citenamefont {Wilson}\ \emph {et~al.}(2009)\citenamefont {Wilson}, \citenamefont {Lu}, \citenamefont {Gygi},\ and\ \citenamefont {Galli}}]{dielectric_wilson_2009}%
  \BibitemOpen
  \bibfield  {author} {\bibinfo {author} {\bibfnamefont {H.~F.}\ \bibnamefont {Wilson}}, \bibinfo {author} {\bibfnamefont {D.}~\bibnamefont {Lu}}, \bibinfo {author} {\bibfnamefont {F.}~\bibnamefont {Gygi}}, \ and\ \bibinfo {author} {\bibfnamefont {G.}~\bibnamefont {Galli}},\ }\href@noop {} {\bibfield  {journal} {\bibinfo  {journal} {Phys. Rev. B}\ }\textbf {\bibinfo {volume} {79}},\ \bibinfo {pages} {245106} (\bibinfo {year} {2009})}\BibitemShut {NoStop}%
\bibitem [{\citenamefont {Malc{\i}o\v{g}lu}\ \emph {et~al.}(2011)\citenamefont {Malc{\i}o\v{g}lu}, \citenamefont {Gebauer}, \citenamefont {Rocca},\ and\ \citenamefont {Baroni}}]{turbotddft_malcioglu_2011}%
  \BibitemOpen
  \bibfield  {author} {\bibinfo {author} {\bibfnamefont {O.~B.}\ \bibnamefont {Malc{\i}o\v{g}lu}}, \bibinfo {author} {\bibfnamefont {R.}~\bibnamefont {Gebauer}}, \bibinfo {author} {\bibfnamefont {D.}~\bibnamefont {Rocca}}, \ and\ \bibinfo {author} {\bibfnamefont {S.}~\bibnamefont {Baroni}},\ }\href@noop {} {\bibfield  {journal} {\bibinfo  {journal} {Comput. Phys. Commun.}\ }\textbf {\bibinfo {volume} {182}},\ \bibinfo {pages} {1744} (\bibinfo {year} {2011})}\BibitemShut {NoStop}%
\bibitem [{\citenamefont {Godby}\ \emph {et~al.}(1988)\citenamefont {Godby}, \citenamefont {Schl\"{u}ter},\ and\ \citenamefont {Sham}}]{contour_godby_1988}%
  \BibitemOpen
  \bibfield  {author} {\bibinfo {author} {\bibfnamefont {R.~W.}\ \bibnamefont {Godby}}, \bibinfo {author} {\bibfnamefont {M.}~\bibnamefont {Schl\"{u}ter}}, \ and\ \bibinfo {author} {\bibfnamefont {L.~J.}\ \bibnamefont {Sham}},\ }\href@noop {} {\bibfield  {journal} {\bibinfo  {journal} {Phys. Rev. B}\ }\textbf {\bibinfo {volume} {37}},\ \bibinfo {pages} {10159} (\bibinfo {year} {1988})}\BibitemShut {NoStop}%
\bibitem [{\citenamefont {Leb\`{e}gue}\ \emph {et~al.}(2003)\citenamefont {Leb\`{e}gue}, \citenamefont {Arnaud}, \citenamefont {Alouani},\ and\ \citenamefont {Bloechl}}]{contour_lebegue_2003}%
  \BibitemOpen
  \bibfield  {author} {\bibinfo {author} {\bibfnamefont {S.}~\bibnamefont {Leb\`{e}gue}}, \bibinfo {author} {\bibfnamefont {B.}~\bibnamefont {Arnaud}}, \bibinfo {author} {\bibfnamefont {M.}~\bibnamefont {Alouani}}, \ and\ \bibinfo {author} {\bibfnamefont {P.~E.}\ \bibnamefont {Bloechl}},\ }\href@noop {} {\bibfield  {journal} {\bibinfo  {journal} {Phys. Rev. B}\ }\textbf {\bibinfo {volume} {67}},\ \bibinfo {pages} {155208} (\bibinfo {year} {2003})}\BibitemShut {NoStop}%
\bibitem [{\citenamefont {Baroni}\ \emph {et~al.}(1987)\citenamefont {Baroni}, \citenamefont {Giannozzi},\ and\ \citenamefont {Testa}}]{dfpt_baroni_1987}%
  \BibitemOpen
  \bibfield  {author} {\bibinfo {author} {\bibfnamefont {S.}~\bibnamefont {Baroni}}, \bibinfo {author} {\bibfnamefont {P.}~\bibnamefont {Giannozzi}}, \ and\ \bibinfo {author} {\bibfnamefont {A.}~\bibnamefont {Testa}},\ }\href@noop {} {\bibfield  {journal} {\bibinfo  {journal} {Phys. Rev. Lett.}\ }\textbf {\bibinfo {volume} {58}},\ \bibinfo {pages} {1861} (\bibinfo {year} {1987})}\BibitemShut {NoStop}%
\bibitem [{\citenamefont {Baroni}\ \emph {et~al.}(2001)\citenamefont {Baroni}, \citenamefont {de~Gironcoli}, \citenamefont {Dal~Corso},\ and\ \citenamefont {Giannozzi}}]{dfpt_baroni_2001}%
  \BibitemOpen
  \bibfield  {author} {\bibinfo {author} {\bibfnamefont {S.}~\bibnamefont {Baroni}}, \bibinfo {author} {\bibfnamefont {S.}~\bibnamefont {de~Gironcoli}}, \bibinfo {author} {\bibfnamefont {A.}~\bibnamefont {Dal~Corso}}, \ and\ \bibinfo {author} {\bibfnamefont {P.}~\bibnamefont {Giannozzi}},\ }\href@noop {} {\bibfield  {journal} {\bibinfo  {journal} {Rev. Mod. Phys.}\ }\textbf {\bibinfo {volume} {73}},\ \bibinfo {pages} {515} (\bibinfo {year} {2001})}\BibitemShut {NoStop}%
\bibitem [{\citenamefont {Sternheimer}(1954)}]{sternheimer_sternheimer_1954}%
  \BibitemOpen
  \bibfield  {author} {\bibinfo {author} {\bibfnamefont {R.~M.}\ \bibnamefont {Sternheimer}},\ }\href@noop {} {\bibfield  {journal} {\bibinfo  {journal} {Phys. Rev.}\ }\textbf {\bibinfo {volume} {96}},\ \bibinfo {pages} {951} (\bibinfo {year} {1954})}\BibitemShut {NoStop}%
\bibitem [{\citenamefont {Adler}(1962)}]{dielectric_adler_1962}%
  \BibitemOpen
  \bibfield  {author} {\bibinfo {author} {\bibfnamefont {S.~L.}\ \bibnamefont {Adler}},\ }\href@noop {} {\bibfield  {journal} {\bibinfo  {journal} {Phys. Rev.}\ }\textbf {\bibinfo {volume} {126}},\ \bibinfo {pages} {413} (\bibinfo {year} {1962})}\BibitemShut {NoStop}%
\bibitem [{\citenamefont {Wiser}(1963)}]{dielectric_wiser_1963}%
  \BibitemOpen
  \bibfield  {author} {\bibinfo {author} {\bibfnamefont {N.}~\bibnamefont {Wiser}},\ }\href@noop {} {\bibfield  {journal} {\bibinfo  {journal} {Phys. Rev.}\ }\textbf {\bibinfo {volume} {129}},\ \bibinfo {pages} {62} (\bibinfo {year} {1963})}\BibitemShut {NoStop}%
\bibitem [{\citenamefont {Walker}\ \emph {et~al.}(2006)\citenamefont {Walker}, \citenamefont {Saitta}, \citenamefont {Gebauer},\ and\ \citenamefont {Baroni}}]{lanczos_walker_2006}%
  \BibitemOpen
  \bibfield  {author} {\bibinfo {author} {\bibfnamefont {B.}~\bibnamefont {Walker}}, \bibinfo {author} {\bibfnamefont {A.~M.}\ \bibnamefont {Saitta}}, \bibinfo {author} {\bibfnamefont {R.}~\bibnamefont {Gebauer}}, \ and\ \bibinfo {author} {\bibfnamefont {S.}~\bibnamefont {Baroni}},\ }\href@noop {} {\bibfield  {journal} {\bibinfo  {journal} {Phys. Rev. Lett.}\ }\textbf {\bibinfo {volume} {96}},\ \bibinfo {pages} {113001} (\bibinfo {year} {2006})}\BibitemShut {NoStop}%
\bibitem [{\citenamefont {Rocca}\ \emph {et~al.}(2008)\citenamefont {Rocca}, \citenamefont {Gebauer}, \citenamefont {Saad},\ and\ \citenamefont {Baroni}}]{lanczos_rocca_2008}%
  \BibitemOpen
  \bibfield  {author} {\bibinfo {author} {\bibfnamefont {D.}~\bibnamefont {Rocca}}, \bibinfo {author} {\bibfnamefont {R.}~\bibnamefont {Gebauer}}, \bibinfo {author} {\bibfnamefont {Y.}~\bibnamefont {Saad}}, \ and\ \bibinfo {author} {\bibfnamefont {S.}~\bibnamefont {Baroni}},\ }\href@noop {} {\bibfield  {journal} {\bibinfo  {journal} {J. Chem. Phys.}\ }\textbf {\bibinfo {volume} {128}},\ \bibinfo {pages} {154105} (\bibinfo {year} {2008})}\BibitemShut {NoStop}%
\bibitem [{\citenamefont {Hirata}\ and\ \citenamefont {Head-Gordon}(1999)}]{tda_hirata_1999}%
  \BibitemOpen
  \bibfield  {author} {\bibinfo {author} {\bibfnamefont {S.}~\bibnamefont {Hirata}}\ and\ \bibinfo {author} {\bibfnamefont {M.}~\bibnamefont {Head-Gordon}},\ }\href@noop {} {\bibfield  {journal} {\bibinfo  {journal} {Chem. Phys. Lett.}\ }\textbf {\bibinfo {volume} {314}},\ \bibinfo {pages} {291} (\bibinfo {year} {1999})}\BibitemShut {NoStop}%
\bibitem [{\citenamefont {Gygi}(2009)}]{bisection_gygi_2009}%
  \BibitemOpen
  \bibfield  {author} {\bibinfo {author} {\bibfnamefont {F.}~\bibnamefont {Gygi}},\ }\href@noop {} {\bibfield  {journal} {\bibinfo  {journal} {Phys. Rev. Lett.}\ }\textbf {\bibinfo {volume} {102}},\ \bibinfo {pages} {166406} (\bibinfo {year} {2009})}\BibitemShut {NoStop}%
\bibitem [{\citenamefont {Giannozzi}\ \emph {et~al.}(2020)\citenamefont {Giannozzi}, \citenamefont {Baseggio}, \citenamefont {Bonf\`{a}}, \citenamefont {Brunato}, \citenamefont {Car}, \citenamefont {Carnimeo}, \citenamefont {Cavazzoni}, \citenamefont {de~Gironcoli}, \citenamefont {Delugas}, \citenamefont {Ruffino}, \citenamefont {Ferretti}, \citenamefont {Marzari}, \citenamefont {Timrov}, \citenamefont {Urru},\ and\ \citenamefont {Baroni}}]{qe_giannozzi_2020}%
  \BibitemOpen
  \bibfield  {author} {\bibinfo {author} {\bibfnamefont {P.}~\bibnamefont {Giannozzi}}, \bibinfo {author} {\bibfnamefont {O.}~\bibnamefont {Baseggio}}, \bibinfo {author} {\bibfnamefont {P.}~\bibnamefont {Bonf\`{a}}}, \bibinfo {author} {\bibfnamefont {D.}~\bibnamefont {Brunato}}, \bibinfo {author} {\bibfnamefont {R.}~\bibnamefont {Car}}, \bibinfo {author} {\bibfnamefont {I.}~\bibnamefont {Carnimeo}}, \bibinfo {author} {\bibfnamefont {C.}~\bibnamefont {Cavazzoni}}, \bibinfo {author} {\bibfnamefont {S.}~\bibnamefont {de~Gironcoli}}, \bibinfo {author} {\bibfnamefont {P.}~\bibnamefont {Delugas}}, \bibinfo {author} {\bibfnamefont {F.~F.}\ \bibnamefont {Ruffino}}, \bibinfo {author} {\bibfnamefont {A.}~\bibnamefont {Ferretti}}, \bibinfo {author} {\bibfnamefont {N.}~\bibnamefont {Marzari}}, \bibinfo {author} {\bibfnamefont {I.}~\bibnamefont {Timrov}}, \bibinfo {author} {\bibfnamefont {A.}~\bibnamefont {Urru}}, \ and\ \bibinfo {author} {\bibfnamefont {S.}~\bibnamefont {Baroni}},\ }\href@noop {} {\bibfield  {journal}
  {\bibinfo  {journal} {J. Chem. Phys.}\ }\textbf {\bibinfo {volume} {152}},\ \bibinfo {pages} {154105} (\bibinfo {year} {2020})}\BibitemShut {NoStop}%
\bibitem [{\citenamefont {Hamann}(2013)}]{oncv_hamann_2013}%
  \BibitemOpen
  \bibfield  {author} {\bibinfo {author} {\bibfnamefont {D.~R.}\ \bibnamefont {Hamann}},\ }\href@noop {} {\bibfield  {journal} {\bibinfo  {journal} {Phys. Rev. B}\ }\textbf {\bibinfo {volume} {88}},\ \bibinfo {pages} {085117} (\bibinfo {year} {2013})}\BibitemShut {NoStop}%
\bibitem [{\citenamefont {Schlipf}\ and\ \citenamefont {Gygi}(2015)}]{oncv_schlipf_2015}%
  \BibitemOpen
  \bibfield  {author} {\bibinfo {author} {\bibfnamefont {M.}~\bibnamefont {Schlipf}}\ and\ \bibinfo {author} {\bibfnamefont {F.}~\bibnamefont {Gygi}},\ }\href@noop {} {\bibfield  {journal} {\bibinfo  {journal} {Comput. Phys. Commun.}\ }\textbf {\bibinfo {volume} {196}},\ \bibinfo {pages} {36} (\bibinfo {year} {2015})}\BibitemShut {NoStop}%
\bibitem [{\citenamefont {Govoni}\ \emph {et~al.}(2019)\citenamefont {Govoni}, \citenamefont {Munakami}, \citenamefont {Tanikanti}, \citenamefont {Skone}, \citenamefont {Runesha}, \citenamefont {Giberti}, \citenamefont {de~Pablo},\ and\ \citenamefont {Galli}}]{qresp_govoni_2019}%
  \BibitemOpen
  \bibfield  {author} {\bibinfo {author} {\bibfnamefont {M.}~\bibnamefont {Govoni}}, \bibinfo {author} {\bibfnamefont {M.}~\bibnamefont {Munakami}}, \bibinfo {author} {\bibfnamefont {A.}~\bibnamefont {Tanikanti}}, \bibinfo {author} {\bibfnamefont {J.~H.}\ \bibnamefont {Skone}}, \bibinfo {author} {\bibfnamefont {H.~B.}\ \bibnamefont {Runesha}}, \bibinfo {author} {\bibfnamefont {F.}~\bibnamefont {Giberti}}, \bibinfo {author} {\bibfnamefont {J.}~\bibnamefont {de~Pablo}}, \ and\ \bibinfo {author} {\bibfnamefont {G.}~\bibnamefont {Galli}},\ }\href@noop {} {\bibfield  {journal} {\bibinfo  {journal} {Sci. Data}\ }\textbf {\bibinfo {volume} {6}},\ \bibinfo {pages} {190002} (\bibinfo {year} {2019})}\BibitemShut {NoStop}%
\end{thebibliography}%

\end{document}